\documentclass[reprint, superscriptaddress, floatfix, amsmath,amssymb, aps, prx, noeprint]{revtex4-1}

\usepackage[utf8]{inputenc}
\usepackage{amsmath,esint}
\usepackage{mathtools}
\usepackage[colorlinks=true,linkcolor=blue,citecolor=blue,urlcolor=blue]{hyperref}
\usepackage{amsfonts}
\usepackage{amssymb}
\usepackage{bm}
\usepackage{textcomp}
\usepackage{empheq}
\usepackage{siunitx}
\usepackage[titletoc,title]{appendix}
\usepackage{graphicx}
\usepackage{dcolumn}
\usepackage{bm}
\usepackage{subcaption}
\usepackage{xcolor,soul}

\DeclareMathOperator{\Tr}{Tr}
\newcommand{\citeasnoun}[1]{Ref.~\cite{#1}}

\newcommand{\Figref}[1]{Figure~\ref{fig:#1}}
\newcommand{\figref}[1]{Fig.~\ref{fig:#1}}
\renewcommand{\eqref}[1]{Eq.~(\ref{eq:#1})}
\newcommand{\Eqref}[1]{Equation~(\ref{eq:#1})}
\newcommand{\eqreftwo}[2]{Eqs.~(\ref{eq:#1},\ref{eq:#2})}

\newcommand{\vect}[1]{\boldsymbol{\mathbf{#1}}}
\newcommand{\tens}[1]{\bm{#1}}

\newcommand{\secref}[1]{Sec.~\ref{sec:#1}}

\newcommand*{\Ev}{\mathbf{E}}
\newcommand*{\Hv}{\mathbf{H}}

\newcommand*{\xv}{\mathbf{x}}

\newcommand*{\Pv}{\mathbf{P}}

\newcommand*{\Kv}{\mathbf{K}}
\newcommand*{\Nv}{\mathbf{N}}

\newcommand*{\Uv}{\mathbf{U}}

\newcommand*{\pol}{\mu}
\newcommand*{\tG}{\tens{\Gamma}}
\newcommand*{\mcC}{\mathcal{C}}
\newcommand*{\SM}{SM}
\newcommand*{\spot}{\zeta_0}
\newcommand*{\OPFWHM}{(OP\textsubscript{FWHM})}
\newcommand*{\OPKLOC}{(OP\textsubscript{kloc})}
\newcommand*{\OPZF}{(OP\textsubscript{ZF})}

\begin{document}

\preprint{APS/123-QED}

\title{Maximal Free-Space Concentration of {Electromagnetic Waves}}

\author{Hyungki Shim}
\affiliation{Department of Applied Physics and Energy Sciences Institute, Yale University, New Haven, Connecticut 06511, USA}
\affiliation{Department of Physics, Yale University, New Haven, Connecticut 06511, USA}
\author{Haejun Chung}
\affiliation{Department of Applied Physics and Energy Sciences Institute, Yale University, New Haven, Connecticut 06511, USA}
\author{Owen D. Miller}
\email{owen.miller@yale.edu}
\affiliation{Department of Applied Physics and Energy Sciences Institute, Yale University, New Haven, Connecticut 06511, USA}

\date{\today}

\begin{abstract}
    We derive upper bounds to free-space concentration of electromagnetic waves, mapping out the limits to maximum intensity for any spot size and optical beam-shaping device. For sub-diffraction-limited optical beams, our bounds suggest the possibility for orders-of-magnitude intensity enhancements compared to existing demonstrations, and we use inverse design to discover metasurfaces operating near these new limits. {We also demonstrate that our bounds may surprisingly describe maximum concentration defined by a wide variety of metrics.} Our bounds require no assumptions about symmetry, scalar waves, or weak scattering, instead relying primarily on the transformation of a quadratic program via orthogonal-projection methods. The bounds and inverse-designed structures presented here can be useful for applications from imaging to 3D printing.
\end{abstract}

\pacs{Valid PACS appear here}
\maketitle

Free-space optical {waves} with large focal-point intensities and arbitrarily small spot sizes---below the diffraction limit---are a long-sought goal~\cite{McCutchen1967,Stelzer2002,Zheludev2008} for applications ranging from imaging~\cite{Gustafsson2005,Betzig2006,Rust2006,Hell2007,Huang2008} to 3D printing~\cite{Lipson2013,Chia2015}, for which nanostructured lenses have enabled recent experimental breakthroughs~\cite{Huang2009,Rogers2012}. In this Letter, we derive {upper bounds} to free-space concentration of electromagnetic waves, revealing the maximum possible focal-point intensity (related to the well-known ``Strehl ratio''~\cite{Strehl,Born2013}) {for a fixed source power and} for any desired spot size. For waves incident from any region of space---generated by scattering structures, spatial light modulators, or light sources of arbitrary complexity---we show that the non-convex beam-concentration problem can be transformed to a quadratic program~\cite{Bertsekas2016} with easily computable global optima. We also extend this approach to derive maximum intensity independent of the exit surface of an incident wave. Our bounds simplify to those derived by Fourier analysis of prolate spheroidal wave functions~\cite{Slepian1961,Landau1961,Ferreira2006} in the scalar 1D limit. By honing in on the two essential degrees of freedom---the field intensity at the focal point, and its average over a ring at the desired spot size---the beam-concentration problem can be further simplified to a rank-two optimization problem, resulting in \emph{analytical} upper bounds in the far zone. For very small spot sizes $G$, which are most desirable for transformative applications, we show that the focal-point intensity must decrease proportional to $G^4$, a dimension-independent scaling law that cannot be overcome through any form of wavefront engineering. The bounds have an intuitive interpretation: the ideal field profile at the exit surface of an optical beam-shaping device must have maximum overlap with the fields radiating from a dipole at the origin yet be orthogonal to the fields emanating from a current loop at the spot size radius. We compare theoretical proposals and experimental demonstrations to our bounds, and we find that there is significant opportunity for order-of-magnitude intensity enhancements at those small spot sizes. We use ``inverse design''~\cite{Jensen2011,miller2012photonic,Bendsoe2013,Molesky2018}, a large-scale computational-optimization technique, to design metasurfaces that generate nearly optimal wavefronts and closely approach our general bounds. {By reformulating the light-concentration problem under alternative spot-size metrics, we show that the ideal field profiles for all these metrics are nearly identical in the far zone, suggesting that our analytical bounds, scaling laws, and ideal field profiles may be even more general than expected.} 

It is now well-understood that the diffraction ``limit,'' which is a critical factor underpinning resolution limits in imaging~\cite{Gustafsson2005,Betzig2006,Rust2006,Hell2007,Huang2008}, photolithography~\cite{Kawata1989,Ito2000,Odom2002}, and more ~\cite{Rugar1984,Indebetouw2007,Kumar2012,Collier2013} {(and is highly pertinent for applications such as surface-enhanced Raman scattering (SERS)~\cite{Moskovits1985,Nie1997,Kneipp1997} and extraordinary optical transmission (EOT)~\cite{Ebbesen1998,Moreno2001,Genet2007}}), is not a strict bound on the size of an optical focal spot, but rather a soft threshold below which beam formation is difficult in some generic sense (e.g., accompanied by high-intensity sidelobes). Although evanescent waves can be leveraged to surpass the diffraction limit~\cite{Fang2005,Kostelak2006,Merlin2007,Zhang2008,Ma2018}, they require structuring in the near field. The possibility of sub-diffraction-limited spot sizes without near-field effects was recognized in 1952 by Toraldo di Francia~\cite{ToraldodiFrancia1952}; stimulated by results on highly-directive antennas~\cite{Schelkunoff1943}, he analytically constructed successively narrower beam profiles with successively larger sidelobe energies (i.e. energies outside the first zero), in a scalar, weak-scattering asymptotic limit. Subsequent studies~\cite{Berry1994a,Berry1994,Berry2006,Lindberg2012} have connected the theory of sub-diffraction-limited beams to ``super-oscillations'' in Fourier analysis~\cite{Kempf2004,Aharonov2011}, i.e. bandlimited functions that oscillate over length/time scales faster than the inverse of their largest Fourier component. For one- and two-dimensional scalar fields, superoscillatory wave solutions have been explicitly constructed~\cite{Berry2006,Wong2010,Chremmos2015,Smith2016}, and in the one-dimensional case energy-concentration bounds have been derived~\cite{Ferreira2006} by the theory of prolate spheroidal wave functions~\cite{Slepian1961,Landau1961}. For optical beams, the only known bounds to focusing (apart from bounds on energy density at a point without considering spot sizes~\cite{Bassett1986,Sheppard1994}) are those derived in Refs.~\cite{Sales1997,Liu2002} (and recently in \citeasnoun{Rogers2018}, albeit with bounds on related but different quantities), which use special-function expansions and/or numerical-optimization techniques to discover computational bounds that apply for weakly scattering, rotationally symmetric filters in a scalar approximation. A bound that does not require weak scattering was developed in \citeasnoun{Liu2002}, but it still assumes rotational symmetry in a scalar diffraction theory. 

{There has been further work towards mapping possible field distributions and the structures that might achieve them. Singular-value decompositions can be used to rigorously identify a basis for all possible ``receiver'' (e.g., image-plane) field solutions~\cite{Miller2019a}. Then, given a \emph{known} feasible solution, one can use the equivalence principle via physical polarizabilities to identify practical metasurfaces that achieve such field distributions~\cite{Pfeiffer2013}. However, neither of these approaches is able to identify among all possible Maxwell solutions which ones are optimal.}

The recent demonstrations~\cite{Huang2007,Huang2007a,Dennis2008,Huang2009,Wang2009, Kitamura2010,Rogers2012,Rogers2013,Huang2013,Huang2014,Qin2015,Wong2015,Wong2017} of complex wavelength-scale surface patterns focusing plane waves to sub-diffraction-limited spot sizes have inspired hope that the previous tradeoffs of large sidelobe energies or small focal-point intensities might be circumvented or ameliorated by strongly scattering media accounting for the vector nature of light~\cite{Wang2009,Huang2013}, as all previous~\cite{Berry2006,Ferreira2006,Sales1997,Liu2002} asymptotic scaling relations and energy bounds require assumptions of rotational symmetry, weak scattering (except \citeasnoun{Liu2002}), and scalar waves. Such possibilities are especially enticing in the context of the broader emergence of ``metasurfaces''~\cite{Kildishev2013,Yu2014,Khorasaninejad2016,Li2018} enabling unprecedented optical response. {Given the importance of polarization filtering and high-numerical-aperture lenses to various imaging modalities, incorporation of the vector nature of light is critical to identifying ultimate resolution limits~\cite{Serrels2008,Jabbour2008,Bauer2014}. Moreover, recent design strategies for (diffraction-limited) metasurface lenses have shown that it is critical to account for strong-scattering physics in order to achieve high-efficiency structures~\cite{Lin2019,Chung2019}, a conclusion that extends to superresolving metalenses as well.} 

In this Article, we derive bounds on the maximum concentration of light that \emph{do} apply in the fully vectorial, strongly scattering regime, without imposing any symmetry constraints. Our derivation starts with the electromagnetic equivalence principle~\cite{Kong1975}, which allows us to consider the effects of any scatterer/modulator/light source as effective currents on some exit surface (\secref{apbound}). The optimal beam-concentration problem is non-convex due to the requirement for a particular spot size, but we use standard transformations from optimization theory to rewrite the problem as a quadratic program amenable to computational solutions for global extrema. We subsequently bound the solution to the quadratic problem by a simpler and more general rank-two optimization problem (\secref{apbound}), and also develop bounds independent of exit surface via modal decomposition (\secref{mobound}). The rank-two bounds reduce to analytic expressions in the far zone, and we compare the ideal field profiles to various theoretical and experimental demonstrations in \secref{farzone}. We show that there is still opportunity for orders-of-magnitude improvements, and design metasurfaces approaching our bounds (\secref{invdesign}). We also investigate how our bounds on maximum intensity perform under alternative spot-size metrics in \secref{metric}. Finally, in \secref{sum}, we discuss extensions of our framework to incorporate metrics other than focal-point intensity, near-field modalities, inhomogeneities, new point-spread functions, and more.
       
\begin{figure*} [t!]
    \includegraphics[width=\linewidth]{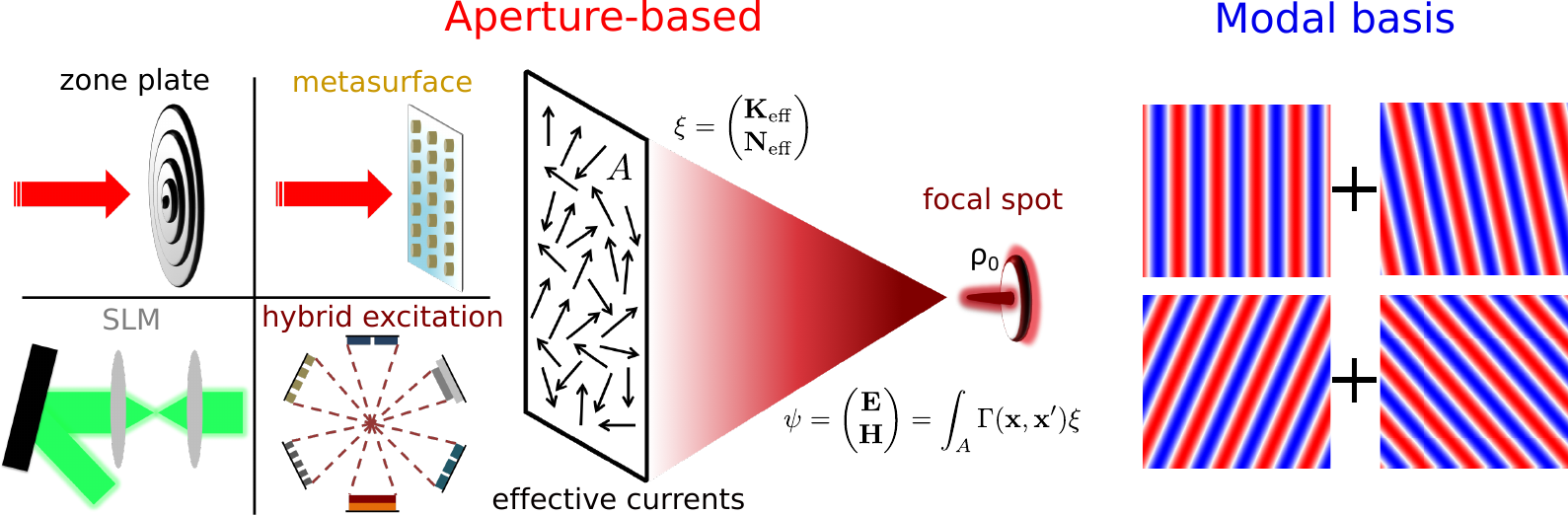} 
    \caption{Our framework establishes maximum light concentration for any zero-field contour, such as a circle. We derive two bounds: one that incorporates the shape of the exit aperture (enabling comparison with the well-known Strehl ratio), while otherwise independent of the beam-generation method, and a second that requires only a modal basis and is independent of aperture.}
    \label{fig:setup} 
\end{figure*} 

\section{General Bounds} 
\subsection{Aperture-Dependent Bounds} \label{sec:apbound}
Consider a beam generated by almost any means, e.g., an incident wave passing through a scatterer with a complex structural profile~\cite{Levy2007,Wei2013,KerenZur2016}, precisely controlled spatial light modulators~\cite{Weiner2000,Chattrapiban2003,Guo2007,Zhu2014}, or a light source with a complex spatial emission profile~\cite{Lodahl2004,Ringler2008,Bleuse2011}. The physics underlying the extent to which such a beam is spatially concentrated in free space is distilled to its essence by the electromagnetic equivalence principle~\cite{Kong1975}: the propagating fields are uniquely defined by their tangential values on any beam-generation exit surface, forming effective surface currents that encapsulate the entire complexity of the field-generation process ({The surface equivalence principle is the starting point for diffraction theory}~\cite{Born2013} {and has been used recently for metasurface-based wavefront shaping}~\cite{Pfeiffer2013,Epstein2016,Radi2017} {and optimal antenna designs}~\cite{Gustafsson2013,Shi2017}). By this principle, the beam-focusing problem is equivalent to asking: what is the maximum spatial concentration of a beam generated by electric and magnetic surface currents radiating in free space? We depict this distillation of the problem in \figref{setup}. We consider fields and currents at a single temporal frequency $\omega$ ($e^{-i\omega t}$ time evolution), and simplify the expressions to follow by encapsulating the electric and magnetic fields ($\Ev$, $\Hv$) and currents ($\Kv_{\rm eff}$, $\Nv_{\rm eff}$) in 6-vectors $\psi$ and $\xi$, respectively:
\begin{align}
\psi = \begin{pmatrix}
                        \Ev \\
                        \Hv
                    \end{pmatrix}, \quad
                    \xi = \begin{pmatrix}
                        \Kv_{\rm{eff}} \\
                        \Nv_{\rm{eff}}
           \end{pmatrix}.
                    \label{eq:defs}
\end{align}
The fields $\psi$ emanating from the effective currents $\xi$ distributed across the ``exit'' surface $A$ are given by the convolution of the currents with $\tens{\Gamma}$, the known $6\times6$ free-space dyadic Green's function~\cite{Chew1995}: 
\begin{align}
\psi(\xv) = \int_{A} \tens{\Gamma}(\xv,\xv') \xi(\xv') \label{eq:psidef}. 
\end{align}
Thus, the currents comprise the degrees of freedom determining the beam shape. As illustrated in \figref{setup}, finding the maximum focal intensity at a single point for any desired focal spot size now reduces to determining the optimal effective currents. We will assume equations such as \eqref{psidef} can be solved by any standard electromagnetic discretization scheme~\cite{Jin2011}, and we will write the matrix versions with the same symbols but without position arguments. For example, $\psi = \tens{\Gamma} \xi$ is the matrix equivalent of \eqref{psidef}, with $\psi$ and $\xi$ vectors and $\tens{\Gamma}$ a matrix. The total intensity at any point in free space, summing electric and magnetic contributions, is given by the squared norm of $\psi$:                         
\begin{align}
I(\xv)  = & \left | \psi(\xv) \right | ^2 =  \int_{A} \int_{A} \xi^{\dagger}(\xv'')\tens{\Gamma}^{\dagger}(\xv,\xv'')\tens{\Gamma}(\xv,\xv') \xi(\xv') \nonumber \\
        = & \xi^\dagger \tens{\Gamma}^{\dagger}\tens{\Gamma} \xi .  \label{eq:psi}
\end{align}

We now formulate the maximal-concentration question as a constrained optimization problem. The ideal optical beam has maximum focal intensity at a point (set at $\xv=0$), zero field along some spot-size contour $\mathcal{C}$, and a total propagating power $P$ not exceeding an input value of $P_0$. In \secref{metric} {we consider alternatives to a zero-field contour as metrics of concentration, thus we denote the optimization problem with the zero-field contour as ``\OPZF.''} Then, the maximum focal intensity, and the ideal effective currents generating it, solve the optimization problem denoted by:
\begin{equation}
    \begin{aligned}
     & {\rm (OP_{ZF})} \\
        & \underset{\xi}{\text{maximize}} & & I(\xv=0) = \xi^\dagger \tG_0^\dagger \tG_0 \xi \\
        & \text{subject to}       & &  \psi(\xv) \big\rvert_{\mathcal{C}} = \tG_\mcC \xi = 0 \quad \textrm{and} \quad P \leq  P_0, \label {eq:optimization}
    \end{aligned}
\end{equation}
where the ``0'' and $\mcC$ subscripts indicate that $\tG$ and $\psi$ are evaluated (in the appropriate basis) at the origin or at the spot-size contour, respectively{. (It does not violate any physical laws to set all electric and magnetic components to zero on a contour. If one wanted to require only a subset of field components to be zero then that could be achieved by only including those components in the definition of $\tG_\mcC$.)} Attempting to directly solve \eqref{optimization} is infeasible: the $\tG_0^\dagger \tG_0$ matrix is positive semidefinite (which is nonconvex under maximization~\cite{Boyd2004}), the equality constraint prevents the use of Rayleigh-quotient-based approaches~\cite{Horn2013}, and the power constraint is difficult to write in a simple {convex} form. 

We can bypass the nonconvexity of the optimization problem, $\rm (OP_{ZF})$, through multiple transformations. First, to simplify the power constraint, we replace it with a constraint on the intensity of the effective currents, normalized such that their total intensity is one: $\xi^\dagger \xi = 1$. We seek the ideal beam, which has all of its intensity generating power in the direction of the maximum-intensity spot, validating this replacement. Second,  {we subsume the equality constraint by considering only the effective currents that satisfy zero field on $\mcC$ by construction}. To this end, we project the currents $\xi$ onto the subspace of all currents that generate zero field on $\mcC$: $\xi = \left(\mathbf{I} - \tG_\mcC^\dagger \left(\tG_\mcC \tG_\mcC^\dagger\right)^{-1} \tG_\mcC\right) \nu = \mathbf{P} \nu$, where $\mathbf{I}$ is the identity matrix, the second term is the orthogonal projection matrix~\cite{Trefethen1997,Bertsekas2016} for $\tG_\mcC$, and the $\mathbf{P}$ matrix projects onto the null space of $\tG_\mcC$ (we have assumed any linearly dependent rows of $\tG_\mcC$ have been removed such that the inverse of $\tG_\mcC \tG_\mcC^\dagger$ exists). By this projection, {the equality constraint, which sets the field to zero on $\mcC$, is satisfied for arbitrary effective currents $\xi$:} $\tG_\mcC \xi = \tG_\mcC \mathbf{P} \nu = \left(\tG_\mcC - \tG_\mcC\right) \nu = 0$. Finally, we simplify the quadratic figure of merit {encoding total intensity at the origin, $\tG_0^\dagger \tG_0$, by projecting it along an arbitrary polarization}. As $\tG_0$ is a 6 $\times$ $6N$ matrix, where $N$ {is the number of degrees of freedom of the effective currents}, $\tG_0^\dagger \tG_0$ is a matrix with rank at most 6, as dictated by the polarizations of the electric and magnetic fields at the origin. Instead of incorporating all intensities, we project the field at the origin onto an arbitrary six-component polarization vector $\pol$ {to obtain the following (scalar) expression: $\pol^\dagger \psi(\xv=0) = \pol^\dagger \tG_0 \Pv \nu$}. The intensity at the origin in this polarization is then given by $\left | \pol^\dagger \psi(\xv=0)  \right | ^2 = \nu^\dagger \Pv \tG_0^\dagger \pol \pol^\dagger \tG_0 \Pv \nu$, where the inner matrix $\tG_0^\dagger \pol \pol^\dagger \tG_0$ is now rank \emph{one} {(the polarization vector $\pol$ should have a fixed norm in order to compare intensities along different polarizations on an equal footing)}. Rank-one quadratic forms are particularly simple, as evidenced here by the fact that we can define a vector $\gamma_\pol = \tG_0^\dagger \pol$ such that the intensity at the origin {simply reduces to the inner product of vector quantities ($\Pv \nu$ is a vector of effective currents $\xi$ satisfying the zero-field constraint),} $\nu^\dagger \Pv \gamma_\pol \gamma_\pol^\dagger \Pv \nu = \left( \gamma_\pol^\dagger \Pv \nu \right)^\dagger \gamma_\pol^\dagger \Pv \nu$.

The above transformations yield the equivalent but now tractable optimization problem:
\begin{equation}
\begin{aligned}
     & \underset{\pol,\nu}{\text{maximize}} & & \nu^\dagger \Pv \gamma_\pol \gamma_\pol^\dagger \Pv \nu \label{eq:qform} \\
     & \text{subject to}               & & \nu^\dagger \Pv \nu \leq 1,
\end{aligned}
\end{equation}
where $\nu$ represents arbitrary effective currents, $\Pv$ projects them to satisfy the zero-field condition, and $\gamma_\pol$ represents the conjugate transpose of the Green's function from the effective currents to the maximum-intensity point. \Eqref{qform} is equivalent to a Rayleigh-quotient maximization, and the solution is therefore given by the largest eigenvalue and corresponding eigenvector of the generalized eigenproblem $\Pv \gamma_\pol \gamma_\pol^\dagger \Pv \nu = \lambda \Pv \nu$. Here, because $\gamma_\pol \gamma_\pol^\dagger$ is rank one, it is straightforward to show (\SM) that the solution can be written analytically, with maximal eigenvector $\nu = \Pv \gamma_\pol / \|\Pv \gamma_\pol\|$ and maximal eigenvalue of $\gamma_\pol^\dagger \Pv \gamma_\pol$. Reinserting the transformed variable definitions from above, the optimal (unnormalized) effective currents are given by $\xi_{\rm opt} = \tG_0^\dagger \pol  - \tG_\mcC^\dagger \left(\tG_\mcC \tG_\mcC^\dagger \right)^{-1} \tG_\mcC \tG_0^\dagger \pol$. Then, we have that the $\pol$-polarized intensity at the origin, for any wavefront-shaping device in any configuration, is bounded above by the expression
\begin{align}
    I \leq \pol^\dagger \left[ \tG_0  \tG_0^\dagger - \tG_0 \tG_\mcC^\dagger \left(\tG_\mcC \tG_\mcC^\dagger \right)^{-1} \tG_\mcC \tG_0^\dagger \right]\pol.
    \label{eq:Ibound}
\end{align}
\Eqref{Ibound} represents a first key theoretical result of our work. Although it may have an abstract appearance, it is a decisive global bound to the optimization problem, requiring only evaluation of the known free-space dyadic Green's function at the maximum-intensity point, the zero-field contour, and the effective-current exit surface. The matrix in the square brackets is a 6 $\times$ 6 matrix, whose eigenvector {with the largest eigenvalue} represents the optimal polarization {for which intensity is maximized}. And the structure of \eqref{Ibound} has simple physical intuition: the maximum intensity of an unconstrained beam would simply focus as much of the effective-current radiation to the origin, as dictated by the term $\tG_0 \tG_0^\dagger$, but the constraint requiring zero field on $\mcC$ necessarily reduces the intensity by an amount proportional to the projection of the spot-size field ($\tG_\mcC$) on the field at the origin ($\tG_0$). 

The transformations leading to \eqref{Ibound} are exact, requiring no approximations nor simplifications. Thus the optimal fields, given by $\tG \xi_{\rm opt}$, are theoretically achievable Maxwell-equation solutions, and the bound of \eqref{Ibound} is \emph{tight}: no smaller upper bound is possible. We find that using a spectral basis~\cite{boyd2001chebyshev} for the zero-field contour and simple collocation~\cite{boyd2001chebyshev} for the aperture plane suffice for rapid convergence and numerical evaluation of \eqref{Ibound} within seconds on a laptop computer.

To simplify the upper bound and gain further physical intuition, we can leverage the fact that the high-interest scenario is small, sub-diffraction-limited spot sizes. The zero-field condition, i.e. $\tG_\mcC \xi = 0$ in \eqref{optimization}, is typically a high-rank matrix due to the arbitrarily large number of degrees of freedom in discretizing the zero-field contour. Yet in a spectral basis, such as Fourier modes on a circular contour or spherical harmonics on a spherical surface, for small spot sizes it will be the lowest-order mode, polarized along the $\pol$ direction, that is most important in constraining the field. If we denote the basis functions as $\phi_i$, then $\phi_0$ would be the prime determinant of the zero-field constraint at small spot sizes.  Instead of constraining the entire field to be zero along the zero-field contour, then, if we only constrain the zeroth-order, $\pol$-polarized mode, we will loosen the bound but gain the advantage that the zero-field constraint is now of the form $\left(\phi_0^\dagger \tG_\mcC\right) \xi = 0$, a vector--vector product with rank one. Then the previous analysis can be applied, with the replacement $\tG_\mcC \rightarrow \phi_0^\dagger \tG_\mcC$. We can introduce two new fields, physically motivated below, by the definitions
\begin{align}
    \psi_0 &= \tG_0^\dagger \pol       \label{eq:f1} \\
    \psi_1 &= \tG_\mcC^\dagger \phi_0. \label{eq:f2}
\end{align}
Given these two fields, algebraic manipulations (\SM) lead to an upper bound on the maximum intensity,
\begin{align}
    I \leq \psi_0^\dagger \psi_0 - \frac{|\psi_0^\dagger \psi_1|^2}{\psi_1^\dagger \psi_1}, \label{eq:Imax}
\end{align}
where the bound comprises a first term that denotes the intensity of a spot-size-unconstrained beam, while the second term accounts for the reduction due to imposition of the spot size constraint.

\begin{figure} [t!]
    \includegraphics[width=1\linewidth]{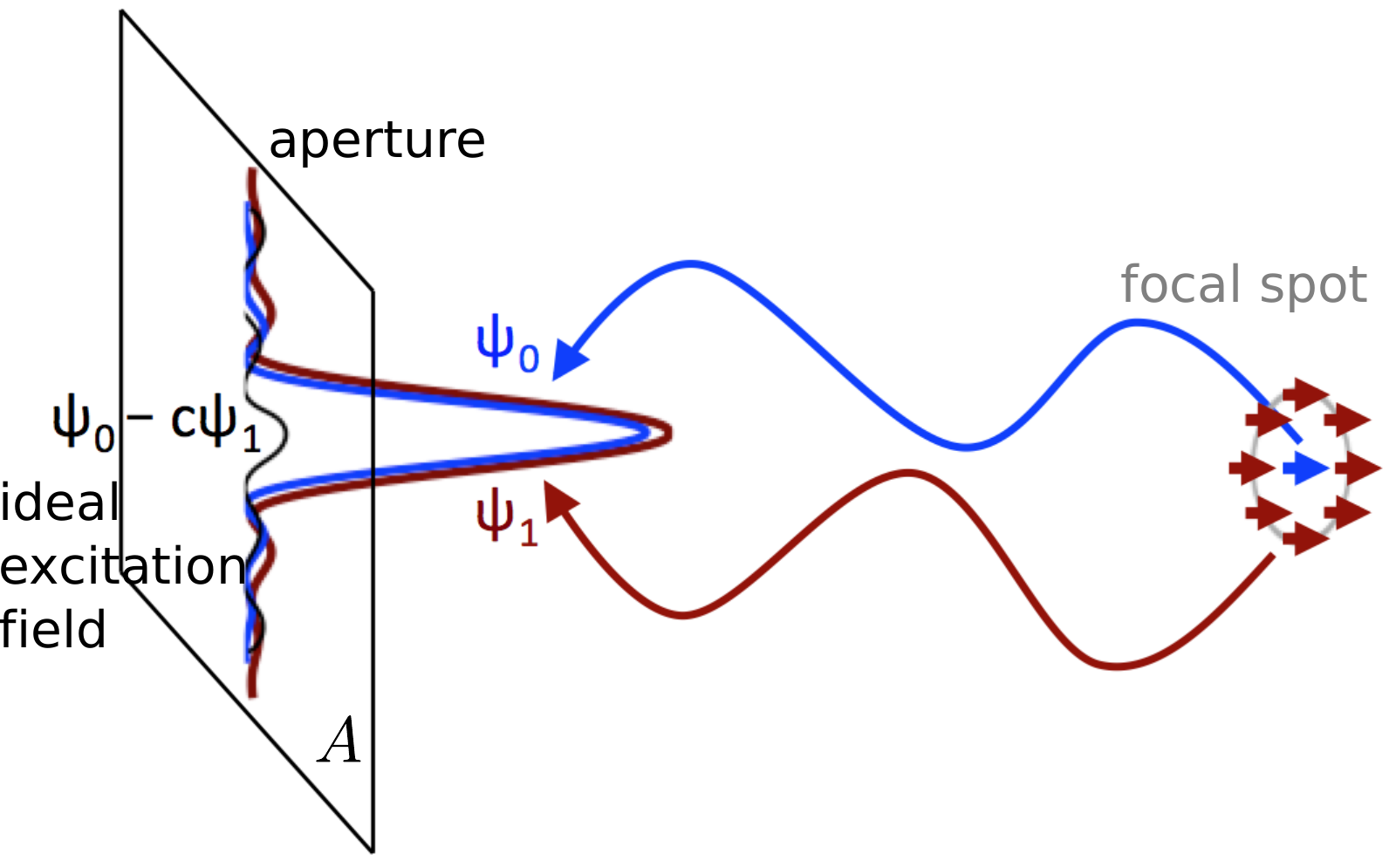} 
    \caption{Reciprocity-based illustration of fields that determine the maximum intensity of \eqref{Imax}. By reversing the source and measurement positions, it is shown that the ideal excitation field on the aperture $A$ maximizes overlap with $\psi_0$ while requiring zero overlap with $\psi_1$ ($\psi_0$ and $\psi_1$ are fields emanating from a point dipole at the origin and dipoles on the spot-size ring respectively).}  
    \label{fig:reciprocity} 
\end{figure} 

The fields of \eqreftwo{f1}{f2} can intuitively explain the bound of \eqref{Imax}. Whereas $\tG_0$ and $\tG_\mcC$ generate fields in the focusing region from currents in the aperture plane, $\tG_0^\dagger$ and $\tG_\mcC^\dagger$ generate fields in the aperture plane from the focusing region. By reciprocity~\cite{Kong1975}, which relates $\tG(\xv,\xv')$ to $\tG(\xv',\xv)$, the field $\psi_0 = \tG_0^\dagger \pol$ is related to the field emanating from dipolar sources at the focal spot back to the aperture plane (it is the conjugate of that field, with the signs of magnetic sources and fields reversed---reciprocity flips the signs of off-diagonal matrices of $\tG$). Similarly, $\psi_1 = \tG_\mcC^\dagger \phi_0$ is related to the field emanating from the zero-field contour back to the aperture plane. As illustrated in \figref{reciprocity}, the bound of \eqref{Imax} states that the maximum focal-spot intensity is given by the norm of the first field (focal point to aperture) minus the overlap of that field with the second (zero-field region to aperture). The smaller a desired spot size is, the closer these fields are to each other, increasing their overlap and reducing the maximum intensity possible. This intuition is furthered by considering the optimal effective currents that would achieve the bound of \eqref{Imax}, which are given by (\SM):
\begin{align}
\xi_{\rm opt}=\xi_0 \Bigg[\psi_0 - \frac{ \psi_1^\dagger \psi_0  }{\psi_1^\dagger \psi_1}\psi_1 \Bigg]. \label{eq:optcurrentgen} 
\end{align}	
where $\xi_0= 1 / \sqrt{\psi_0^\dagger \psi_0 - |\psi_0^\dagger \psi_1|^2 / \psi_1^\dagger \psi_1}$ is a normalization factor such that $\xi_{\rm opt}^\dagger \xi_{\rm opt} = 1$. \Eqref{optcurrentgen} demonstrates that the ideal field on the exit surface should maximize overlap with $\psi_0$ while being orthogonal to $\psi_1$. For small spot sizes, these two fields are almost identical, resulting in a significantly reduced maximum intensity.

\subsection{Modal-Decomposition Bounds} \label{sec:mobound}
Alternatively, one might ask about maximum spatial concentration of light \emph{independent} of exit surface, simply enforcing the condition that the light field comprises propagating waves. For example, in a plane, what combination of plane waves (or any other modal basis~\cite{Levy2016}) offers maximum concentration? In this case, the formulation is very similar to that of \eqref{optimization}, except that now the field $\psi$ is given as a linear combination of modal fields: $\psi = \vect{U} \vect{c}$, where $\vect{U}$ is a modal basis matrix (after appropriate discretization) and $\vect{c}$ is a vector of modal-decomposition coefficients. By analogy with $\tG_0$ and $\tG_\mcC$, we can define the field at zero and on the zero-field contour in the modal basis as $\Uv_0$ and $\Uv_\mcC$, respectively. Then, the bounds of \eqreftwo{Ibound}{Imax} and the definitions of \eqreftwo{f1}{f2} apply directly to the modal-decomposition case with the replacements $\tG_0 \rightarrow \Uv_0$ and $\tG_\mcC \rightarrow \Uv_\mcC$. For completeness, we can write here the general bounds:
\begin{align}
    I \leq \pol^\dagger \left[ \Uv_0  \Uv_0^\dagger - \Uv_0 \Uv_\mcC^\dagger \left(\Uv_\mcC \Uv_\mcC^\dagger \right)^{-1} \Uv_\mcC \Uv_0^\dagger \right]\pol.
    \label{eq:IboundU}
\end{align}
\Eqref{IboundU} represents the second key general theoretical result; as for \eqref{Ibound}, it appears abstract, but it is a simple-to-compute global bound on the intensity via the $6\times6$ matrix in square brackets, which again has clear physical intuition as the maximum unconstrained intensity (from $\Uv_0 \Uv_0^\dagger$) minus the projection of that field onto the representation of a constant field along the zero contour projected onto the modal basis (the second term). 

\begin{figure} [t!]
    \includegraphics[width=1\linewidth]{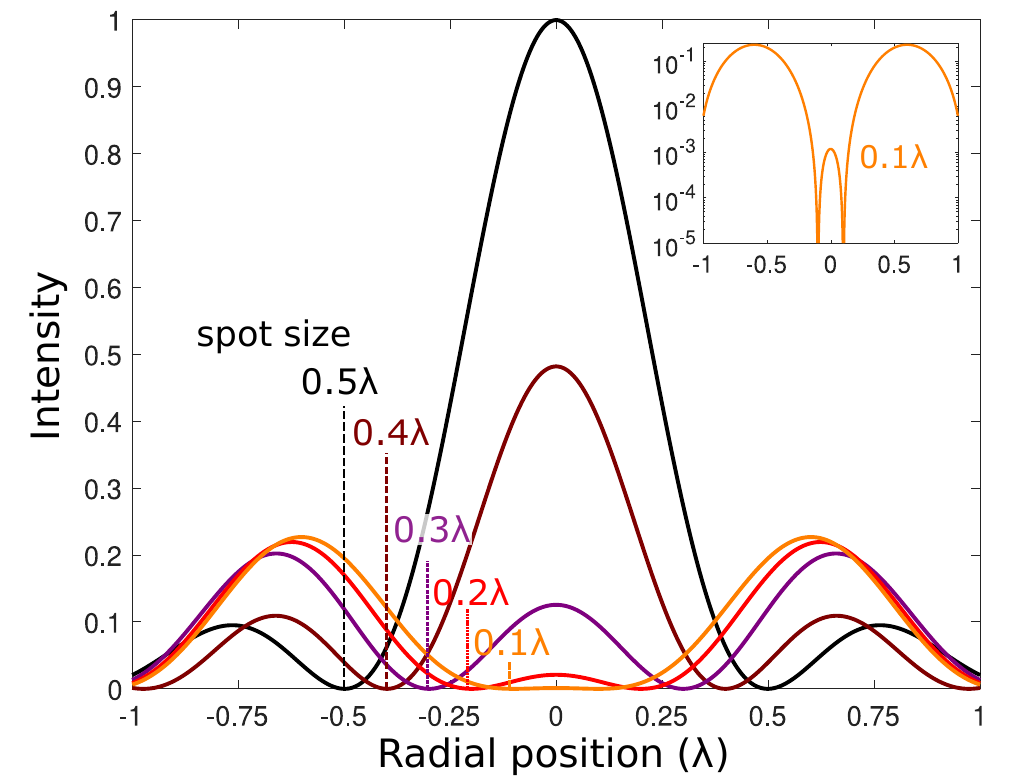} 
    \caption{Ideal field intensities in a plane (along any radial direction), determined by \eqref{IboundU}, for spot size $G$ from $0.1\lambda$ to $0.5\lambda$. The intensity for all spot sizes is normalized to the peak intensity for $G=0.5$. Inset: intensity profile for $G = 0.1\lambda$ in logarithmic scale.}
    \label{fig:genbound} 
\end{figure}

The bound of \eqref{IboundU} applies generally to any modal basis and zero-field contour. For the prototypical case of plane-wave modes and a circular zero-field contour in the plane, one can find a semi-analytical expression for \eqref{IboundU}, with ideal field profiles shown in \figref{genbound}. If one defines the maximal \emph{unconstrained} intensity as $I_0$ (which is $3k^2 / 16\pi$ given appropriate normalizations) then, as we show in the {\SM}, the maximum focusing intensity for spot size $R$ is given by a straightforward though tedious combination of zeroth, first, and second-order Bessel functions; in the small-spot-size limit ($kR \ll 1$), the asymptotic bound is
\begin{align}
    I \leq \frac{13}{13824\pi} \left(kR\right)^4 I_0.
    \label{eq:IboundUssz}
\end{align}
The maximum intensity must fall {off} at least as the fourth power of spot size, identical to the dependence of the aperture-dependent bounds in the far field, which has important ramifications for practical design, as we show in the next section.

Our bounds share a common origin with those of an ``optical eigenmode'' approach~\cite{Mazilu2011}: the quadratic nature of power and momentum flows in electromagnetism. A key difference appears to be the choice of figure of merit, as well as the purely computational nature of the optical-eigenmode approach~\cite{Mazilu2011,Kosmeier2011}, using computational projections onto numerical subspaces. Above, we have shown that orthogonal projections and physically-motivated Fourier decompositions lead to analytical and semi-analytical bound expressions.

We show in the SM that, for scalar waves in one dimension, our modal-decomposition bounds coincide exactly with those derived by a combination of Fourier analysis and interpolation theory~\cite{Ferreira2006,Levi1965}. In fact, if in the 1D case one were to stack $\Uv_0$ and $\Uv_\mcC$ in a single matrix and multiply by its conjugate transpose then the resulting matrix, $\begin{pmatrix} \Uv_0 \Uv_0^\dagger & \Uv_0 \Uv_\mcC^\dagger \\ \Uv_\mcC \Uv_0^\dagger & \Uv_\mcC \Uv_\mcC^\dagger \end{pmatrix}$, is exactly the matrix of sinc functions that defines the eigenproblem for which prolate spheroidal wave functions (PSWFs) are the eigenvectors~\cite{Slepian1978}. Thus, our modal-basis approach can be understood as a vector-valued, multi-dimensional generalization of the PSWF-based Fourier analysis of minimum-energy superoscillatory signals.

\section{Optical Beams in the Far Zone} \label{sec:farzone}
The bounds of \eqreftwo{Ibound}{Imax} allow arbitrary shapes for the exit surface and the zero-field contour. The prototypical case of interest, for many applications across imaging and 3D printing, for example, involves a beam of light shaped or created within a planar aperture, or more generally within any half space where the exit surface can be chosen to be a plane, and propagating along one direction, with concentration measured by the spot size in a transverse two-dimensional plane. Hence the exit surface is an aperture plane and the zero-field contour is a spot-size circle. A dimensionless concentration metric known as the ``Strehl ratio''~\cite{Strehl,Born2013} quantifies focusing in the far zone of such beams, where diffraction effects can be accounted for in the normalization.

In the far zone, with the focusing--aperture distance much larger than the aperture radius and the wavelength of light, the six electric and magnetic polarizations decouple, reducing the response for any one to a scalar problem. As we show in the {\SM}, for any aperture-plane polarization, the focal-point field $\psi_0$ of \eqref{f1} is proportional to $e^{-ikz}/z$, for propagation direction $z$ and wavenumber $k = \omega/c$, while the zero-contour field $\psi_1$ of \eqref{f2} is proportional to the same factor multiplied by the zeroth-order Bessel function $J_0$, i.e., $\psi_1 \sim J_0(kr\rho_0/z) e^{-ikz}/z$, where $\rho_0$ is the spot-size radius and $r$ is the radial position in the aperture plane. These are the Green's-function solutions and require no assumptions about the symmetries of the optimal fields. The evaluation of the overlap integrals $\psi_0^\dagger \psi_0$, $\psi_0^\dagger \psi_1$, and $\psi_1^\dagger \psi_1$ in the aperture plane are integrals of constants and Bessel functions. For any aperture shape, we can find an analytical bound on the maximal focusing intensity by evaluating the bound for the circumscribing circle of radius $R$. Performing the integrals (\SM), \eqref{Imax} becomes
\begin{align}
    I \leq \frac{k^2 R^2}{16 \pi z^2} - \frac{1}{4 \pi \rho_0^2} \frac{\left[J_1(kR\rho_0/z)\right]^2}{\left[J_0(kR\rho_0/z)\right]^2 + \left[J_1(kR\rho_0/z)\right]^2}.
    \label{eq:ImaxFZ}
\end{align}
\Eqref{ImaxFZ} provides a general bound for any aperture--focus separation distance $z$ and spot-size radius $\rho_0$. The dependence on $kR/z$ and related quantities is characteristic of any far-zone beam, and can be divided out for a separation-distance-\emph{independent} bound. The Strehl ratio accounts for this dependence in circularly-symmetric beams by dividing the focal-point intensity by that of an Airy disk, which is the diffraction-limited pattern produced by a circular aperture. Within the Strehl ratio is a normalized spot-size radius, $\spot = k R \rho_0 / z$, which equals the Airy-pattern spot size multiplied by a normalized spot size $G$ between 0 and 1. We can generalize the Strehl definition beyond the Airy pattern: instead, divide the maximum intensity, \eqref{Imax}, by the intensity of an unconstrained focused beam (without the zero-field condition), which is simply $\psi_1^\dagger \psi_1$ (which conforms to the usual Airy definition for a circular aperture). Thus $S = I / I_{\rm max} = 1 - |\psi_0^\dagger \psi_1|^2 / \sqrt{\psi_0^\dagger \psi_0 \psi_1^\dagger\psi_1}$. By this definition, the Strehl ratio $S_{\rm max}$ of the optimal-intensity beam of \eqref{ImaxFZ} is given by
\begin{align}
    S_{\rm max} = 1 - \frac{4}{\spot^2} \frac{[J_1(\spot)]^2}{[J_0(\spot)]^2 + [J_1(\spot)]^2}.\label{eq:optS}
\end{align}
A diffraction-limited Airy beam occurs for $\spot$ equaling the first zero of $J_1$, in which case $S_{\rm max} = 1$. As $\spot$ decreases below the first zero of $J_1$, the second term of \eqref{optS} increases from zero, reducing $S_{\rm max}$. Although we arrived at \eqreftwo{ImaxFZ}{optS} from \eqref{Imax}, the bound derived from loosening the constraints and solving the rank-two optimization problem, our numerical results show that in the far zone, the full-rank optimization problem of \eqref{qform} that is bounded above by \eqref{Ibound} has exactly the same solution (the equivalence is not exact for non-circular apertures, but even then the discrepancy practically vanishes for spot size $G\ll1$). Physically, this means that in the far zone, maximally focused beams are symmetric under rotations around the propagation axis, for any spot size, such that only their first Fourier coefficient is nonzero on the spot-size ring. From a design perspective, this equivalence implies that the bound of \eqref{optS} is physically achievable, and that the corresponding Maxwell field exhibits the largest possible intensity for a given spot size.

\begin{figure*} [t!]
    \includegraphics[width=1\linewidth]{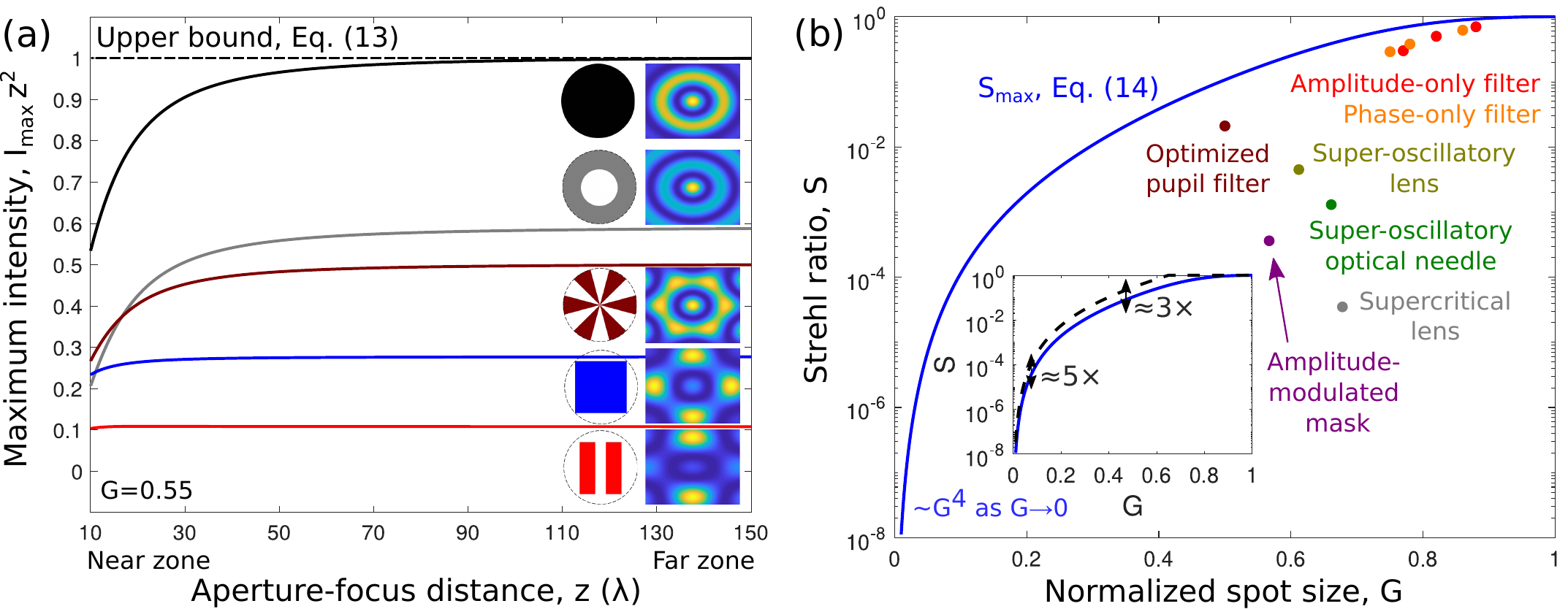} 
    \caption{(a) Intensity bounds for various aperture shapes (normalized to far-zone bound, \eqref{ImaxFZ}). The bounds, scaled by $z^2$ for aperture--focus distance $z$, are largest in the far zone, where the optimal field profiles (inset images) are highly dependent on the aperture shape. \Eqref{ImaxFZ} is the general bound of the circumscribing circle for each shape, represented in the dashed line. (b) Maximum Strehl ratio (\eqref{optS}) compared to previous designs, including amplitude~\cite{Gundu2005}/phase~\cite{Dejuana2003}-only pupil filters, optimized pupil filters~\cite{Kosmeier2011}, a super-oscillatory lens~\cite{Rogers2012}, a super-oscillatory optical needle~\cite{Rogers2013}, a supercritical lens~\cite{Qin2017}, and an amplitude-modulated mask~\cite{Dong2017}. Inset: Our bound (blue), despite allowing for arbitrary diffractive optical elements, is \emph{smaller} than the bound of Ref.~\cite{Sales1997} (black dashed), which requires rotation symmetry in the scalar and weak-scattering limit.}
    \label{fig:bound} 
\end{figure*} 

\Figref{bound}(a) plots the intensity bound, \eqref{Imax}, for a variety of exit-aperture shapes and from the near zone to the far zone, with a generic spot size $G=0.55$. The bound is scaled by $z^2$, the square of the aperture--focus distance, to account for the quadratic power decay. In each case the far-zone bound is larger than that of the near zone or mid zone, suggesting that the far-zone bounds of \eqref{ImaxFZ} and \eqref{optS} may be global bounds at any distance. For each case, the maximum intensity is bounded above by the bound for the circumscribing circle (solid black line), while the optimal field profiles are highly dependent on the aperture shapes (inset images).

\Figref{bound}(b) plots the far-zone bound $S_{\max}$ as a function of spot size $G$ (blue curve), and compares various theoretical results from the literature~\cite{Dejuana2003,Gundu2005,Kosmeier2011,Rogers2012,Rogers2013,Qin2017,Dong2017}. (Many of the references include experimental results; for fair comparison and to exclude experimental errors, we use either their \emph{simulated} $G$ and $S$ values, or reconstruct them with our own simulations as detailed in the \SM.) For relatively large spot sizes ($G > 0.7$), theoretical proposals for amplitude~\cite{Gundu2005} or phase~\cite{Dejuana2003} filters can closely approach the limits, though embedded in the proposals is a weak-scattering assumption that may be difficult to achieve in practice. (The key reason they fall short of the bound is that they do not allow for multiple scattering to redistribute energy in the exit plane.) For small spot sizes, on the other hand, the maximum Strehl ratio decreases rapidly. A Taylor expansion of \eqref{optS} reveals the asymptotic bound (\SM),
\begin{align}
    S_{\rm max} = \spot^4 /192, \quad \spot \ll 1, \label{eq:optSsmall}
\end{align}
which represents a severe restriction---halving the spot size costs a sixteenfold decrease in maximum focal intensity. This fundamental limit suggests that extremely small spot sizes are impractical, from both power-consumption and fabrication-tolerance perspectives. The quartic dependence is independent of dimensionality (in the {\SM} we show that the same dependence arises for a focal sphere, as well as for focal points in 2D problems) and can be explained generally: for small enough spot sizes, the zero-contour field will always have a maximum at the origin and thus all odd powers in a Taylor expansion around the origin must be zero. The first non-constant field dependence in the expansion is quadratic, and since the overlap quantities in the intensity bound are themselves quadratic in the field, the general intensity dependence on spot size always results in an $\sim \spot^4$ scaling law. {(While this is reminiscent of the quartic dependence between the transmission through a subwavelength hole in a conducting screen and the hole size~\cite{Bethe1944}, we show in the {\SM} that their physical origins are unrelated.)}

Perhaps the most important region of the figure is for intermediate values of spot size ($0.1 \lesssim G \lesssim 0.7$), where it is possible to meaningfully shrink the spot size below the diffraction limit without an overwhelming sacrifice of intensity. This is the region that recent designs~\cite{Kosmeier2011,Rogers2012,Rogers2013,Qin2017,Dong2017} have targeted (especially $0.5 < G < 0.7$) with a variety of approaches, including super-oscillatory lenses/needles and optimized pupil filters. Yet as seen in \figref{bound}(a), these designs mostly fall dramatically short of the bounds. The best result by this metric is the ``optimized pupil filter'' of \citeasnoun{Kosmeier2011}, whose quadratic-programming approach comes within a factor of 5 of the bound, and demonstrates the utility of computational-design approaches for maximum intensity. The other designs fall short of the bounds by factors of 100-1000X or more, offering the possibility for significant improvement by judicious design of the diffractive optical element(s). 

The inset of \figref{bound}(b) compares our analytical bound of \eqref{optS} to the computational bounds of \citeasnoun{Sales1997}, which used special-function expansions to identify upper limits to intensity as a function of spot size for scalar, rotation-symmetric waves in the weak-scattering limit. Perhaps surprisingly, despite allowing for far more general optical setups and for vector waves without any symmetry or weak-scattering assumptions, our bounds are 3-5X \emph{smaller} than those of \citeasnoun{Sales1997}. (An analytical bound coinciding with \eqref{optS} has been derived in \citeasnoun{Liu2002}, albeit assuming rotational symmetry in a scalar approximation and without the other general results herein.) Note conversely, however, that our approach is also prescriptive, in the sense that it identifies the exact field profiles that can reach our bounds.

We can also characterize the effective currents on the aperture that achieve the bound in \eqref{optS}. For spot sizes close to the diffraction limit ($G=1$), the currents are maximally concentrated around the aperture rim and decreases towards the center, where the amplitude is close to zero. This is because small, localized spots require large transverse wavevector components, which originate from the currents around the rim. As spot size decreases, those edge currents are partially redistributed to the center, to create the interference effects giving rise to sub-diffraction-limited spots.

{Given the feasibility of achieving sub-diffraction-limited waves, it is important to contextualize the recent work of Ref.~\cite[Sec.~5.5]{Miller2019a}, which uses a singular-value-decomposition approach to identify possible field patterns that can be generated. In that work, it is shown that sub-diffraction-limited Gaussian field profiles require exponentially large input powers, and suggested that therefore sub-diffraction-limited focusing is essentially impossible. The key distinction our work makes is loosening the requirement for a Gaussian field profile, instead imposing only spot-size constraints on the field without reference to any desired profile. This more general problem has feasible, sub-diffraction-limited solutions whose input powers scale only polynomially with spot size.}

As discussed above, our bounds, both in the general case of \eqref{Ibound} and in the optical-beam case of \eqref{ImaxFZ}, are tight in the sense that they are achievable by fields that are solutions of Maxwell's equations, given by $\psi = \tG \xi_{\rm opt}$. Yet as shown in \figref{bound}, theoretical designs for sub-diffraction-limited beams have fallen far short of the bounds. A natural question, then, is whether realistic material patterning and designs can generate the requisite fields to achieve the bounds?

\section{Inverse-designed metasurfaces} \label{sec:invdesign}
\begin{figure*} [t!]
    \includegraphics[width=1\linewidth]{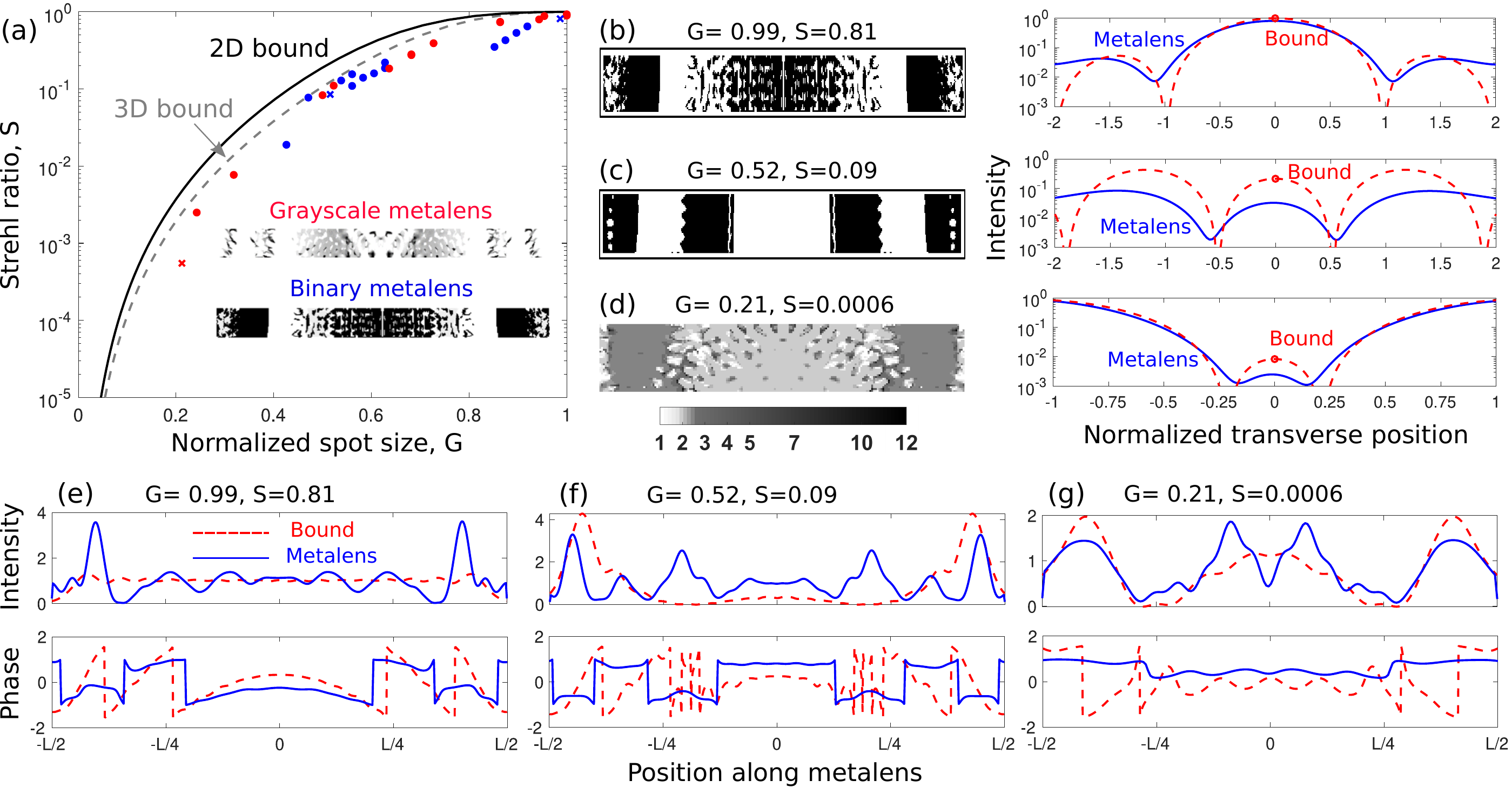} 
    \caption{(a) Inverse-designed metasurfaces, with grayscale (gradient-index, red circles) and binary (blue circles) material distributions, approach the 2D bounds (black line) of \eqref{Smax2D} for a variety of spot sizes. The similar 3D bound of \eqref{optS} is included for comparison with \figref{bound}. Three specific designs, highlighted with an ``$\times$'' marker in (a), are shown in (b,c,d), along with their intensity profiles {on the focal plane} (blue line) and those achieving the 2D bounds (red dashed line), normalized relative to the diffraction-limited intensity for each case. (e,f,g) {Field intensity (normalized to the incident intensity)} {and phase profiles near the exit surface ($2\lambda$ from metasurface) for the corresponding designs in (b,c,d), compared to those achieving the 2D bounds. The focal distance is set to $50\lambda$, and the metasurfaces have widths $L=23\lambda$ for $G=0.99, 0.52$ and $L=10\lambda$ for $G=0.21$.}}
    \label{fig:metalens} 
\end{figure*} 
We use ``inverse design'' to discover refractive-index profiles that can approach the concentration bounds. Inverse design~\cite{Jensen2011,miller2012photonic,Bendsoe2013,Molesky2018} is a large-scale computational-optimization technique, mathematically equivalent to backpropagation in neural networks~\cite{Rumelhart1986,LeCun1989,Werbos1994}, that enables rapid computation of sensitivities with respect to arbitrarily many structural/material degrees of freedom. Given such sensitivities, standard optimization techniques~\cite{Nocedal2006} such as gradient descent (employed here) can be used to discover locally optimal structures, often exhibiting orders-of-magnitude better performance~\cite{Lalau-Keraly2013,Ganapati2014} than structures with few parameters designed by hand or brute force.

For some target wavelength $\lambda$, we consider metasurfaces with thicknesses of $1.9\lambda$ and widths (diameters) ranging from $10\lambda$ to $23\lambda$, equivalent to films with thicknesses on the order of 1${\mu}$m and widths in the dozens of microns for visible-frequency light. We consider two-dimensional scattering (i.e. metasurfaces that are translation-invariant along one dimension), to reduce the computational cost and demonstrate the design principle. Dimensionality has only a small effect on the bounds; in the {\SM}, we show that the 2D equivalent of \eqref{optS} is
\begin{align}
    S_{\rm max}^{2D} = 1 - \frac{4}{\spot} \frac{\sin^2(\spot)}{\sin(2\spot) + 2\spot}.
    \label{eq:Smax2D}
\end{align}

For the design variables, we allow the permittivity at every point on the metasurface to vary (i.e., ``topology optimization''~\cite{Jensen2011,Bendsoe2013}), and we generate two types of designs (depicted in \figref{metalens}): binary metasurfaces, in which the permittivity must take one of two values (chosen here as 1 and 12), and grayscale metasurfaces, in which the permittivity can vary smoothly between two values as in gradient-index optics~\cite{marchand1978gradient}.

The optical figure of merit $F$ that we use to discover optimal design is not exactly the zero-point intensity of \eqref{optimization}, as the zero-field constraint is difficult to implement numerically. Instead, for a desired spot size $G$, we subtract a constant $\alpha$ times the field intensity at the points $\pm G$ away from the origin:
\begin{align}
    F = |\psi(0)|^2 - \alpha \left( \left|\psi(-G)\right|^2 + \left|\psi(G)\right|^2 \right). 
    \label{eq:FOM}
\end{align}
This is a penalty method~\cite{bertsekas2014constrained} that can enforce arbitrarily small field intensities (with sufficiently accurate simulations) by increasing the constant $\alpha$ over the course of the optimization. We take the electric field polarized out of the plane, such that $\psi$ can be simplified to a scalar field solution of the Helmholtz equation, which we solve by the finite-difference time-domain (FDTD) method ~\cite{Taflove2013,Oskooi2010}. Adjoint-based sensitivities are computed for every structural iteration via two computations: the ``direct'' fields propagating through the metasurface, and ``adjoint'' fields that emanate from the maximum-intensity and zero-field locations, with phases and amplitudes of the exciting currents determined by the derivatives of \eqref{FOM} with respect to the field.

\Figref{metalens} depicts the results of many inverse-designed metasurfaces. \Figref{metalens}(a) compares the bound of \eqref{Smax2D} (black; the nearly identical 3D bound is in grey) to the computed Strehl ratio of unique optimal designs at spot sizes ranging from 0.21 to 1; strikingly, the designed metalenses closely approach the bound for all spot sizes, with the best designs achieving 90\% of the maximum possible intensity. In \figref{metalens}(b,c,d) three specific designs are shown alongside the resulting field profiles in their focal planes {(we provide a full list of permittivity values at each grid point of these three metasurface designs in the {\SM})}. The intensity does not perfectly reach zero but is forced to be significantly smaller than the peak intensity through the penalty constant $\alpha$ in \eqref{FOM}. {The field intensities and phases on the exit surface (2$\lambda$ from metasurface) are shown in \figref{metalens}(e,f,g) corresponding to the three designs. \figref{metalens}(e) shows good qualitative agreement with the ideal field profiles achieving the 2D bounds, and even for smaller spot sizes the designed metasurfaces achieve the intensity redistributions---exhibited by local intensities larger than one---required to approach the bounds (\figref{metalens}(f,g))}. It is difficult to explain exactly how the computationally designed metasurface patterns achieve nearly optimal focusing; for spot sizes close to 1, the variations in material density suggest an effective gradient-index-like profile that offers lens-like phase variations across the device width, though the scattering effects of the front and rear surfaces renders such explanations necessarily incomplete. The depicted design with $G=0.21$ exhibits to our knowledge the smallest spot size of any theoretical proposal to date.

{
\section{Alternative spot-size metrics} \label{sec:metric}
To this point, we have considered the problem of maximizing field intensity subject to a spot size defined by the field equaling zero around a prescribed contour. Of course, this is not the only metric one might consider. In this section, we consider two alternative definitions of spot size which do not require the field to go to zero anywhere, and which enforce ``concentration'' through other characteristics of an optical beam. The two properties that we consider are the full-width at half-maximum (FWHM) of the beam, a commonly used metric for experimental measurements~\cite{Huang2007,Kitamura2010,Rogers2012,Rogers2013,Qin2015}, and the local wavenumber (as measured by the spatial variations in the field), whose maximization was an impetus for early work in super-oscillations~\cite{Berry1994a,Berry1994,Kempf2004,Berry2006,Aharonov2011,Lindberg2012}. We show that using these constraints, the corresponding field-concentration problems can be formulated as nonconvex, quadratically constrained quadratic programs (QCQPs) whose global bounds can be found computationally in reasonable (polynomial) time. For the canonical far-zone scenario considered in \secref{farzone}, we compare the optimal beams for the three metrics: a zero-field contour, a FWHM spot-size, and a minimum local wavevector, and we find that the beams are nearly identical. These numerical experiments suggest that the intensities and optimal-beam profiles found by our analytical bounds for the zero-field-contour metric may describe optimal beams across a wide swath of possible ``concentration'' metrics.

\begin{figure*} [t!]
    \includegraphics[width=1\linewidth]{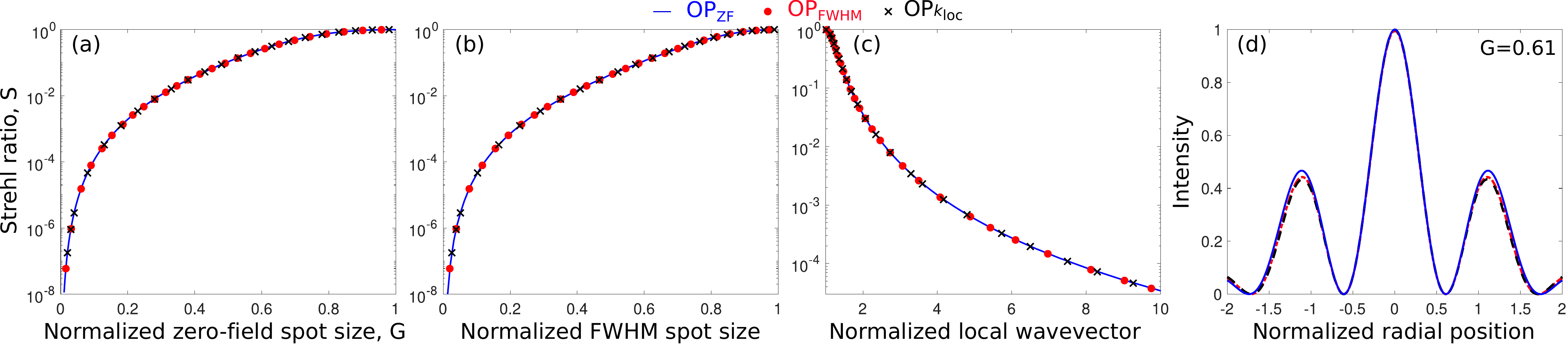} 
    \caption{Optimal beam characteristics for three concentration metrics: $OP_{\textrm{ZF}}$ (\eqref{optimization}, blue lines), $OP_{\textrm{FWHM}}$ (\eqref{opt_fwhm}, red circles) and $OP_{\textrm{kloc}}$ (\eqref{opt_kcut}, black crosses) refer to the optimization problems with constraints on the zero-field spot size $G$, FWHM spot size, and the local wavevector respectively. (a,b,c) Bounds plotted against spot-size radius, FWHM, and local wavevector at the focal point, $k_{\textrm{loc}}(\xv=0)$, respectively (all normalized to those of a diffraction-limited beam). (d) Optimal beam profiles with zero-field spot size $G=0.61$ for the three different optimization problems, showing nearly identical beam profiles, especially around the central peak.}
    \label{fig:metric} 
\end{figure*} 

For each of the alternative metrics we consider, we retain the objective of maximizing intensity at the origin, and replace the zero-field-contour constraint with a constraint on another property. As a first alternative, we impose a constraint that requires the average field intensity to fall to its FWHM value within a prescribed, sub-diffraction-limited contour $\mathcal{C}$. The average field value around the contour is given by $\frac{1}{\left|\mathcal{C}\right|} \int_\mathcal{C}\,| \psi(\xv) |^2 \,{\rm d}\xv = \xi^\dagger \left[\left(1/|\mathcal{C}|\right) \int_\mcC \tG^\dagger \tG\right] \xi$, where $\left|\mathcal{C}\right|$ denotes the perimeter of contour $\mathcal{C}$. We denote this optimization problem ``(OP\textsubscript{FWHM}),'' which is written:
\begin{equation}
    \begin{aligned}
        & {\rm (OP_{FWHM})} \\
        & \underset{\xi}{\text{maximize}} & & I(\xv=0) = \xi^\dagger \tG_0^\dagger \tG_0 \xi \\
        & \text{subject to} & & \xi^\dagger \left[\frac{1}{|\mathcal{C}|} \int_\mcC \tG^\dagger \tG \right] \xi \leq \frac{1}{2} \xi^\dagger \tG_0^\dagger \tG_0 \xi \\
        & & & \xi^\dagger \xi \leq 1,
        \label{eq:opt_fwhm}
    \end{aligned}
\end{equation}
where the objective represents the intensity at the origin, the first constraint enforces the FWHM to occur within the contour $\mcC$, and the second constraint is the power normalization.

As a second alternative, we consider one of the early definitions of ``super-oscillations:'' a signal comprising spatial frequencies up to some maximum $k$ can have a ``local,'' position-dependent wavenumber $k_{\rm loc}(\xv)$, as measured by spatial variations in the field, that is in fact larger than $k$ by an arbitrary extent~\cite{Berry2006}. To define $k_{\rm loc}$, one could use a gradient-based expression like $-i\nabla$, yet the resulting operator would not be Hermitian, limiting its viability. Instead, we use the negative Laplacian, a positive-definite real-symmetric operator (for fields that decay sufficiently fast), normalized by the field intensity at the origin: $k_{\rm loc}(\xv) = \sqrt{-\nabla^2 |\psi(\xv)|^2 / |\psi(0)|^2}$. For a plane wave $e^{i\vect{k}\cdot \xv}$, the local wavenumber equals that of the plane wave, i.e., $k_{\rm loc}(\xv) = |\vect{k}|$. Thus, an alternative way to define a tightly confined beam would be to require the local wavenumber at the origin to be at least as large as some $k_{\rm loc,min}$ that is much larger than the free-space wavenumber $\omega/c$, i.e. $k_{\rm loc,min} \gg \omega/c$. In this optimization-problem formulation, denoted by ``(OP\textsubscript{kloc}),'' we now use the constraint $k_{\rm loc}^2(\xv=0) \geq k_{\rm loc,min}^2$:
\begin{equation}
    \begin{aligned}
        & {\rm (OP_{kloc})} \\
        & \underset{\xi}{\text{maximize}} & & I(\xv=0) = \xi^\dagger \tG_0^\dagger \tG_0 \xi \\
        & \text{subject to}       & &  \xi^\dagger \left[ -\nabla^2 \left(\tG^\dagger \tG\right)\big\rvert_{\xv=0}\right] \xi \geq k^2_{\textrm{loc,min}} \xi^\dagger \tG_0^\dagger \tG_0 \xi \\
        & & & \xi^\dagger \xi \leq 1,
        \label{eq:opt_kcut}
    \end{aligned}
\end{equation}
where in the $k_{\rm loc}$ constraint we have moved the field normalization $|\psi(0)|^2$ to the right-hand side, such that the quadratic nature of the problem is readily apparent.

The two alternative optimization problems, {\OPFWHM} and {\OPKLOC}, both comprise nonconvex quadratic objectives subject to two quadratic constraints (one convex and one nonconvex in each case). It is not possible (to our knowledge) to identify analytical bounds for such objectives. However, one can use semidefinite relaxation (SDR), a now-standard technique~\cite{Luo2010} in optimization theory, to ``relax'' the problem from a quadratic one for variables $\xi$ of size $N$ to a \emph{linear} one in a much larger space of dimension $N^2$, for which interior-point methods can find global optima~\cite{Nocedal2006}. Typically, such relaxations lead to bounds that may be ``loose,'' i.e. unattainable, but it is known in QCQP theory~\cite{Huang2007b} that for complex-variable quadratic problems with three or fewer constraints, SDR bounds are ``tight,'' i.e. attainable (due to a duality gap that is strictly zero). Thus, we can use SDR to compute exact bounds and field profiles for problems {\OPFWHM} and {\OPKLOC}.

We can briefly summarize the application of SDR to problems {\OPFWHM} and {\OPKLOC}. Each problem is of the form
\begin{equation}
    \begin{aligned}
        & \underset{\xi}{\text{maximize}} & & \xi^\dagger \vect{A} \xi \\
        & \text{subject to}       & &  \xi^\dagger \vect{B} \xi \leq 0 \\
        & & & \xi^\dagger \xi \leq 1.
        \label{eq:opt_gen}
    \end{aligned}
\end{equation}
The first idea in SDR is to rewrite terms of the form $\xi^\dagger \vect{A} \xi$ as $\Tr\left( \vect{A} \xi \xi^\dagger\right)$, and then define the positive semidefinite, rank-one matrix $\vect{X}$ by $\vect{X} = \xi\xi^\dagger$, such that $\xi^\dagger \vect{A} \xi = \Tr\left( \vect{A} \vect{X}\right)$. Then \eqref{opt_gen} is equivalent to
\begin{equation}
    \begin{aligned}
        & \underset{\xi}{\text{maximize}} & & \Tr\left(\vect{A} \vect{X} \right) \\
        & \text{subject to}       & &  \Tr\left(\vect{B}\vect{X}\right)  \leq 0 \\
        & & & \Tr\left(\vect{X}\right) \leq 1 \\
        & & & \vect{X} \succeq 0, \quad \operatorname{rank}(\vect{X}) = 1.
        \label{eq:opt_gen2}
    \end{aligned}
\end{equation}
The objective and constraints in \eqref{opt_gen2} are all linear in $\vect{X}$ except for the constraint $\operatorname{rank}(\vect{X}) = 1$. The ``relaxation'' of SDR denotes dropping this rank-one constraint, leaving a linear program whose solution represents an upper bound (for our maximization problem) for the original quadratic program. By the QCQP theory for fewer than four constraints, as discussed above, this bound is tight.

\Figref{metric} compares the analytical bounds of the zero-field optimization problem, {\OPZF}, to the computational SDR bounds for the alternative optimization problems, {\OPFWHM} and {\OPKLOC}. Each bound is computed for many sub-diffraction-limited values of a different parameter: {\OPZF} with respect to the zero-field spot size, {\OPFWHM} with respect to the FWHM spot size, and {\OPKLOC} with respect to the local wavevector (all normalized relative to the respective values for a diffraction-limited beam, cf. \SM). There is no simple mapping between these three parameters, but we can measure those parameters for the optimal beams of each of the three optimization problems, and \figref{metric}(a--c) shows the results. One can see that the optimal beams for the three different problems exhibit nearly identical parameters. This suggests that the beams themselves might be very similar, and this is confirmed in \figref{metric}(d), which shows the optimal beams for the three metrics that all exhibit a normalized zero-field spot size $G=0.61$, and from which it is clear that the beams themselves are nearly identical, especially near the origin and up to the first zero. In this region, the relative difference between each pair of the beams is less than 1\%. This striking similarity suggests two conclusions: first, the optimal beam profiles are ``robust,'' in the sense that changing the concentration definition may result in nearly equivalent beams, and second, that our analytical bounds may describe even the optimal concentration that is possible as defined by a wide variety of metrics. (Of course, one can concoct metrics that must have different bounds; e.g., if the objective is maximum intensity subject to small intensity within some ``window'' that represents a field of view, then certainly the corresponding bounds must depend on the size of the window. However, our analytical expressions would still almost certainly represent upper bounds, since they would represent the limit of a very small window.)
}

\section{Summary and Extensions} \label{sec:sum}
We have established bounds on the maximal concentration of free-space vector electromagnetic waves. We derive bounds for any desired zero-field contour, either incorporating an aperture as in Eqs.~(\ref{eq:Ibound},\ref{eq:Imax},\ref{eq:optS}) or dependent only on a modal basis, as in \eqreftwo{Ibound}{IboundUssz}. By a suitable transformation of the light-concentration problem, we obtain analytic bounds in multiple regimes (small spot sizes, far zone, etc.),  providing insight into the ideal excitation field as well as revealing a dimension-independent quartic spot-size scaling law. Using inverse design, we have theoretically proposed optimal metasurface designs approaching these bounds. {We have also demonstrated that the ideal beam profiles under alternative spot-size metrics are nearly identical to those for the zero-field-contour metric in the far zone.}

Looking forward, there are a number of related questions and application areas where this approach can be applied. {One may further explore other field-concentration metrics. For example, we can ask if it is feasible to specify certain target Maxwell fields on a contour.} We show in the {\SM} that the problem of maximizing focal intensity given a target field profile on some contour can be reduced to a complex-variable QCQP with one constraint. As explained in the previous section, such problems can be solved in polynomial time by SDR. In this way, we obtain exact, global bounds and physically attainable fields. One may also be interested in metrics other than focal-point intensity. One common metric, a minimal-energy metric for a fixed focal-point intensity~\cite{Levi1965,Ferreira2006}, in fact is equivalent to focal-point maximization metric under constraints of fixed energy, as can be shown by comparing the Lagrangian functions of each. Other metrics, potentially focusing on minimizing energy within a specific region (e.g., the field of view), can be seamlessly incorporated into the approach developed here. 

We have considered the medium through which light propagates to be free space (or any homogeneous material, which simply modifies the speed of light in the medium), but our results in fact apply directly to any \emph{inhomogeneous} background by using the corresponding Green's function in \eqref{Ibound}. For near-field imaging with the image plane near some scattering medium, our approach can identify the optimal resolution. It can also potentially be applied to random media~\cite{Mosk2012}, where the Green's function can be appropriately modified through ensemble averaging, to potentially identify optimal concentration within complex disordered media.

{One can also explore bounds to other related phenomena such as superdirectivity~\cite{Schelkunoff1943,Hansen1981}---directivity greater than that obtained with the same antenna configuration uniformly excited. Just as achieving sub-diffraction-limited beams relies on a delicate interference of propagating waves, superdirectivity requires structured antenna arrays with finely tuned excitation amplitudes. Superdirective antennas suffer from diminished efficiency and large sidelobe energies~\cite{Hansen1981}, and the bound techniques developed herein may provide deeper insight or generalized bounds for such phenomena.}

Finally, we can expect that the bounds here can be used as a family of potential point-spread functions across imaging applications. Various emerging techniques at the intersection of quantum optics, metrology, and parameter estimation theory~\cite{Tsang2016,Zhou2019} suggest the possibility for imaging with resolution improvements beyond the classical Rayleigh limit and Airy disk. A tandem of the sub-diffraction-limited point-spread functions provided here with quantum measurement theory may yield even further improvements.



\section{Acknowledgments}
We thank Michael Fiddy and Lawrence Domash for helpful conversations. H. S., H. C., and O. D. M. were partially supported by DARPA and Triton Systems under award number 140D6318C0011, and were partially supported by the Air Force Office of Scientific Research under award number FA9550-17-1-0093.


\begin{thebibliography}{125}%
\makeatletter
\providecommand \@ifxundefined [1]{%
 \@ifx{#1\undefined}
}%
\providecommand \@ifnum [1]{%
 \ifnum #1\expandafter \@firstoftwo
 \else \expandafter \@secondoftwo
 \fi
}%
\providecommand \@ifx [1]{%
 \ifx #1\expandafter \@firstoftwo
 \else \expandafter \@secondoftwo
 \fi
}%
\providecommand \natexlab [1]{#1}%
\providecommand \enquote  [1]{``#1''}%
\providecommand \bibnamefont  [1]{#1}%
\providecommand \bibfnamefont [1]{#1}%
\providecommand \citenamefont [1]{#1}%
\providecommand \href@noop [0]{\@secondoftwo}%
\providecommand \href [0]{\begingroup \@sanitize@url \@href}%
\providecommand \@href[1]{\@@startlink{#1}\@@href}%
\providecommand \@@href[1]{\endgroup#1\@@endlink}%
\providecommand \@sanitize@url [0]{\catcode `\\12\catcode `\$12\catcode
  `\&12\catcode `\#12\catcode `\^12\catcode `\_12\catcode `\%12\relax}%
\providecommand \@@startlink[1]{}%
\providecommand \@@endlink[0]{}%
\providecommand \url  [0]{\begingroup\@sanitize@url \@url }%
\providecommand \@url [1]{\endgroup\@href {#1}{\urlprefix }}%
\providecommand \urlprefix  [0]{URL }%
\providecommand \Eprint [0]{\href }%
\providecommand \doibase [0]{https://doi.org/}%
\providecommand \selectlanguage [0]{\@gobble}%
\providecommand \bibinfo  [0]{\@secondoftwo}%
\providecommand \bibfield  [0]{\@secondoftwo}%
\providecommand \translation [1]{[#1]}%
\providecommand \BibitemOpen [0]{}%
\providecommand \bibitemStop [0]{}%
\providecommand \bibitemNoStop [0]{.\EOS\space}%
\providecommand \EOS [0]{\spacefactor3000\relax}%
\providecommand \BibitemShut  [1]{\csname bibitem#1\endcsname}%
\let\auto@bib@innerbib\@empty
\bibitem [{\citenamefont {McCutchen}(1967)}]{McCutchen1967}%
  \BibitemOpen
  \bibfield  {author} {\bibinfo {author} {\bibfnamefont {C.~W.}\ \bibnamefont
  {McCutchen}},\ }\bibfield  {title} {\bibinfo {title} {{Superresolution in
  Microscopy and the Abbe Resolution Limit}},\ }\href
  {https://doi.org/10.1364/josa.57.001190} {\bibfield  {journal} {\bibinfo
  {journal} {Journal of the Optical Society of America}\ }\textbf {\bibinfo
  {volume} {57}},\ \bibinfo {pages} {1190} (\bibinfo {year}
  {1967})}\BibitemShut {NoStop}%
\bibitem [{\citenamefont {Stelzer}(2002)}]{Stelzer2002}%
  \BibitemOpen
  \bibfield  {author} {\bibinfo {author} {\bibfnamefont {E.~H.~K.}\
  \bibnamefont {Stelzer}},\ }\bibfield  {title} {\bibinfo {title} {{Beyond the
  diffraction limit?}},\ }\href {https://doi.org/10.1038/417806a} {\bibfield
  {journal} {\bibinfo  {journal} {Nature}\ }\textbf {\bibinfo {volume} {417}},\
  \bibinfo {pages} {806} (\bibinfo {year} {2002})}\BibitemShut {NoStop}%
\bibitem [{\citenamefont {Zheludev}(2008)}]{Zheludev2008}%
  \BibitemOpen
  \bibfield  {author} {\bibinfo {author} {\bibfnamefont {N.~I.}\ \bibnamefont
  {Zheludev}},\ }\bibfield  {title} {\bibinfo {title} {{What diffraction
  limit?}},\ }\href {https://doi.org/10.1038/nmat2163} {\bibfield  {journal}
  {\bibinfo  {journal} {Nature Materials}\ }\textbf {\bibinfo {volume} {7}},\
  \bibinfo {pages} {420} (\bibinfo {year} {2008})}\BibitemShut {NoStop}%
\bibitem [{\citenamefont {Gustafsson}(2005)}]{Gustafsson2005}%
  \BibitemOpen
  \bibfield  {author} {\bibinfo {author} {\bibfnamefont {M.~G.~L.}\
  \bibnamefont {Gustafsson}},\ }\bibfield  {title} {\bibinfo {title}
  {{Nonlinear structured-illumination microscopy: Wide-field fluorescence
  imaging with theoretically unlimited resolution}},\ }\href
  {https://doi.org/10.1073/pnas.0406877102} {\bibfield  {journal} {\bibinfo
  {journal} {Proceedings of the National Academy of Sciences}\ }\textbf
  {\bibinfo {volume} {102}},\ \bibinfo {pages} {13081} (\bibinfo {year}
  {2005})}\BibitemShut {NoStop}%
\bibitem [{\citenamefont {Betzig}\ \emph {et~al.}(2006)\citenamefont {Betzig},
  \citenamefont {Patterson}, \citenamefont {Sougrat}, \citenamefont
  {Lindwasser}, \citenamefont {Olenych}, \citenamefont {Bonifacino},
  \citenamefont {Davidson}, \citenamefont {Lippincott-Schwartz},\ and\
  \citenamefont {Hess}}]{Betzig2006}%
  \BibitemOpen
  \bibfield  {author} {\bibinfo {author} {\bibfnamefont {E.}~\bibnamefont
  {Betzig}}, \bibinfo {author} {\bibfnamefont {G.~H.}\ \bibnamefont
  {Patterson}}, \bibinfo {author} {\bibfnamefont {R.}~\bibnamefont {Sougrat}},
  \bibinfo {author} {\bibfnamefont {O.~W.}\ \bibnamefont {Lindwasser}},
  \bibinfo {author} {\bibfnamefont {S.}~\bibnamefont {Olenych}}, \bibinfo
  {author} {\bibfnamefont {J.~S.}\ \bibnamefont {Bonifacino}}, \bibinfo
  {author} {\bibfnamefont {M.~W.}\ \bibnamefont {Davidson}}, \bibinfo {author}
  {\bibfnamefont {J.}~\bibnamefont {Lippincott-Schwartz}},\ and\ \bibinfo
  {author} {\bibfnamefont {H.~F.}\ \bibnamefont {Hess}},\ }\bibfield  {title}
  {\bibinfo {title} {{Imaging Intracellular Fluorescent Proteins at Nanometer
  Resolution}},\ }\href {https://doi.org/10.1126/science.1127344} {\bibfield
  {journal} {\bibinfo  {journal} {Science}\ }\textbf {\bibinfo {volume}
  {313}},\ \bibinfo {pages} {1642} (\bibinfo {year} {2006})}\BibitemShut
  {NoStop}%
\bibitem [{\citenamefont {Rust}\ \emph {et~al.}(2006)\citenamefont {Rust},
  \citenamefont {Bates},\ and\ \citenamefont {Zhuang}}]{Rust2006}%
  \BibitemOpen
  \bibfield  {author} {\bibinfo {author} {\bibfnamefont {M.~J.}\ \bibnamefont
  {Rust}}, \bibinfo {author} {\bibfnamefont {M.}~\bibnamefont {Bates}},\ and\
  \bibinfo {author} {\bibfnamefont {X.}~\bibnamefont {Zhuang}},\ }\bibfield
  {title} {\bibinfo {title} {{Sub-diffraction-limit imaging by stochastic
  optical reconstruction microscopy (STORM)}},\ }\href
  {https://doi.org/10.1038/nmeth929} {\bibfield  {journal} {\bibinfo  {journal}
  {Nature Methods}\ }\textbf {\bibinfo {volume} {3}},\ \bibinfo {pages} {793}
  (\bibinfo {year} {2006})}\BibitemShut {NoStop}%
\bibitem [{\citenamefont {Hell}(2007)}]{Hell2007}%
  \BibitemOpen
  \bibfield  {author} {\bibinfo {author} {\bibfnamefont {S.~W.}\ \bibnamefont
  {Hell}},\ }\bibfield  {title} {\bibinfo {title} {{Far-field optical
  nanoscopy}},\ }\href {https://doi.org/10.1126/science.1137395} {\bibfield
  {journal} {\bibinfo  {journal} {Science}\ }\textbf {\bibinfo {volume}
  {316}},\ \bibinfo {pages} {1153} (\bibinfo {year} {2007})}\BibitemShut
  {NoStop}%
\bibitem [{\citenamefont {Huang}\ \emph {et~al.}(2008)\citenamefont {Huang},
  \citenamefont {Wang}, \citenamefont {Bates},\ and\ \citenamefont
  {Zhuang}}]{Huang2008}%
  \BibitemOpen
  \bibfield  {author} {\bibinfo {author} {\bibfnamefont {B.}~\bibnamefont
  {Huang}}, \bibinfo {author} {\bibfnamefont {W.}~\bibnamefont {Wang}},
  \bibinfo {author} {\bibfnamefont {M.}~\bibnamefont {Bates}},\ and\ \bibinfo
  {author} {\bibfnamefont {X.}~\bibnamefont {Zhuang}},\ }\bibfield  {title}
  {\bibinfo {title} {{Three-Dimensional Super-Resolution Imaging by Stochastic
  Optical Reconstruction Microscopy}},\ }\href
  {https://doi.org/10.1126/science.1153529} {\bibfield  {journal} {\bibinfo
  {journal} {Science}\ }\textbf {\bibinfo {volume} {319}},\ \bibinfo {pages}
  {810} (\bibinfo {year} {2008})}\BibitemShut {NoStop}%
\bibitem [{\citenamefont {Lipson}\ and\ \citenamefont
  {Kurman}(2013)}]{Lipson2013}%
  \BibitemOpen
  \bibfield  {author} {\bibinfo {author} {\bibfnamefont {H.}~\bibnamefont
  {Lipson}}\ and\ \bibinfo {author} {\bibfnamefont {M.}~\bibnamefont
  {Kurman}},\ }\href@noop {} {\emph {\bibinfo {title} {{Fabricated: The new
  world of 3D printing}}}}\ (\bibinfo  {publisher} {John Wiley {\&} Sons},\
  \bibinfo {year} {2013})\BibitemShut {NoStop}%
\bibitem [{\citenamefont {Chia}\ and\ \citenamefont {Wu}(2015)}]{Chia2015}%
  \BibitemOpen
  \bibfield  {author} {\bibinfo {author} {\bibfnamefont {H.~N.}\ \bibnamefont
  {Chia}}\ and\ \bibinfo {author} {\bibfnamefont {B.~M.}\ \bibnamefont {Wu}},\
  }\bibfield  {title} {\bibinfo {title} {{Recent advances in 3D printing of
  biomaterials}},\ }\href {https://doi.org/10.1186/s13036-015-0001-4}
  {\bibfield  {journal} {\bibinfo  {journal} {Journal of Biological
  Engineering}\ }\textbf {\bibinfo {volume} {9}},\ \bibinfo {pages} {4}
  (\bibinfo {year} {2015})}\BibitemShut {NoStop}%
\bibitem [{\citenamefont {Huang}\ and\ \citenamefont
  {Zheludev}(2009)}]{Huang2009}%
  \BibitemOpen
  \bibfield  {author} {\bibinfo {author} {\bibfnamefont {F.~M.}\ \bibnamefont
  {Huang}}\ and\ \bibinfo {author} {\bibfnamefont {N.~I.}\ \bibnamefont
  {Zheludev}},\ }\bibfield  {title} {\bibinfo {title} {{Super-resolution
  without evanescent waves}},\ }\href {https://doi.org/10.1021/nl9002014}
  {\bibfield  {journal} {\bibinfo  {journal} {Nano Letters}\ }\textbf {\bibinfo
  {volume} {9}},\ \bibinfo {pages} {1249} (\bibinfo {year} {2009})}\BibitemShut
  {NoStop}%
\bibitem [{\citenamefont {Rogers}\ \emph {et~al.}(2012)\citenamefont {Rogers},
  \citenamefont {Lindberg}, \citenamefont {Roy}, \citenamefont {Savo},
  \citenamefont {Chad}, \citenamefont {Dennis},\ and\ \citenamefont
  {Zheludev}}]{Rogers2012}%
  \BibitemOpen
  \bibfield  {author} {\bibinfo {author} {\bibfnamefont {E.~T.}\ \bibnamefont
  {Rogers}}, \bibinfo {author} {\bibfnamefont {J.}~\bibnamefont {Lindberg}},
  \bibinfo {author} {\bibfnamefont {T.}~\bibnamefont {Roy}}, \bibinfo {author}
  {\bibfnamefont {S.}~\bibnamefont {Savo}}, \bibinfo {author} {\bibfnamefont
  {J.~E.}\ \bibnamefont {Chad}}, \bibinfo {author} {\bibfnamefont {M.~R.}\
  \bibnamefont {Dennis}},\ and\ \bibinfo {author} {\bibfnamefont {N.~I.}\
  \bibnamefont {Zheludev}},\ }\bibfield  {title} {\bibinfo {title} {{A
  super-oscillatory lens optical microscope for subwavelength imaging}},\
  }\href {https://doi.org/10.1038/nmat3280} {\bibfield  {journal} {\bibinfo
  {journal} {Nature Materials}\ }\textbf {\bibinfo {volume} {11}},\ \bibinfo
  {pages} {432} (\bibinfo {year} {2012})}\BibitemShut {NoStop}%
\bibitem [{\citenamefont {Strehl}(1902)}]{Strehl}%
  \BibitemOpen
  \bibfield  {author} {\bibinfo {author} {\bibfnamefont {K.}~\bibnamefont
  {Strehl}},\ }\bibfield  {title} {\bibinfo {title} {{{\"{U}}ber luftschlieren
  und zonenfehler}},\ }\href@noop {} {\bibfield  {journal} {\bibinfo  {journal}
  {Zeitschrift f{\"{u}}r Instrumentenkunde}\ }\textbf {\bibinfo {volume}
  {22}},\ \bibinfo {pages} {213} (\bibinfo {year} {1902})}\BibitemShut
  {NoStop}%
\bibitem [{\citenamefont {Born}\ and\ \citenamefont {Wolf}(2013)}]{Born2013}%
  \BibitemOpen
  \bibfield  {author} {\bibinfo {author} {\bibfnamefont {M.}~\bibnamefont
  {Born}}\ and\ \bibinfo {author} {\bibfnamefont {E.}~\bibnamefont {Wolf}},\
  }\href@noop {} {\emph {\bibinfo {title} {{Principles of optics:
  electromagnetic theory of propagation, interference and diffraction of
  light}}}}\ (\bibinfo  {publisher} {Elsevier},\ \bibinfo {year}
  {2013})\BibitemShut {NoStop}%
\bibitem [{\citenamefont {Bertsekas}(2016)}]{Bertsekas2016}%
  \BibitemOpen
  \bibfield  {author} {\bibinfo {author} {\bibfnamefont {D.~P.}\ \bibnamefont
  {Bertsekas}},\ }\href@noop {} {\emph {\bibinfo {title} {{Nonlinear
  programming}}}},\ \bibinfo {edition} {3rd}\ ed.\ (\bibinfo  {publisher}
  {Athena Scientific},\ \bibinfo {address} {Belmont, MA},\ \bibinfo {year}
  {2016})\BibitemShut {NoStop}%
\bibitem [{\citenamefont {Slepian}(1961)}]{Slepian1961}%
  \BibitemOpen
  \bibfield  {author} {\bibinfo {author} {\bibfnamefont {D.}~\bibnamefont
  {Slepian}},\ }\bibfield  {title} {\bibinfo {title} {{Prolate Spheroidal Wave
  Functions, Fourier Analysis and Uncertainty — I}},\ }\href
  {https://doi.org/10.1002/j.1538-7305.1961.tb03976.x} {\bibfield  {journal}
  {\bibinfo  {journal} {Bell System Technical Journal}\ }\textbf {\bibinfo
  {volume} {40}},\ \bibinfo {pages} {43} (\bibinfo {year} {1961})}\BibitemShut
  {NoStop}%
\bibitem [{\citenamefont {Landau}\ and\ \citenamefont
  {Pollak}(1961)}]{Landau1961}%
  \BibitemOpen
  \bibfield  {author} {\bibinfo {author} {\bibfnamefont {H.~J.}\ \bibnamefont
  {Landau}}\ and\ \bibinfo {author} {\bibfnamefont {H.~O.}\ \bibnamefont
  {Pollak}},\ }\bibfield  {title} {\bibinfo {title} {{Prolate Spheroidal Wave
  Functions, Fourier Analysis and Uncertainty — II}},\ }\href@noop {}
  {\bibfield  {journal} {\bibinfo  {journal} {Bell System Technical Journal}\
  }\textbf {\bibinfo {volume} {40}},\ \bibinfo {pages} {65} (\bibinfo {year}
  {1961})}\BibitemShut {NoStop}%
\bibitem [{\citenamefont {Ferreira}\ and\ \citenamefont
  {Kempf}(2006)}]{Ferreira2006}%
  \BibitemOpen
  \bibfield  {author} {\bibinfo {author} {\bibfnamefont {P.~J.}\ \bibnamefont
  {Ferreira}}\ and\ \bibinfo {author} {\bibfnamefont {A.}~\bibnamefont
  {Kempf}},\ }\bibfield  {title} {\bibinfo {title} {{Superoscillations: Faster
  than the Nyquist rate}},\ }\href {https://doi.org/10.1109/TSP.2006.877642}
  {\bibfield  {journal} {\bibinfo  {journal} {IEEE Transactions on Signal
  Processing}\ }\textbf {\bibinfo {volume} {54}},\ \bibinfo {pages} {3732}
  (\bibinfo {year} {2006})}\BibitemShut {NoStop}%
\bibitem [{\citenamefont {Jensen}\ and\ \citenamefont
  {Sigmund}(2011)}]{Jensen2011}%
  \BibitemOpen
  \bibfield  {author} {\bibinfo {author} {\bibfnamefont {J.~S.}\ \bibnamefont
  {Jensen}}\ and\ \bibinfo {author} {\bibfnamefont {O.}~\bibnamefont
  {Sigmund}},\ }\bibfield  {title} {\bibinfo {title} {{Topology optimization
  for nano-photonics}},\ }\href {https://doi.org/10.1002/lpor.201000014}
  {\bibfield  {journal} {\bibinfo  {journal} {Laser and Photonics Reviews}\
  }\textbf {\bibinfo {volume} {5}},\ \bibinfo {pages} {308} (\bibinfo {year}
  {2011})}\BibitemShut {NoStop}%
\bibitem [{\citenamefont {Miller}(2012)}]{miller2012photonic}%
  \BibitemOpen
  \bibfield  {author} {\bibinfo {author} {\bibfnamefont {O.~D.}\ \bibnamefont
  {Miller}},\ }\emph {\bibinfo {title} {{Photonic Design: From Fundamental
  Solar Cell Physics to Computational Inverse Design}}},\ \href@noop {} {Ph.D.
  thesis},\ \bibinfo  {school} {University of California, Berkeley} (\bibinfo
  {year} {2012})\BibitemShut {NoStop}%
\bibitem [{\citenamefont {Bendsoe}\ and\ \citenamefont
  {Sigmund}(2013)}]{Bendsoe2013}%
  \BibitemOpen
  \bibfield  {author} {\bibinfo {author} {\bibfnamefont {M.~P.}\ \bibnamefont
  {Bendsoe}}\ and\ \bibinfo {author} {\bibfnamefont {O.}~\bibnamefont
  {Sigmund}},\ }\href@noop {} {\emph {\bibinfo {title} {{Topology optimization:
  theory, methods, and applications}}}}\ (\bibinfo  {publisher} {Springer
  Science {\&} Business Media},\ \bibinfo {year} {2013})\BibitemShut {NoStop}%
\bibitem [{\citenamefont {Molesky}\ \emph {et~al.}(2018)\citenamefont
  {Molesky}, \citenamefont {Lin}, \citenamefont {Piggott}, \citenamefont {Jin},
  \citenamefont {Vuckovi{\'{c}}},\ and\ \citenamefont
  {Rodriguez}}]{Molesky2018}%
  \BibitemOpen
  \bibfield  {author} {\bibinfo {author} {\bibfnamefont {S.}~\bibnamefont
  {Molesky}}, \bibinfo {author} {\bibfnamefont {Z.}~\bibnamefont {Lin}},
  \bibinfo {author} {\bibfnamefont {A.~Y.}\ \bibnamefont {Piggott}}, \bibinfo
  {author} {\bibfnamefont {W.}~\bibnamefont {Jin}}, \bibinfo {author}
  {\bibfnamefont {J.}~\bibnamefont {Vuckovi{\'{c}}}},\ and\ \bibinfo {author}
  {\bibfnamefont {A.~W.}\ \bibnamefont {Rodriguez}},\ }\bibfield  {title}
  {\bibinfo {title} {{Inverse design in nanophotonics}},\ }\href
  {https://doi.org/10.1038/s41566-018-0246-9} {\bibfield  {journal} {\bibinfo
  {journal} {Nature Photonics}\ }\textbf {\bibinfo {volume} {12}},\ \bibinfo
  {pages} {659} (\bibinfo {year} {2018})}\BibitemShut {NoStop}%
\bibitem [{\citenamefont {Kawata}\ \emph {et~al.}(1989)\citenamefont {Kawata},
  \citenamefont {Carter}, \citenamefont {Yen},\ and\ \citenamefont
  {Smith}}]{Kawata1989}%
  \BibitemOpen
  \bibfield  {author} {\bibinfo {author} {\bibfnamefont {H.}~\bibnamefont
  {Kawata}}, \bibinfo {author} {\bibfnamefont {J.~M.}\ \bibnamefont {Carter}},
  \bibinfo {author} {\bibfnamefont {A.}~\bibnamefont {Yen}},\ and\ \bibinfo
  {author} {\bibfnamefont {H.~I.}\ \bibnamefont {Smith}},\ }\bibfield  {title}
  {\bibinfo {title} {{Optical projection lithography using lenses with
  numerical apertures greater than unity}},\ }\href
  {https://doi.org/10.1016/0167-9317(89)90008-7} {\bibfield  {journal}
  {\bibinfo  {journal} {Microelectronic Engineering}\ }\textbf {\bibinfo
  {volume} {9}},\ \bibinfo {pages} {31} (\bibinfo {year} {1989})}\BibitemShut
  {NoStop}%
\bibitem [{\citenamefont {Ito}\ and\ \citenamefont {Okazaki}(2000)}]{Ito2000}%
  \BibitemOpen
  \bibfield  {author} {\bibinfo {author} {\bibfnamefont {T.}~\bibnamefont
  {Ito}}\ and\ \bibinfo {author} {\bibfnamefont {S.}~\bibnamefont {Okazaki}},\
  }\bibfield  {title} {\bibinfo {title} {{Pushing the limits of lithography}},\
  }\href {https://doi.org/10.1038/35023233} {\bibfield  {journal} {\bibinfo
  {journal} {Nature}\ }\textbf {\bibinfo {volume} {406}},\ \bibinfo {pages}
  {1027} (\bibinfo {year} {2000})}\BibitemShut {NoStop}%
\bibitem [{\citenamefont {Odom}\ \emph {et~al.}(2002)\citenamefont {Odom},
  \citenamefont {Thalladi}, \citenamefont {Love},\ and\ \citenamefont
  {Whitesides}}]{Odom2002}%
  \BibitemOpen
  \bibfield  {author} {\bibinfo {author} {\bibfnamefont {T.~W.}\ \bibnamefont
  {Odom}}, \bibinfo {author} {\bibfnamefont {V.~R.}\ \bibnamefont {Thalladi}},
  \bibinfo {author} {\bibfnamefont {J.~C.}\ \bibnamefont {Love}},\ and\
  \bibinfo {author} {\bibfnamefont {G.~M.}\ \bibnamefont {Whitesides}},\
  }\bibfield  {title} {\bibinfo {title} {{Generation of 30-50 nm structures
  using easily fabricated, composite PDMS masks}},\ }\href
  {https://doi.org/10.1021/ja0209464} {\bibfield  {journal} {\bibinfo
  {journal} {Journal of the American Chemical Society}\ }\textbf {\bibinfo
  {volume} {124}},\ \bibinfo {pages} {12112} (\bibinfo {year}
  {2002})}\BibitemShut {NoStop}%
\bibitem [{\citenamefont {Rugar}(1984)}]{Rugar1984}%
  \BibitemOpen
  \bibfield  {author} {\bibinfo {author} {\bibfnamefont {D.}~\bibnamefont
  {Rugar}},\ }\bibfield  {title} {\bibinfo {title} {{Resolution beyond the
  diffraction limit in the acoustic microscope: A nonlinear effect}},\ }\href
  {https://doi.org/10.1063/1.334124} {\bibfield  {journal} {\bibinfo  {journal}
  {Journal of Applied Physics}\ }\textbf {\bibinfo {volume} {56}},\ \bibinfo
  {pages} {1338} (\bibinfo {year} {1984})}\BibitemShut {NoStop}%
\bibitem [{\citenamefont {Indebetouw}\ \emph {et~al.}(2007)\citenamefont
  {Indebetouw}, \citenamefont {Tada}, \citenamefont {Rosen},\ and\
  \citenamefont {Brooker}}]{Indebetouw2007}%
  \BibitemOpen
  \bibfield  {author} {\bibinfo {author} {\bibfnamefont {G.}~\bibnamefont
  {Indebetouw}}, \bibinfo {author} {\bibfnamefont {Y.}~\bibnamefont {Tada}},
  \bibinfo {author} {\bibfnamefont {J.}~\bibnamefont {Rosen}},\ and\ \bibinfo
  {author} {\bibfnamefont {G.}~\bibnamefont {Brooker}},\ }\bibfield  {title}
  {\bibinfo {title} {{Scanning holographic microscopy with resolution exceeding
  the Rayleigh limit of the objective by superposition of off-axis
  holograms}},\ }\href {https://doi.org/10.1364/ao.46.000993} {\bibfield
  {journal} {\bibinfo  {journal} {Applied Optics}\ }\textbf {\bibinfo {volume}
  {46}},\ \bibinfo {pages} {993} (\bibinfo {year} {2007})}\BibitemShut
  {NoStop}%
\bibitem [{\citenamefont {Kumar}\ \emph {et~al.}(2012)\citenamefont {Kumar},
  \citenamefont {Duan}, \citenamefont {Hegde}, \citenamefont {Koh},
  \citenamefont {Wei},\ and\ \citenamefont {Yang}}]{Kumar2012}%
  \BibitemOpen
  \bibfield  {author} {\bibinfo {author} {\bibfnamefont {K.}~\bibnamefont
  {Kumar}}, \bibinfo {author} {\bibfnamefont {H.}~\bibnamefont {Duan}},
  \bibinfo {author} {\bibfnamefont {R.~S.}\ \bibnamefont {Hegde}}, \bibinfo
  {author} {\bibfnamefont {S.~C.}\ \bibnamefont {Koh}}, \bibinfo {author}
  {\bibfnamefont {J.~N.}\ \bibnamefont {Wei}},\ and\ \bibinfo {author}
  {\bibfnamefont {J.~K.}\ \bibnamefont {Yang}},\ }\bibfield  {title} {\bibinfo
  {title} {{Printing colour at the optical diffraction limit}},\ }\href
  {https://doi.org/10.1038/nnano.2012.128} {\bibfield  {journal} {\bibinfo
  {journal} {Nature Nanotechnology}\ }\textbf {\bibinfo {volume} {7}},\
  \bibinfo {pages} {557} (\bibinfo {year} {2012})}\BibitemShut {NoStop}%
\bibitem [{\citenamefont {Collier}(2013)}]{Collier2013}%
  \BibitemOpen
  \bibfield  {author} {\bibinfo {author} {\bibfnamefont {R.}~\bibnamefont
  {Collier}},\ }\href@noop {} {\emph {\bibinfo {title} {{Optical
  holography}}}}\ (\bibinfo  {publisher} {Elsevier},\ \bibinfo {year}
  {2013})\BibitemShut {NoStop}%
\bibitem [{\citenamefont {Moskovits}(1985)}]{Moskovits1985}%
  \BibitemOpen
  \bibfield  {author} {\bibinfo {author} {\bibfnamefont {M.}~\bibnamefont
  {Moskovits}},\ }\bibfield  {title} {\bibinfo {title} {{Surface-enhanced
  spectroscopy}},\ }\href {https://doi.org/10.1103/RevModPhys.57.783}
  {\bibfield  {journal} {\bibinfo  {journal} {Reviews of Modern Physics}\
  }\textbf {\bibinfo {volume} {57}},\ \bibinfo {pages} {783} (\bibinfo {year}
  {1985})}\BibitemShut {NoStop}%
\bibitem [{\citenamefont {Nie}\ and\ \citenamefont {Emory}(1997)}]{Nie1997}%
  \BibitemOpen
  \bibfield  {author} {\bibinfo {author} {\bibfnamefont {S.}~\bibnamefont
  {Nie}}\ and\ \bibinfo {author} {\bibfnamefont {S.~R.}\ \bibnamefont
  {Emory}},\ }\bibfield  {title} {\bibinfo {title} {{Probing single molecules
  and single nanoparticles by surface-enhanced Raman scattering}},\ }\href
  {https://doi.org/10.1126/science.275.5303.1102} {\bibfield  {journal}
  {\bibinfo  {journal} {Science}\ }\textbf {\bibinfo {volume} {275}},\ \bibinfo
  {pages} {1102} (\bibinfo {year} {1997})}\BibitemShut {NoStop}%
\bibitem [{\citenamefont {Kneipp}\ \emph {et~al.}(1997)\citenamefont {Kneipp},
  \citenamefont {Wang}, \citenamefont {Kneipp}, \citenamefont {Perelman},
  \citenamefont {Itzkan}, \citenamefont {Dasari},\ and\ \citenamefont
  {Feld}}]{Kneipp1997}%
  \BibitemOpen
  \bibfield  {author} {\bibinfo {author} {\bibfnamefont {K.}~\bibnamefont
  {Kneipp}}, \bibinfo {author} {\bibfnamefont {Y.}~\bibnamefont {Wang}},
  \bibinfo {author} {\bibfnamefont {H.}~\bibnamefont {Kneipp}}, \bibinfo
  {author} {\bibfnamefont {L.~T.}\ \bibnamefont {Perelman}}, \bibinfo {author}
  {\bibfnamefont {I.}~\bibnamefont {Itzkan}}, \bibinfo {author} {\bibfnamefont
  {R.~R.}\ \bibnamefont {Dasari}},\ and\ \bibinfo {author} {\bibfnamefont
  {M.~S.}\ \bibnamefont {Feld}},\ }\bibfield  {title} {\bibinfo {title}
  {{Single Molecule Detection Using Surface-Enhanced Raman Scattering
  (SERS)}},\ }\href {https://doi.org/10.1103/PhysRevLett.78.1667} {\bibfield
  {journal} {\bibinfo  {journal} {Physical Review Letters}\ }\textbf {\bibinfo
  {volume} {78}},\ \bibinfo {pages} {1667} (\bibinfo {year}
  {1997})}\BibitemShut {NoStop}%
\bibitem [{\citenamefont {Ebbesen}\ \emph {et~al.}(1998)\citenamefont
  {Ebbesen}, \citenamefont {Lezec}, \citenamefont {Ghaemi}, \citenamefont
  {Thio},\ and\ \citenamefont {Wolff}}]{Ebbesen1998}%
  \BibitemOpen
  \bibfield  {author} {\bibinfo {author} {\bibfnamefont {T.~W.}\ \bibnamefont
  {Ebbesen}}, \bibinfo {author} {\bibfnamefont {H.~J.}\ \bibnamefont {Lezec}},
  \bibinfo {author} {\bibfnamefont {H.~F.}\ \bibnamefont {Ghaemi}}, \bibinfo
  {author} {\bibfnamefont {T.}~\bibnamefont {Thio}},\ and\ \bibinfo {author}
  {\bibfnamefont {P.~A.}\ \bibnamefont {Wolff}},\ }\bibfield  {title} {\bibinfo
  {title} {{Extraordinary optical transmission through sub-wavelength hole
  arrays}},\ }\href {https://doi.org/10.1038/35570} {\bibfield  {journal}
  {\bibinfo  {journal} {Nature}\ }\textbf {\bibinfo {volume} {391}},\ \bibinfo
  {pages} {667} (\bibinfo {year} {1998})}\BibitemShut {NoStop}%
\bibitem [{\citenamefont {Mart{\'{i}}n-Moreno}\ \emph
  {et~al.}(2001)\citenamefont {Mart{\'{i}}n-Moreno}, \citenamefont
  {Garc{\'{i}}a-Vidal}, \citenamefont {Lezec}, \citenamefont {Pellerin},
  \citenamefont {Thio}, \citenamefont {Pendry},\ and\ \citenamefont
  {Ebbesen}}]{Moreno2001}%
  \BibitemOpen
  \bibfield  {author} {\bibinfo {author} {\bibfnamefont {L.}~\bibnamefont
  {Mart{\'{i}}n-Moreno}}, \bibinfo {author} {\bibfnamefont {F.~J.}\
  \bibnamefont {Garc{\'{i}}a-Vidal}}, \bibinfo {author} {\bibfnamefont {H.~J.}\
  \bibnamefont {Lezec}}, \bibinfo {author} {\bibfnamefont {K.~M.}\ \bibnamefont
  {Pellerin}}, \bibinfo {author} {\bibfnamefont {T.}~\bibnamefont {Thio}},
  \bibinfo {author} {\bibfnamefont {J.~B.}\ \bibnamefont {Pendry}},\ and\
  \bibinfo {author} {\bibfnamefont {T.~W.}\ \bibnamefont {Ebbesen}},\
  }\bibfield  {title} {\bibinfo {title} {{Theory of extraordinary optical
  transmission through subwavelength hole arrays}},\ }\href
  {https://doi.org/10.1103/PhysRevLett.86.1114} {\bibfield  {journal} {\bibinfo
   {journal} {Physical Review Letters}\ }\textbf {\bibinfo {volume} {86}},\
  \bibinfo {pages} {1114} (\bibinfo {year} {2001})}\BibitemShut {NoStop}%
\bibitem [{\citenamefont {Genet}\ and\ \citenamefont
  {Ebbesen}(2007)}]{Genet2007}%
  \BibitemOpen
  \bibfield  {author} {\bibinfo {author} {\bibfnamefont {C.}~\bibnamefont
  {Genet}}\ and\ \bibinfo {author} {\bibfnamefont {T.~W.}\ \bibnamefont
  {Ebbesen}},\ }\bibfield  {title} {\bibinfo {title} {{Light in tiny holes}},\
  }\href {https://doi.org/10.1038/nature05350} {\bibfield  {journal} {\bibinfo
  {journal} {Nature}\ }\textbf {\bibinfo {volume} {445}},\ \bibinfo {pages}
  {39} (\bibinfo {year} {2007})}\BibitemShut {NoStop}%
\bibitem [{\citenamefont {Fang}\ \emph {et~al.}(2005)\citenamefont {Fang},
  \citenamefont {Lee}, \citenamefont {Sun},\ and\ \citenamefont
  {Zhang}}]{Fang2005}%
  \BibitemOpen
  \bibfield  {author} {\bibinfo {author} {\bibfnamefont {N.}~\bibnamefont
  {Fang}}, \bibinfo {author} {\bibfnamefont {H.}~\bibnamefont {Lee}}, \bibinfo
  {author} {\bibfnamefont {C.}~\bibnamefont {Sun}},\ and\ \bibinfo {author}
  {\bibfnamefont {X.}~\bibnamefont {Zhang}},\ }\bibfield  {title} {\bibinfo
  {title} {{Sub–Diffraction-Limited Optical Imaging with a Silver
  Superlens}},\ }\href {https://doi.org/10.1364/OE.14.008247} {\bibfield
  {journal} {\bibinfo  {journal} {Science}\ }\textbf {\bibinfo {volume}
  {308}},\ \bibinfo {pages} {534} (\bibinfo {year} {2005})}\BibitemShut
  {NoStop}%
\bibitem [{\citenamefont {Kostelak}\ \emph {et~al.}(1991)\citenamefont
  {Kostelak}, \citenamefont {Weiner}, \citenamefont {Betzig}, \citenamefont
  {Harris},\ and\ \citenamefont {Trautman}}]{Kostelak2006}%
  \BibitemOpen
  \bibfield  {author} {\bibinfo {author} {\bibfnamefont {R.~L.}\ \bibnamefont
  {Kostelak}}, \bibinfo {author} {\bibfnamefont {J.~S.}\ \bibnamefont
  {Weiner}}, \bibinfo {author} {\bibfnamefont {E.}~\bibnamefont {Betzig}},
  \bibinfo {author} {\bibfnamefont {T.~D.}\ \bibnamefont {Harris}},\ and\
  \bibinfo {author} {\bibfnamefont {J.~K.}\ \bibnamefont {Trautman}},\
  }\bibfield  {title} {\bibinfo {title} {{Breaking the Diffraction Barrier:
  Optical Microscopy on a Nanometric Scale}},\ }\href
  {https://doi.org/10.1126/science.251.5000.1468} {\bibfield  {journal}
  {\bibinfo  {journal} {Science}\ }\textbf {\bibinfo {volume} {251}},\ \bibinfo
  {pages} {1468} (\bibinfo {year} {1991})}\BibitemShut {NoStop}%
\bibitem [{\citenamefont {Merlin}(2007)}]{Merlin2007}%
  \BibitemOpen
  \bibfield  {author} {\bibinfo {author} {\bibfnamefont {R.}~\bibnamefont
  {Merlin}},\ }\bibfield  {title} {\bibinfo {title} {{Radiationless
  Electromagnetic Interference: Evanescent-Field Lenses and Perfect
  Focusing}},\ }\href {https://doi.org/10.1126/science.1143884} {\bibfield
  {journal} {\bibinfo  {journal} {Science}\ }\textbf {\bibinfo {volume}
  {317}},\ \bibinfo {pages} {927} (\bibinfo {year} {2007})}\BibitemShut
  {NoStop}%
\bibitem [{\citenamefont {Zhang}\ and\ \citenamefont {Liu}(2008)}]{Zhang2008}%
  \BibitemOpen
  \bibfield  {author} {\bibinfo {author} {\bibfnamefont {X.}~\bibnamefont
  {Zhang}}\ and\ \bibinfo {author} {\bibfnamefont {Z.}~\bibnamefont {Liu}},\
  }\bibfield  {title} {\bibinfo {title} {{Superlenses to overcome the
  diffraction limit}},\ }\href {https://doi.org/10.1038/nmat2141} {\bibfield
  {journal} {\bibinfo  {journal} {Nature Materials}\ }\textbf {\bibinfo
  {volume} {7}},\ \bibinfo {pages} {435} (\bibinfo {year} {2008})}\BibitemShut
  {NoStop}%
\bibitem [{\citenamefont {Ma}\ \emph {et~al.}(2018)\citenamefont {Ma},
  \citenamefont {Fan}, \citenamefont {Ma}, \citenamefont {{De Rosny}},
  \citenamefont {Sheng},\ and\ \citenamefont {Fink}}]{Ma2018}%
  \BibitemOpen
  \bibfield  {author} {\bibinfo {author} {\bibfnamefont {G.}~\bibnamefont
  {Ma}}, \bibinfo {author} {\bibfnamefont {X.}~\bibnamefont {Fan}}, \bibinfo
  {author} {\bibfnamefont {F.}~\bibnamefont {Ma}}, \bibinfo {author}
  {\bibfnamefont {J.}~\bibnamefont {{De Rosny}}}, \bibinfo {author}
  {\bibfnamefont {P.}~\bibnamefont {Sheng}},\ and\ \bibinfo {author}
  {\bibfnamefont {M.}~\bibnamefont {Fink}},\ }\bibfield  {title} {\bibinfo
  {title} {{Towards anti-causal Green's function for three-dimensional
  sub-diffraction focusing}},\ }\href
  {https://doi.org/10.1038/s41567-018-0082-3} {\bibfield  {journal} {\bibinfo
  {journal} {Nature Physics}\ }\textbf {\bibinfo {volume} {14}},\ \bibinfo
  {pages} {608} (\bibinfo {year} {2018})}\BibitemShut {NoStop}%
\bibitem [{\citenamefont {{Toraldo di Francia}}(1952)}]{ToraldodiFrancia1952}%
  \BibitemOpen
  \bibfield  {author} {\bibinfo {author} {\bibfnamefont {G.}~\bibnamefont
  {{Toraldo di Francia}}},\ }\bibfield  {title} {\bibinfo {title} {{Super-gain
  antennas and optical resolving power}},\ }\href
  {https://doi.org/10.1007/BF02903413} {\bibfield  {journal} {\bibinfo
  {journal} {Il Nuovo Cimento}\ }\textbf {\bibinfo {volume} {9}},\ \bibinfo
  {pages} {426} (\bibinfo {year} {1952})}\BibitemShut {NoStop}%
\bibitem [{\citenamefont {Schelkunoff}(1943)}]{Schelkunoff1943}%
  \BibitemOpen
  \bibfield  {author} {\bibinfo {author} {\bibfnamefont {S.~A.}\ \bibnamefont
  {Schelkunoff}},\ }\bibfield  {title} {\bibinfo {title} {{A mathematical
  theory of linear arrays}},\ }\href@noop {} {\bibfield  {journal} {\bibinfo
  {journal} {The Bell System Technical Journal}\ }\textbf {\bibinfo {volume}
  {22}},\ \bibinfo {pages} {80} (\bibinfo {year} {1943})}\BibitemShut {NoStop}%
\bibitem [{\citenamefont {Berry}(1994{\natexlab{a}})}]{Berry1994a}%
  \BibitemOpen
  \bibfield  {author} {\bibinfo {author} {\bibfnamefont {M.}~\bibnamefont
  {Berry}},\ }\href@noop {} {\bibinfo {title} {{Faster than Fourier in quantum
  coherence and reality Celebration of the 60th Birthday of Yakir Aharonov ed
  JS Anandan and JL Safko}}} (\bibinfo {year} {1994}{\natexlab{a}})\BibitemShut
  {NoStop}%
\bibitem [{\citenamefont {Berry}(1994{\natexlab{b}})}]{Berry1994}%
  \BibitemOpen
  \bibfield  {author} {\bibinfo {author} {\bibfnamefont {M.~V.}\ \bibnamefont
  {Berry}},\ }\bibfield  {title} {\bibinfo {title} {{Evanescent and real waves
  in quantum billiards and Gaussian beams}},\ }\href
  {https://doi.org/10.1088/0305-4470/27/11/008} {\bibfield  {journal} {\bibinfo
   {journal} {Journal of Physics A: Mathematical and General}\ }\textbf
  {\bibinfo {volume} {27}},\ \bibinfo {pages} {L391} (\bibinfo {year}
  {1994}{\natexlab{b}})}\BibitemShut {NoStop}%
\bibitem [{\citenamefont {Berry}\ and\ \citenamefont
  {Popescu}(2006)}]{Berry2006}%
  \BibitemOpen
  \bibfield  {author} {\bibinfo {author} {\bibfnamefont {M.~V.}\ \bibnamefont
  {Berry}}\ and\ \bibinfo {author} {\bibfnamefont {S.}~\bibnamefont
  {Popescu}},\ }\bibfield  {title} {\bibinfo {title} {{Evolution of quantum
  superoscillations and optical superresolution without evanescent waves}},\
  }\href {https://doi.org/10.1088/0305-4470/39/22/011} {\bibfield  {journal}
  {\bibinfo  {journal} {Journal of Physics A: Mathematical and General}\
  }\textbf {\bibinfo {volume} {39}},\ \bibinfo {pages} {6965} (\bibinfo {year}
  {2006})}\BibitemShut {NoStop}%
\bibitem [{\citenamefont {Lindberg}(2012)}]{Lindberg2012}%
  \BibitemOpen
  \bibfield  {author} {\bibinfo {author} {\bibfnamefont {J.}~\bibnamefont
  {Lindberg}},\ }\bibfield  {title} {\bibinfo {title} {{Mathematical concepts
  of optical superresolution}},\ }\href
  {https://doi.org/10.1088/2040-8978/14/8/083001} {\bibfield  {journal}
  {\bibinfo  {journal} {Journal of Optics}\ }\textbf {\bibinfo {volume} {14}},\
  \bibinfo {pages} {083001} (\bibinfo {year} {2012})}\BibitemShut {NoStop}%
\bibitem [{\citenamefont {Kempf}\ and\ \citenamefont
  {Ferreira}(2004)}]{Kempf2004}%
  \BibitemOpen
  \bibfield  {author} {\bibinfo {author} {\bibfnamefont {A.}~\bibnamefont
  {Kempf}}\ and\ \bibinfo {author} {\bibfnamefont {P.~J.}\ \bibnamefont
  {Ferreira}},\ }\bibfield  {title} {\bibinfo {title} {{Unusual properties of
  superoscillating particles}},\ }\href
  {https://doi.org/10.1088/0305-4470/37/50/009} {\bibfield  {journal} {\bibinfo
   {journal} {Journal of Physics A: Mathematical and General}\ }\textbf
  {\bibinfo {volume} {37}},\ \bibinfo {pages} {12067} (\bibinfo {year}
  {2004})}\BibitemShut {NoStop}%
\bibitem [{\citenamefont {Aharonov}\ \emph {et~al.}(2011)\citenamefont
  {Aharonov}, \citenamefont {Colombo}, \citenamefont {Sabadini}, \citenamefont
  {Struppa},\ and\ \citenamefont {Tollaksen}}]{Aharonov2011}%
  \BibitemOpen
  \bibfield  {author} {\bibinfo {author} {\bibfnamefont {Y.}~\bibnamefont
  {Aharonov}}, \bibinfo {author} {\bibfnamefont {F.}~\bibnamefont {Colombo}},
  \bibinfo {author} {\bibfnamefont {I.}~\bibnamefont {Sabadini}}, \bibinfo
  {author} {\bibfnamefont {D.~C.}\ \bibnamefont {Struppa}},\ and\ \bibinfo
  {author} {\bibfnamefont {J.}~\bibnamefont {Tollaksen}},\ }\bibfield  {title}
  {\bibinfo {title} {{Some mathematical properties of superoscillations}},\
  }\href {https://doi.org/10.1088/1751-8113/44/36/365304} {\bibfield  {journal}
  {\bibinfo  {journal} {Journal of Physics A: Mathematical and Theoretical}\
  }\textbf {\bibinfo {volume} {44}},\ \bibinfo {pages} {365304} (\bibinfo
  {year} {2011})}\BibitemShut {NoStop}%
\bibitem [{\citenamefont {Wong}\ and\ \citenamefont
  {Eleftheriades}(2010)}]{Wong2010}%
  \BibitemOpen
  \bibfield  {author} {\bibinfo {author} {\bibfnamefont {A.~M.~H.}\
  \bibnamefont {Wong}}\ and\ \bibinfo {author} {\bibfnamefont {G.~V.}\
  \bibnamefont {Eleftheriades}},\ }\bibfield  {title} {\bibinfo {title}
  {{Adaptation of Schelkunoff's Superdirective Antenna Theory for the
  Realization of Superoscillatory Antenna Arrays}},\ }\href
  {https://doi.org/10.1109/LAWP.2010.2047710} {\bibfield  {journal} {\bibinfo
  {journal} {IEEE Antennas and Wireless Propagation Letters}\ }\textbf
  {\bibinfo {volume} {9}},\ \bibinfo {pages} {315} (\bibinfo {year}
  {2010})}\BibitemShut {NoStop}%
\bibitem [{\citenamefont {Chremmos}\ and\ \citenamefont
  {Fikioris}(2015)}]{Chremmos2015}%
  \BibitemOpen
  \bibfield  {author} {\bibinfo {author} {\bibfnamefont {I.}~\bibnamefont
  {Chremmos}}\ and\ \bibinfo {author} {\bibfnamefont {G.}~\bibnamefont
  {Fikioris}},\ }\bibfield  {title} {\bibinfo {title} {{Superoscillations with
  arbitrary polynomial shape}},\ }\href
  {https://doi.org/10.1088/1751-8113/48/26/265204} {\bibfield  {journal}
  {\bibinfo  {journal} {Journal of Physics A: Mathematical and Theoretical}\
  }\textbf {\bibinfo {volume} {48}},\ \bibinfo {pages} {265204} (\bibinfo
  {year} {2015})}\BibitemShut {NoStop}%
\bibitem [{\citenamefont {Smith}\ and\ \citenamefont {Gbur}(2016)}]{Smith2016}%
  \BibitemOpen
  \bibfield  {author} {\bibinfo {author} {\bibfnamefont {M.~K.}\ \bibnamefont
  {Smith}}\ and\ \bibinfo {author} {\bibfnamefont {G.~J.}\ \bibnamefont
  {Gbur}},\ }\bibfield  {title} {\bibinfo {title} {{Construction of arbitrary
  vortex and superoscillatory fields}},\ }\href
  {https://doi.org/10.1364/OL.41.004979} {\bibfield  {journal} {\bibinfo
  {journal} {Optics Letters}\ }\textbf {\bibinfo {volume} {41}},\ \bibinfo
  {pages} {4979} (\bibinfo {year} {2016})}\BibitemShut {NoStop}%
\bibitem [{\citenamefont {Bassett}(1986)}]{Bassett1986}%
  \BibitemOpen
  \bibfield  {author} {\bibinfo {author} {\bibfnamefont {I.~M.}\ \bibnamefont
  {Bassett}},\ }\bibfield  {title} {\bibinfo {title} {{Limit to concentration
  by focusing}},\ }\href {https://doi.org/10.1080/713821943} {\bibfield
  {journal} {\bibinfo  {journal} {Optica Acta}\ }\textbf {\bibinfo {volume}
  {33}},\ \bibinfo {pages} {279} (\bibinfo {year} {1986})}\BibitemShut
  {NoStop}%
\bibitem [{\citenamefont {Sheppard}\ and\ \citenamefont
  {Larkin}(1994)}]{Sheppard1994}%
  \BibitemOpen
  \bibfield  {author} {\bibinfo {author} {\bibfnamefont {C.~J.~R.}\
  \bibnamefont {Sheppard}}\ and\ \bibinfo {author} {\bibfnamefont {K.~G.}\
  \bibnamefont {Larkin}},\ }\bibfield  {title} {\bibinfo {title} {{Optimal
  concentration of electromagnetic radiation}},\ }\href
  {https://doi.org/10.1080/09500349414551421} {\bibfield  {journal} {\bibinfo
  {journal} {Journal of Modern Optics}\ }\textbf {\bibinfo {volume} {41}},\
  \bibinfo {pages} {1495} (\bibinfo {year} {1994})}\BibitemShut {NoStop}%
\bibitem [{\citenamefont {Sales}\ and\ \citenamefont
  {Morris}(1997)}]{Sales1997}%
  \BibitemOpen
  \bibfield  {author} {\bibinfo {author} {\bibfnamefont {T.~R.~M.}\
  \bibnamefont {Sales}}\ and\ \bibinfo {author} {\bibfnamefont {G.~M.}\
  \bibnamefont {Morris}},\ }\bibfield  {title} {\bibinfo {title} {{Fundamental
  limits of optical superresolution}},\ }\href
  {https://doi.org/10.1364/OL.22.000582} {\bibfield  {journal} {\bibinfo
  {journal} {Optics Letters}\ }\textbf {\bibinfo {volume} {22}},\ \bibinfo
  {pages} {582} (\bibinfo {year} {1997})}\BibitemShut {NoStop}%
\bibitem [{\citenamefont {Liu}\ \emph {et~al.}(2002)\citenamefont {Liu},
  \citenamefont {Yan}, \citenamefont {Tan},\ and\ \citenamefont
  {Jin}}]{Liu2002}%
  \BibitemOpen
  \bibfield  {author} {\bibinfo {author} {\bibfnamefont {H.}~\bibnamefont
  {Liu}}, \bibinfo {author} {\bibfnamefont {Y.}~\bibnamefont {Yan}}, \bibinfo
  {author} {\bibfnamefont {Q.}~\bibnamefont {Tan}},\ and\ \bibinfo {author}
  {\bibfnamefont {G.}~\bibnamefont {Jin}},\ }\bibfield  {title} {\bibinfo
  {title} {{Theories for the design of diffractive superresolution elements and
  limits of optical superresolution}},\ }\href
  {https://doi.org/10.1364/josaa.19.002185} {\bibfield  {journal} {\bibinfo
  {journal} {Journal of the Optical Society of America A}\ }\textbf {\bibinfo
  {volume} {19}},\ \bibinfo {pages} {2185} (\bibinfo {year}
  {2002})}\BibitemShut {NoStop}%
\bibitem [{\citenamefont {Rogers}\ \emph {et~al.}(2018)\citenamefont {Rogers},
  \citenamefont {Bourdakos}, \citenamefont {Yuan}, \citenamefont {Mahajan},\
  and\ \citenamefont {Rogers}}]{Rogers2018}%
  \BibitemOpen
  \bibfield  {author} {\bibinfo {author} {\bibfnamefont {K.~S.}\ \bibnamefont
  {Rogers}}, \bibinfo {author} {\bibfnamefont {K.~N.}\ \bibnamefont
  {Bourdakos}}, \bibinfo {author} {\bibfnamefont {G.~H.}\ \bibnamefont {Yuan}},
  \bibinfo {author} {\bibfnamefont {S.}~\bibnamefont {Mahajan}},\ and\ \bibinfo
  {author} {\bibfnamefont {E.~T.~F.}\ \bibnamefont {Rogers}},\ }\bibfield
  {title} {\bibinfo {title} {{Optimising superoscillatory spots for far-field
  super-resolution imaging}},\ }\href {https://doi.org/10.1364/oe.26.008095}
  {\bibfield  {journal} {\bibinfo  {journal} {Optics Express}\ }\textbf
  {\bibinfo {volume} {26}},\ \bibinfo {pages} {8095} (\bibinfo {year}
  {2018})}\BibitemShut {NoStop}%
\bibitem [{\citenamefont {Miller}(2019)}]{Miller2019a}%
  \BibitemOpen
  \bibfield  {author} {\bibinfo {author} {\bibfnamefont {D.~A.~B.}\
  \bibnamefont {Miller}},\ }\bibfield  {title} {\bibinfo {title} {{Waves,
  modes, communications, and optics: a tutorial}},\ }\href
  {https://doi.org/10.1364/AOP.11.000679} {\bibfield  {journal} {\bibinfo
  {journal} {Advances in Optics and Photonics}\ }\textbf {\bibinfo {volume}
  {11}},\ \bibinfo {pages} {679} (\bibinfo {year} {2019})}\BibitemShut
  {NoStop}%
\bibitem [{\citenamefont {Pfeiffer}\ and\ \citenamefont
  {Grbic}(2013)}]{Pfeiffer2013}%
  \BibitemOpen
  \bibfield  {author} {\bibinfo {author} {\bibfnamefont {C.}~\bibnamefont
  {Pfeiffer}}\ and\ \bibinfo {author} {\bibfnamefont {A.}~\bibnamefont
  {Grbic}},\ }\bibfield  {title} {\bibinfo {title} {{Metamaterial Huygens'
  Surfaces: Tailoring Wave Fronts with Reflectionless Sheets}},\ }\href
  {https://doi.org/10.1103/PhysRevLett.110.197401} {\bibfield  {journal}
  {\bibinfo  {journal} {Physical Review Letters}\ }\textbf {\bibinfo {volume}
  {110}},\ \bibinfo {pages} {197401} (\bibinfo {year} {2013})}\BibitemShut
  {NoStop}%
\bibitem [{\citenamefont {Huang}\ \emph
  {et~al.}(2007{\natexlab{a}})\citenamefont {Huang}, \citenamefont {Zheludev},
  \citenamefont {Chen},\ and\ \citenamefont {{Javier Garcia De
  Abajo}}}]{Huang2007}%
  \BibitemOpen
  \bibfield  {author} {\bibinfo {author} {\bibfnamefont {F.~M.}\ \bibnamefont
  {Huang}}, \bibinfo {author} {\bibfnamefont {N.}~\bibnamefont {Zheludev}},
  \bibinfo {author} {\bibfnamefont {Y.}~\bibnamefont {Chen}},\ and\ \bibinfo
  {author} {\bibfnamefont {F.}~\bibnamefont {{Javier Garcia De Abajo}}},\
  }\bibfield  {title} {\bibinfo {title} {{Focusing of light by a nanohole
  array}},\ }\href {https://doi.org/10.1063/1.2710775} {\bibfield  {journal}
  {\bibinfo  {journal} {Applied Physics Letters}\ }\textbf {\bibinfo {volume}
  {90}},\ \bibinfo {pages} {091119} (\bibinfo {year}
  {2007}{\natexlab{a}})}\BibitemShut {NoStop}%
\bibitem [{\citenamefont {Huang}\ \emph
  {et~al.}(2007{\natexlab{b}})\citenamefont {Huang}, \citenamefont {Chen},
  \citenamefont {{Garcia De Abajo}},\ and\ \citenamefont
  {Zheludev}}]{Huang2007a}%
  \BibitemOpen
  \bibfield  {author} {\bibinfo {author} {\bibfnamefont {F.~M.}\ \bibnamefont
  {Huang}}, \bibinfo {author} {\bibfnamefont {Y.}~\bibnamefont {Chen}},
  \bibinfo {author} {\bibfnamefont {F.~J.}\ \bibnamefont {{Garcia De Abajo}}},\
  and\ \bibinfo {author} {\bibfnamefont {N.~I.}\ \bibnamefont {Zheludev}},\
  }\bibfield  {title} {\bibinfo {title} {{Optical super-resolution through
  super-oscillations}},\ }\href {https://doi.org/10.1088/1464-4258/9/9/S01}
  {\bibfield  {journal} {\bibinfo  {journal} {Journal of Optics A: Pure and
  Applied Optics}\ }\textbf {\bibinfo {volume} {9}},\ \bibinfo {pages} {S285}
  (\bibinfo {year} {2007}{\natexlab{b}})}\BibitemShut {NoStop}%
\bibitem [{\citenamefont {Dennis}\ \emph {et~al.}(2008)\citenamefont {Dennis},
  \citenamefont {Hamilton},\ and\ \citenamefont {Courtial}}]{Dennis2008}%
  \BibitemOpen
  \bibfield  {author} {\bibinfo {author} {\bibfnamefont {M.~R.}\ \bibnamefont
  {Dennis}}, \bibinfo {author} {\bibfnamefont {A.~C.}\ \bibnamefont
  {Hamilton}},\ and\ \bibinfo {author} {\bibfnamefont {J.}~\bibnamefont
  {Courtial}},\ }\bibfield  {title} {\bibinfo {title} {{Superoscillation in
  speckle patterns}},\ }\href {https://doi.org/10.1364/OL.33.002976} {\ \textbf
  {\bibinfo {volume} {33}},\ \bibinfo {pages} {2976} (\bibinfo {year}
  {2008})}\BibitemShut {NoStop}%
\bibitem [{\citenamefont {Wang}\ \emph {et~al.}(2009)\citenamefont {Wang},
  \citenamefont {Fu}, \citenamefont {Liu},\ and\ \citenamefont
  {Tong}}]{Wang2009}%
  \BibitemOpen
  \bibfield  {author} {\bibinfo {author} {\bibfnamefont {X.}~\bibnamefont
  {Wang}}, \bibinfo {author} {\bibfnamefont {J.}~\bibnamefont {Fu}}, \bibinfo
  {author} {\bibfnamefont {X.}~\bibnamefont {Liu}},\ and\ \bibinfo {author}
  {\bibfnamefont {L.-M.}\ \bibnamefont {Tong}},\ }\bibfield  {title} {\bibinfo
  {title} {{Subwavelength focusing by a micro/nanofiber array}},\ }\href
  {https://doi.org/10.1364/josaa.26.001827} {\bibfield  {journal} {\bibinfo
  {journal} {Journal of the Optical Society of America A}\ }\textbf {\bibinfo
  {volume} {26}},\ \bibinfo {pages} {1827} (\bibinfo {year}
  {2009})}\BibitemShut {NoStop}%
\bibitem [{\citenamefont {Kitamura}\ \emph {et~al.}(2010)\citenamefont
  {Kitamura}, \citenamefont {Sakai},\ and\ \citenamefont
  {Noda}}]{Kitamura2010}%
  \BibitemOpen
  \bibfield  {author} {\bibinfo {author} {\bibfnamefont {K.}~\bibnamefont
  {Kitamura}}, \bibinfo {author} {\bibfnamefont {K.}~\bibnamefont {Sakai}},\
  and\ \bibinfo {author} {\bibfnamefont {S.}~\bibnamefont {Noda}},\ }\bibfield
  {title} {\bibinfo {title} {{Sub-wavelength focal spot with long depth of
  focus generated by radially polarized, narrow-width annular beam}},\ }\href
  {https://doi.org/10.1364/OE.18.004518} {\bibfield  {journal} {\bibinfo
  {journal} {Optics Express}\ }\textbf {\bibinfo {volume} {18}},\ \bibinfo
  {pages} {4518} (\bibinfo {year} {2010})}\BibitemShut {NoStop}%
\bibitem [{\citenamefont {Rogers}\ \emph {et~al.}(2013)\citenamefont {Rogers},
  \citenamefont {Savo}, \citenamefont {Lindberg}, \citenamefont {Roy},
  \citenamefont {Dennis},\ and\ \citenamefont {Zheludev}}]{Rogers2013}%
  \BibitemOpen
  \bibfield  {author} {\bibinfo {author} {\bibfnamefont {E.~T.}\ \bibnamefont
  {Rogers}}, \bibinfo {author} {\bibfnamefont {S.}~\bibnamefont {Savo}},
  \bibinfo {author} {\bibfnamefont {J.}~\bibnamefont {Lindberg}}, \bibinfo
  {author} {\bibfnamefont {T.}~\bibnamefont {Roy}}, \bibinfo {author}
  {\bibfnamefont {M.~R.}\ \bibnamefont {Dennis}},\ and\ \bibinfo {author}
  {\bibfnamefont {N.~I.}\ \bibnamefont {Zheludev}},\ }\bibfield  {title}
  {\bibinfo {title} {{Super-oscillatory optical needle}},\ }\href
  {https://doi.org/10.1063/1.4774385} {\bibfield  {journal} {\bibinfo
  {journal} {Applied Physics Letters}\ }\textbf {\bibinfo {volume} {102}},\
  \bibinfo {pages} {031108} (\bibinfo {year} {2013})}\BibitemShut {NoStop}%
\bibitem [{\citenamefont {Huang}\ \emph {et~al.}(2013)\citenamefont {Huang},
  \citenamefont {Chen}, \citenamefont {M{\"{u}}hlenbernd}, \citenamefont
  {Zhang}, \citenamefont {Chen}, \citenamefont {Bai}, \citenamefont {Tan},
  \citenamefont {Jin}, \citenamefont {Cheah}, \citenamefont {Qiu},
  \citenamefont {Li}, \citenamefont {Zentgraf},\ and\ \citenamefont
  {Zhang}}]{Huang2013}%
  \BibitemOpen
  \bibfield  {author} {\bibinfo {author} {\bibfnamefont {L.}~\bibnamefont
  {Huang}}, \bibinfo {author} {\bibfnamefont {X.}~\bibnamefont {Chen}},
  \bibinfo {author} {\bibfnamefont {H.}~\bibnamefont {M{\"{u}}hlenbernd}},
  \bibinfo {author} {\bibfnamefont {H.}~\bibnamefont {Zhang}}, \bibinfo
  {author} {\bibfnamefont {S.}~\bibnamefont {Chen}}, \bibinfo {author}
  {\bibfnamefont {B.}~\bibnamefont {Bai}}, \bibinfo {author} {\bibfnamefont
  {Q.}~\bibnamefont {Tan}}, \bibinfo {author} {\bibfnamefont {G.}~\bibnamefont
  {Jin}}, \bibinfo {author} {\bibfnamefont {K.~W.}\ \bibnamefont {Cheah}},
  \bibinfo {author} {\bibfnamefont {C.~W.}\ \bibnamefont {Qiu}}, \bibinfo
  {author} {\bibfnamefont {J.}~\bibnamefont {Li}}, \bibinfo {author}
  {\bibfnamefont {T.}~\bibnamefont {Zentgraf}},\ and\ \bibinfo {author}
  {\bibfnamefont {S.}~\bibnamefont {Zhang}},\ }\bibfield  {title} {\bibinfo
  {title} {{Three-dimensional optical holography using a plasmonic
  metasurface}},\ }\href {https://doi.org/10.1038/ncomms3808} {\bibfield
  {journal} {\bibinfo  {journal} {Nature Communications}\ }\textbf {\bibinfo
  {volume} {4}},\ \bibinfo {pages} {2808} (\bibinfo {year} {2013})}\BibitemShut
  {NoStop}%
\bibitem [{\citenamefont {Huang}\ \emph {et~al.}(2014)\citenamefont {Huang},
  \citenamefont {Ye}, \citenamefont {Teng}, \citenamefont {Yeo}, \citenamefont
  {Luk'yanchuk},\ and\ \citenamefont {Qiu}}]{Huang2014}%
  \BibitemOpen
  \bibfield  {author} {\bibinfo {author} {\bibfnamefont {K.}~\bibnamefont
  {Huang}}, \bibinfo {author} {\bibfnamefont {H.}~\bibnamefont {Ye}}, \bibinfo
  {author} {\bibfnamefont {J.}~\bibnamefont {Teng}}, \bibinfo {author}
  {\bibfnamefont {S.~P.}\ \bibnamefont {Yeo}}, \bibinfo {author} {\bibfnamefont
  {B.}~\bibnamefont {Luk'yanchuk}},\ and\ \bibinfo {author} {\bibfnamefont
  {C.~W.}\ \bibnamefont {Qiu}},\ }\bibfield  {title} {\bibinfo {title}
  {{Optimization-free superoscillatory lens using phase and amplitude masks}},\
  }\href {https://doi.org/10.1002/lpor.201300123} {\bibfield  {journal}
  {\bibinfo  {journal} {Laser and Photonics Reviews}\ }\textbf {\bibinfo
  {volume} {8}},\ \bibinfo {pages} {152} (\bibinfo {year} {2014})}\BibitemShut
  {NoStop}%
\bibitem [{\citenamefont {Qin}\ \emph {et~al.}(2015)\citenamefont {Qin},
  \citenamefont {Huang}, \citenamefont {Wu}, \citenamefont {Jiao},
  \citenamefont {Luo}, \citenamefont {Qiu},\ and\ \citenamefont
  {Hong}}]{Qin2015}%
  \BibitemOpen
  \bibfield  {author} {\bibinfo {author} {\bibfnamefont {F.}~\bibnamefont
  {Qin}}, \bibinfo {author} {\bibfnamefont {K.}~\bibnamefont {Huang}}, \bibinfo
  {author} {\bibfnamefont {J.}~\bibnamefont {Wu}}, \bibinfo {author}
  {\bibfnamefont {J.}~\bibnamefont {Jiao}}, \bibinfo {author} {\bibfnamefont
  {X.}~\bibnamefont {Luo}}, \bibinfo {author} {\bibfnamefont {C.}~\bibnamefont
  {Qiu}},\ and\ \bibinfo {author} {\bibfnamefont {M.}~\bibnamefont {Hong}},\
  }\bibfield  {title} {\bibinfo {title} {{Shaping a subwavelength needle with
  ultra-long focal length by focusing azimuthally polarized light}},\ }\href
  {https://doi.org/10.1038/srep09977} {\bibfield  {journal} {\bibinfo
  {journal} {Scientific Reports}\ }\textbf {\bibinfo {volume} {5}},\ \bibinfo
  {pages} {9977} (\bibinfo {year} {2015})}\BibitemShut {NoStop}%
\bibitem [{\citenamefont {Wong}\ and\ \citenamefont
  {Eleftheriades}(2015)}]{Wong2015}%
  \BibitemOpen
  \bibfield  {author} {\bibinfo {author} {\bibfnamefont {A.~M.}\ \bibnamefont
  {Wong}}\ and\ \bibinfo {author} {\bibfnamefont {G.~V.}\ \bibnamefont
  {Eleftheriades}},\ }\bibfield  {title} {\bibinfo {title} {{Superoscillations
  without Sidebands: Power-Efficient Sub-Diffraction Imaging with Propagating
  Waves}},\ }\href {https://doi.org/10.1038/srep08449} {\bibfield  {journal}
  {\bibinfo  {journal} {Scientific Reports}\ }\textbf {\bibinfo {volume} {5}},\
  \bibinfo {pages} {8449} (\bibinfo {year} {2015})}\BibitemShut {NoStop}%
\bibitem [{\citenamefont {Wong}\ and\ \citenamefont
  {Eleftheriades}(2017)}]{Wong2017}%
  \BibitemOpen
  \bibfield  {author} {\bibinfo {author} {\bibfnamefont {A.~M.}\ \bibnamefont
  {Wong}}\ and\ \bibinfo {author} {\bibfnamefont {G.~V.}\ \bibnamefont
  {Eleftheriades}},\ }\bibfield  {title} {\bibinfo {title} {{Broadband
  superoscillation brings a wave into perfect three-dimensional focus}},\
  }\href {https://doi.org/10.1103/PhysRevB.95.075148} {\bibfield  {journal}
  {\bibinfo  {journal} {Physical Review B}\ }\textbf {\bibinfo {volume} {95}},\
  \bibinfo {pages} {075148} (\bibinfo {year} {2017})}\BibitemShut {NoStop}%
\bibitem [{\citenamefont {Kildishev}\ \emph {et~al.}(2013)\citenamefont
  {Kildishev}, \citenamefont {Boltasseva},\ and\ \citenamefont
  {Shalaev}}]{Kildishev2013}%
  \BibitemOpen
  \bibfield  {author} {\bibinfo {author} {\bibfnamefont {A.~V.}\ \bibnamefont
  {Kildishev}}, \bibinfo {author} {\bibfnamefont {A.}~\bibnamefont
  {Boltasseva}},\ and\ \bibinfo {author} {\bibfnamefont {V.~M.}\ \bibnamefont
  {Shalaev}},\ }\bibfield  {title} {\bibinfo {title} {{Planar photonics with
  metasurfaces}},\ }\href {https://doi.org/10.1038/nmat3431} {\bibfield
  {journal} {\bibinfo  {journal} {Science}\ }\textbf {\bibinfo {volume}
  {339}},\ \bibinfo {pages} {1232009} (\bibinfo {year} {2013})}\BibitemShut
  {NoStop}%
\bibitem [{\citenamefont {Yu}\ and\ \citenamefont {Capasso}(2014)}]{Yu2014}%
  \BibitemOpen
  \bibfield  {author} {\bibinfo {author} {\bibfnamefont {N.}~\bibnamefont
  {Yu}}\ and\ \bibinfo {author} {\bibfnamefont {F.}~\bibnamefont {Capasso}},\
  }\bibfield  {title} {\bibinfo {title} {{Flat optics with designer
  metasurfaces}},\ }\href {https://doi.org/10.1038/nmat3839} {\bibfield
  {journal} {\bibinfo  {journal} {Nature Materials}\ }\textbf {\bibinfo
  {volume} {13}},\ \bibinfo {pages} {139} (\bibinfo {year} {2014})}\BibitemShut
  {NoStop}%
\bibitem [{\citenamefont {Khorasaninejad}\ \emph {et~al.}(2016)\citenamefont
  {Khorasaninejad}, \citenamefont {Chen}, \citenamefont {Devlin}, \citenamefont
  {Oh}, \citenamefont {Zhu},\ and\ \citenamefont
  {Capasso}}]{Khorasaninejad2016}%
  \BibitemOpen
  \bibfield  {author} {\bibinfo {author} {\bibfnamefont {M.}~\bibnamefont
  {Khorasaninejad}}, \bibinfo {author} {\bibfnamefont {W.~T.}\ \bibnamefont
  {Chen}}, \bibinfo {author} {\bibfnamefont {R.~C.}\ \bibnamefont {Devlin}},
  \bibinfo {author} {\bibfnamefont {J.}~\bibnamefont {Oh}}, \bibinfo {author}
  {\bibfnamefont {A.~Y.}\ \bibnamefont {Zhu}},\ and\ \bibinfo {author}
  {\bibfnamefont {F.}~\bibnamefont {Capasso}},\ }\bibfield  {title} {\bibinfo
  {title} {{Metalenses at visible wavelengths: Diffraction-limited focusing and
  subwavelength resolution imaging}},\ }\href
  {https://doi.org/10.1126/science.aaf6644} {\bibfield  {journal} {\bibinfo
  {journal} {Science}\ }\textbf {\bibinfo {volume} {352}},\ \bibinfo {pages}
  {1190} (\bibinfo {year} {2016})}\BibitemShut {NoStop}%
\bibitem [{\citenamefont {Li}\ \emph {et~al.}(2018)\citenamefont {Li},
  \citenamefont {Singh},\ and\ \citenamefont {Sievenpiper}}]{Li2018}%
  \BibitemOpen
  \bibfield  {author} {\bibinfo {author} {\bibfnamefont {A.}~\bibnamefont
  {Li}}, \bibinfo {author} {\bibfnamefont {S.}~\bibnamefont {Singh}},\ and\
  \bibinfo {author} {\bibfnamefont {D.}~\bibnamefont {Sievenpiper}},\
  }\bibfield  {title} {\bibinfo {title} {{Metasurfaces and their
  applications}},\ }\href {https://doi.org/10.1515/nanoph-2017-0120} {\bibfield
   {journal} {\bibinfo  {journal} {Nanophotonics}\ }\textbf {\bibinfo {volume}
  {7}},\ \bibinfo {pages} {989} (\bibinfo {year} {2018})}\BibitemShut {NoStop}%
\bibitem [{\citenamefont {Serrels}\ \emph {et~al.}(2008)\citenamefont
  {Serrels}, \citenamefont {Ramsay}, \citenamefont {Warburton},\ and\
  \citenamefont {Reid}}]{Serrels2008}%
  \BibitemOpen
  \bibfield  {author} {\bibinfo {author} {\bibfnamefont {K.~A.}\ \bibnamefont
  {Serrels}}, \bibinfo {author} {\bibfnamefont {E.}~\bibnamefont {Ramsay}},
  \bibinfo {author} {\bibfnamefont {R.~J.}\ \bibnamefont {Warburton}},\ and\
  \bibinfo {author} {\bibfnamefont {D.~T.}\ \bibnamefont {Reid}},\ }\bibfield
  {title} {\bibinfo {title} {{Nanoscale optical microscopy in the vectorial
  focusing regime}},\ }\href {https://doi.org/10.1038/nphoton.2008.29}
  {\bibfield  {journal} {\bibinfo  {journal} {Nature Photonics}\ }\textbf
  {\bibinfo {volume} {2}},\ \bibinfo {pages} {311} (\bibinfo {year}
  {2008})}\BibitemShut {NoStop}%
\bibitem [{\citenamefont {Jabbour}\ and\ \citenamefont
  {Kuebler}(2008)}]{Jabbour2008}%
  \BibitemOpen
  \bibfield  {author} {\bibinfo {author} {\bibfnamefont {T.~G.}\ \bibnamefont
  {Jabbour}}\ and\ \bibinfo {author} {\bibfnamefont {S.~M.}\ \bibnamefont
  {Kuebler}},\ }\bibfield  {title} {\bibinfo {title} {{Vectorial beam
  shaping}},\ }\href {https://doi.org/10.1364/OE.16.007203} {\bibfield
  {journal} {\bibinfo  {journal} {Optics Express}\ }\textbf {\bibinfo {volume}
  {16}},\ \bibinfo {pages} {7203} (\bibinfo {year} {2008})}\BibitemShut
  {NoStop}%
\bibitem [{\citenamefont {Bauer}\ \emph {et~al.}(2014)\citenamefont {Bauer},
  \citenamefont {Orlov}, \citenamefont {Peschel}, \citenamefont {Banzer},\ and\
  \citenamefont {Leuchs}}]{Bauer2014}%
  \BibitemOpen
  \bibfield  {author} {\bibinfo {author} {\bibfnamefont {T.}~\bibnamefont
  {Bauer}}, \bibinfo {author} {\bibfnamefont {S.}~\bibnamefont {Orlov}},
  \bibinfo {author} {\bibfnamefont {U.}~\bibnamefont {Peschel}}, \bibinfo
  {author} {\bibfnamefont {P.}~\bibnamefont {Banzer}},\ and\ \bibinfo {author}
  {\bibfnamefont {G.}~\bibnamefont {Leuchs}},\ }\bibfield  {title} {\bibinfo
  {title} {{Nanointerferometric amplitude and phase reconstruction of tightly
  focused vector beams}},\ }\href {https://doi.org/10.1038/nphoton.2013.289}
  {\bibfield  {journal} {\bibinfo  {journal} {Nature Photonics}\ }\textbf
  {\bibinfo {volume} {8}},\ \bibinfo {pages} {23} (\bibinfo {year}
  {2014})}\BibitemShut {NoStop}%
\bibitem [{\citenamefont {Lin}\ \emph {et~al.}(2019)\citenamefont {Lin},
  \citenamefont {Liu}, \citenamefont {Pestourie},\ and\ \citenamefont
  {Johnson}}]{Lin2019}%
  \BibitemOpen
  \bibfield  {author} {\bibinfo {author} {\bibfnamefont {Z.}~\bibnamefont
  {Lin}}, \bibinfo {author} {\bibfnamefont {V.}~\bibnamefont {Liu}}, \bibinfo
  {author} {\bibfnamefont {R.}~\bibnamefont {Pestourie}},\ and\ \bibinfo
  {author} {\bibfnamefont {S.~G.}\ \bibnamefont {Johnson}},\ }\bibfield
  {title} {\bibinfo {title} {{Topology optimization of freeform large-area
  metasurfaces}},\ }\href {https://doi.org/10.1364/OE.27.015765} {\bibfield
  {journal} {\bibinfo  {journal} {Optics Express}\ }\textbf {\bibinfo {volume}
  {27}},\ \bibinfo {pages} {15765} (\bibinfo {year} {2019})}\BibitemShut
  {NoStop}%
\bibitem [{\citenamefont {Chung}\ and\ \citenamefont
  {Miller}(2019)}]{Chung2019}%
  \BibitemOpen
  \bibfield  {author} {\bibinfo {author} {\bibfnamefont {H.}~\bibnamefont
  {Chung}}\ and\ \bibinfo {author} {\bibfnamefont {O.~D.}\ \bibnamefont
  {Miller}},\ }\bibfield  {title} {\bibinfo {title} {{High-NA, Achromatic,
  Visible-Frequency Metalenses by Inverse Design}},\ }\href
  {https://arxiv.org/abs/1905.09213} {\bibfield  {journal} {\bibinfo  {journal}
  {arXiv preprint arXiv:1905.09213}\ } (\bibinfo {year} {2019})}\BibitemShut
  {NoStop}%
\bibitem [{\citenamefont {Kong}(1975)}]{Kong1975}%
  \BibitemOpen
  \bibfield  {author} {\bibinfo {author} {\bibfnamefont {J.~A.}\ \bibnamefont
  {Kong}},\ }\href@noop {} {\emph {\bibinfo {title} {{Theory of electromagnetic
  waves}}}}\ (\bibinfo  {publisher} {Wiley-Interscience},\ \bibinfo {address}
  {New York},\ \bibinfo {year} {1975})\BibitemShut {NoStop}%
\bibitem [{\citenamefont {Levy}\ \emph {et~al.}(2007)\citenamefont {Levy},
  \citenamefont {Abashin}, \citenamefont {Ikeda}, \citenamefont
  {Krishnamoorthy}, \citenamefont {Cunningham},\ and\ \citenamefont
  {Fainman}}]{Levy2007}%
  \BibitemOpen
  \bibfield  {author} {\bibinfo {author} {\bibfnamefont {U.}~\bibnamefont
  {Levy}}, \bibinfo {author} {\bibfnamefont {M.}~\bibnamefont {Abashin}},
  \bibinfo {author} {\bibfnamefont {K.}~\bibnamefont {Ikeda}}, \bibinfo
  {author} {\bibfnamefont {A.}~\bibnamefont {Krishnamoorthy}}, \bibinfo
  {author} {\bibfnamefont {J.}~\bibnamefont {Cunningham}},\ and\ \bibinfo
  {author} {\bibfnamefont {Y.}~\bibnamefont {Fainman}},\ }\bibfield  {title}
  {\bibinfo {title} {{Inhomogenous dielectric metamaterials with space-variant
  polarizability}},\ }\href {https://doi.org/10.1103/PhysRevLett.98.243901}
  {\bibfield  {journal} {\bibinfo  {journal} {Physical Review Letters}\
  }\textbf {\bibinfo {volume} {98}},\ \bibinfo {pages} {243901} (\bibinfo
  {year} {2007})}\BibitemShut {NoStop}%
\bibitem [{\citenamefont {Wei}\ \emph {et~al.}(2013)\citenamefont {Wei},
  \citenamefont {Long}, \citenamefont {Gong}, \citenamefont {Li}, \citenamefont
  {Su},\ and\ \citenamefont {Cao}}]{Wei2013}%
  \BibitemOpen
  \bibfield  {author} {\bibinfo {author} {\bibfnamefont {Z.}~\bibnamefont
  {Wei}}, \bibinfo {author} {\bibfnamefont {Y.}~\bibnamefont {Long}}, \bibinfo
  {author} {\bibfnamefont {Z.}~\bibnamefont {Gong}}, \bibinfo {author}
  {\bibfnamefont {H.}~\bibnamefont {Li}}, \bibinfo {author} {\bibfnamefont
  {X.}~\bibnamefont {Su}},\ and\ \bibinfo {author} {\bibfnamefont
  {Y.}~\bibnamefont {Cao}},\ }\bibfield  {title} {\bibinfo {title} {{Highly
  efficient beam steering with a transparent metasurface}},\ }\href
  {https://doi.org/10.1364/oe.21.010739} {\bibfield  {journal} {\bibinfo
  {journal} {Optics Express}\ }\textbf {\bibinfo {volume} {21}},\ \bibinfo
  {pages} {10739} (\bibinfo {year} {2013})}\BibitemShut {NoStop}%
\bibitem [{\citenamefont {Keren-Zur}\ \emph {et~al.}(2016)\citenamefont
  {Keren-Zur}, \citenamefont {Avayu}, \citenamefont {Michaeli},\ and\
  \citenamefont {Ellenbogen}}]{KerenZur2016}%
  \BibitemOpen
  \bibfield  {author} {\bibinfo {author} {\bibfnamefont {S.}~\bibnamefont
  {Keren-Zur}}, \bibinfo {author} {\bibfnamefont {O.}~\bibnamefont {Avayu}},
  \bibinfo {author} {\bibfnamefont {L.}~\bibnamefont {Michaeli}},\ and\
  \bibinfo {author} {\bibfnamefont {T.}~\bibnamefont {Ellenbogen}},\ }\bibfield
   {title} {\bibinfo {title} {{Nonlinear Beam Shaping with Plasmonic
  Metasurfaces}},\ }\href {https://doi.org/10.1021/acsphotonics.5b00528}
  {\bibfield  {journal} {\bibinfo  {journal} {ACS Photonics}\ }\textbf
  {\bibinfo {volume} {3}},\ \bibinfo {pages} {117} (\bibinfo {year}
  {2016})}\BibitemShut {NoStop}%
\bibitem [{\citenamefont {Weiner}(2000)}]{Weiner2000}%
  \BibitemOpen
  \bibfield  {author} {\bibinfo {author} {\bibfnamefont {A.~M.}\ \bibnamefont
  {Weiner}},\ }\bibfield  {title} {\bibinfo {title} {{Femtosecond pulse shaping
  using spatial light modulators}},\ }\href {https://doi.org/10.1063/1.1150614}
  {\bibfield  {journal} {\bibinfo  {journal} {Review of Scientific
  Instruments}\ }\textbf {\bibinfo {volume} {71}},\ \bibinfo {pages} {1929}
  (\bibinfo {year} {2000})}\BibitemShut {NoStop}%
\bibitem [{\citenamefont {Chattrapiban}\ \emph {et~al.}(2003)\citenamefont
  {Chattrapiban}, \citenamefont {Rogers}, \citenamefont {Cofield},
  \citenamefont {{Hill, III}},\ and\ \citenamefont {Roy}}]{Chattrapiban2003}%
  \BibitemOpen
  \bibfield  {author} {\bibinfo {author} {\bibfnamefont {N.}~\bibnamefont
  {Chattrapiban}}, \bibinfo {author} {\bibfnamefont {E.~A.}\ \bibnamefont
  {Rogers}}, \bibinfo {author} {\bibfnamefont {D.}~\bibnamefont {Cofield}},
  \bibinfo {author} {\bibfnamefont {W.~T.}\ \bibnamefont {{Hill, III}}},\ and\
  \bibinfo {author} {\bibfnamefont {R.}~\bibnamefont {Roy}},\ }\bibfield
  {title} {\bibinfo {title} {{Generation of nondiffracting Bessel beams by use
  of a spatial light modulator}},\ }\href
  {https://doi.org/10.1364/OL.28.002183} {\bibfield  {journal} {\bibinfo
  {journal} {Optics Letters}\ }\textbf {\bibinfo {volume} {28}},\ \bibinfo
  {pages} {2183} (\bibinfo {year} {2003})}\BibitemShut {NoStop}%
\bibitem [{\citenamefont {Guo}\ \emph {et~al.}(2007)\citenamefont {Guo},
  \citenamefont {Wang}, \citenamefont {Ni}, \citenamefont {Wang},\ and\
  \citenamefont {Ding}}]{Guo2007}%
  \BibitemOpen
  \bibfield  {author} {\bibinfo {author} {\bibfnamefont {C.-S.}\ \bibnamefont
  {Guo}}, \bibinfo {author} {\bibfnamefont {X.-L.}\ \bibnamefont {Wang}},
  \bibinfo {author} {\bibfnamefont {W.-J.}\ \bibnamefont {Ni}}, \bibinfo
  {author} {\bibfnamefont {H.-T.}\ \bibnamefont {Wang}},\ and\ \bibinfo
  {author} {\bibfnamefont {J.}~\bibnamefont {Ding}},\ }\bibfield  {title}
  {\bibinfo {title} {{Generation of arbitrary vector beams with a spatial light
  modulator and a common path interferometric arrangement}},\ }\href
  {https://doi.org/10.1364/ol.32.003549} {\bibfield  {journal} {\bibinfo
  {journal} {Optics Letters}\ }\textbf {\bibinfo {volume} {32}},\ \bibinfo
  {pages} {3549} (\bibinfo {year} {2007})}\BibitemShut {NoStop}%
\bibitem [{\citenamefont {Zhu}\ and\ \citenamefont {Wang}(2014)}]{Zhu2014}%
  \BibitemOpen
  \bibfield  {author} {\bibinfo {author} {\bibfnamefont {L.}~\bibnamefont
  {Zhu}}\ and\ \bibinfo {author} {\bibfnamefont {J.}~\bibnamefont {Wang}},\
  }\bibfield  {title} {\bibinfo {title} {{Arbitrary manipulation of spatial
  amplitude and phase using phase-only spatial light modulators}},\ }\href
  {https://doi.org/10.1038/srep07441} {\bibfield  {journal} {\bibinfo
  {journal} {Scientific Reports}\ }\textbf {\bibinfo {volume} {4}},\ \bibinfo
  {pages} {7441} (\bibinfo {year} {2014})}\BibitemShut {NoStop}%
\bibitem [{\citenamefont {Lodahl}\ \emph {et~al.}(2004)\citenamefont {Lodahl},
  \citenamefont {{Van Driel}}, \citenamefont {Nikolaev}, \citenamefont {Irman},
  \citenamefont {Overgaag}, \citenamefont {Vanmaekelbergh},\ and\ \citenamefont
  {Vos}}]{Lodahl2004}%
  \BibitemOpen
  \bibfield  {author} {\bibinfo {author} {\bibfnamefont {P.}~\bibnamefont
  {Lodahl}}, \bibinfo {author} {\bibfnamefont {A.~F.}\ \bibnamefont {{Van
  Driel}}}, \bibinfo {author} {\bibfnamefont {I.~S.}\ \bibnamefont {Nikolaev}},
  \bibinfo {author} {\bibfnamefont {A.}~\bibnamefont {Irman}}, \bibinfo
  {author} {\bibfnamefont {K.}~\bibnamefont {Overgaag}}, \bibinfo {author}
  {\bibfnamefont {D.}~\bibnamefont {Vanmaekelbergh}},\ and\ \bibinfo {author}
  {\bibfnamefont {W.~L.}\ \bibnamefont {Vos}},\ }\bibfield  {title} {\bibinfo
  {title} {{Controlling the dynamics of spontaneous emission from quantum dots
  by photonic crystals}},\ }\href {https://doi.org/10.1038/nature02772}
  {\bibfield  {journal} {\bibinfo  {journal} {Nature}\ }\textbf {\bibinfo
  {volume} {430}},\ \bibinfo {pages} {654} (\bibinfo {year}
  {2004})}\BibitemShut {NoStop}%
\bibitem [{\citenamefont {Ringler}\ \emph {et~al.}(2008)\citenamefont
  {Ringler}, \citenamefont {Schwemer}, \citenamefont {Wunderlich},
  \citenamefont {Nichtl}, \citenamefont {K{\"{u}}rzinger}, \citenamefont
  {Klar},\ and\ \citenamefont {Feldmann}}]{Ringler2008}%
  \BibitemOpen
  \bibfield  {author} {\bibinfo {author} {\bibfnamefont {M.}~\bibnamefont
  {Ringler}}, \bibinfo {author} {\bibfnamefont {A.}~\bibnamefont {Schwemer}},
  \bibinfo {author} {\bibfnamefont {M.}~\bibnamefont {Wunderlich}}, \bibinfo
  {author} {\bibfnamefont {A.}~\bibnamefont {Nichtl}}, \bibinfo {author}
  {\bibfnamefont {K.}~\bibnamefont {K{\"{u}}rzinger}}, \bibinfo {author}
  {\bibfnamefont {T.~A.}\ \bibnamefont {Klar}},\ and\ \bibinfo {author}
  {\bibfnamefont {J.}~\bibnamefont {Feldmann}},\ }\bibfield  {title} {\bibinfo
  {title} {{Shaping Emission Spectra of Fluorescent Molecules with Single
  Plasmonic Nanoresonators}},\ }\href
  {https://doi.org/10.1103/PhysRevLett.100.203002} {\bibfield  {journal}
  {\bibinfo  {journal} {Physical Review Letters}\ }\textbf {\bibinfo {volume}
  {100}},\ \bibinfo {pages} {203002} (\bibinfo {year} {2008})}\BibitemShut
  {NoStop}%
\bibitem [{\citenamefont {Bleuse}\ \emph {et~al.}(2011)\citenamefont {Bleuse},
  \citenamefont {Claudon}, \citenamefont {Creasey}, \citenamefont {Malik},
  \citenamefont {G{\'{e}}rard}, \citenamefont {Maksymov}, \citenamefont
  {Hugonin},\ and\ \citenamefont {Lalanne}}]{Bleuse2011}%
  \BibitemOpen
  \bibfield  {author} {\bibinfo {author} {\bibfnamefont {J.}~\bibnamefont
  {Bleuse}}, \bibinfo {author} {\bibfnamefont {J.}~\bibnamefont {Claudon}},
  \bibinfo {author} {\bibfnamefont {M.}~\bibnamefont {Creasey}}, \bibinfo
  {author} {\bibfnamefont {N.~S.}\ \bibnamefont {Malik}}, \bibinfo {author}
  {\bibfnamefont {J.-M.}\ \bibnamefont {G{\'{e}}rard}}, \bibinfo {author}
  {\bibfnamefont {I.}~\bibnamefont {Maksymov}}, \bibinfo {author}
  {\bibfnamefont {J.-P.}\ \bibnamefont {Hugonin}},\ and\ \bibinfo {author}
  {\bibfnamefont {P.}~\bibnamefont {Lalanne}},\ }\bibfield  {title} {\bibinfo
  {title} {{Inhibition, Enhancement, and Control of Spontaneous Emission in
  Photonic Nanowires}},\ }\href
  {https://doi.org/10.1103/PhysRevLett.106.103601} {\bibfield  {journal}
  {\bibinfo  {journal} {Physical Review Letters}\ }\textbf {\bibinfo {volume}
  {106}},\ \bibinfo {pages} {103601} (\bibinfo {year} {2011})}\BibitemShut
  {NoStop}%
\bibitem [{\citenamefont {Epstein}\ and\ \citenamefont
  {Eleftheriades}(2016)}]{Epstein2016}%
  \BibitemOpen
  \bibfield  {author} {\bibinfo {author} {\bibfnamefont {A.}~\bibnamefont
  {Epstein}}\ and\ \bibinfo {author} {\bibfnamefont {G.~V.}\ \bibnamefont
  {Eleftheriades}},\ }\bibfield  {title} {\bibinfo {title} {{Huygens'
  metasurfaces via the equivalence principle: design and applications}},\
  }\href {https://doi.org/10.1364/josab.33.000a31} {\bibfield  {journal}
  {\bibinfo  {journal} {Journal of the Optical Society of America B}\ }\textbf
  {\bibinfo {volume} {33}},\ \bibinfo {pages} {A31} (\bibinfo {year}
  {2016})}\BibitemShut {NoStop}%
\bibitem [{\citenamefont {Ra'di}\ \emph {et~al.}(2017)\citenamefont {Ra'di},
  \citenamefont {Sounas},\ and\ \citenamefont {Al{\`{u}}}}]{Radi2017}%
  \BibitemOpen
  \bibfield  {author} {\bibinfo {author} {\bibfnamefont {Y.}~\bibnamefont
  {Ra'di}}, \bibinfo {author} {\bibfnamefont {D.~L.}\ \bibnamefont {Sounas}},\
  and\ \bibinfo {author} {\bibfnamefont {A.}~\bibnamefont {Al{\`{u}}}},\
  }\bibfield  {title} {\bibinfo {title} {{Metagratings: Beyond the Limits of
  Graded Metasurfaces for Wave Front Control}},\ }\href
  {https://doi.org/10.1103/PhysRevLett.119.067404} {\bibfield  {journal}
  {\bibinfo  {journal} {Physical Review Letters}\ }\textbf {\bibinfo {volume}
  {119}},\ \bibinfo {pages} {067404} (\bibinfo {year} {2017})}\BibitemShut
  {NoStop}%
\bibitem [{\citenamefont {Gustafsson}\ and\ \citenamefont
  {Nordebo}(2013)}]{Gustafsson2013}%
  \BibitemOpen
  \bibfield  {author} {\bibinfo {author} {\bibfnamefont {M.}~\bibnamefont
  {Gustafsson}}\ and\ \bibinfo {author} {\bibfnamefont {S.}~\bibnamefont
  {Nordebo}},\ }\bibfield  {title} {\bibinfo {title} {{Optimal antenna currents
  for Q, superdirectivity, and radiation patterns using convex optimization}},\
  }\href {https://doi.org/10.1109/TAP.2012.2227656} {\bibfield  {journal}
  {\bibinfo  {journal} {IEEE Transactions on Antennas and Propagation}\
  }\textbf {\bibinfo {volume} {61}},\ \bibinfo {pages} {1109} (\bibinfo {year}
  {2013})}\BibitemShut {NoStop}%
\bibitem [{\citenamefont {Shi}\ \emph {et~al.}(2017)\citenamefont {Shi},
  \citenamefont {Wang},\ and\ \citenamefont {Jonsson}}]{Shi2017}%
  \BibitemOpen
  \bibfield  {author} {\bibinfo {author} {\bibfnamefont {S.}~\bibnamefont
  {Shi}}, \bibinfo {author} {\bibfnamefont {L.}~\bibnamefont {Wang}},\ and\
  \bibinfo {author} {\bibfnamefont {B.~L.~G.}\ \bibnamefont {Jonsson}},\
  }\bibfield  {title} {\bibinfo {title} {{Antenna Current Optimization and
  Realizations for Far-Field Pattern Shaping}},\ }\href
  {http://arxiv.org/abs/1711.09709} {\bibfield  {journal} {\bibinfo  {journal}
  {arXiv preprint arXiv:1711.09709v2}\ } (\bibinfo {year} {2017})}\BibitemShut
  {NoStop}%
\bibitem [{\citenamefont {Chew}(1995)}]{Chew1995}%
  \BibitemOpen
  \bibfield  {author} {\bibinfo {author} {\bibfnamefont {W.~C.}\ \bibnamefont
  {Chew}},\ }\href@noop {} {\emph {\bibinfo {title} {{Waves and fields in
  inhomogeneous media}}}}\ (\bibinfo  {publisher} {IEEE press},\ \bibinfo
  {year} {1995})\BibitemShut {NoStop}%
\bibitem [{\citenamefont {Jin}(2011)}]{Jin2011}%
  \BibitemOpen
  \bibfield  {author} {\bibinfo {author} {\bibfnamefont {J.-M.}\ \bibnamefont
  {Jin}},\ }\href@noop {} {\emph {\bibinfo {title} {{Theory and computation of
  electromagnetic fields}}}}\ (\bibinfo  {publisher} {John Wiley {\&} Sons},\
  \bibinfo {year} {2011})\BibitemShut {NoStop}%
\bibitem [{\citenamefont {Boyd}\ and\ \citenamefont
  {Vandenberghe}(2004)}]{Boyd2004}%
  \BibitemOpen
  \bibfield  {author} {\bibinfo {author} {\bibfnamefont {S.}~\bibnamefont
  {Boyd}}\ and\ \bibinfo {author} {\bibfnamefont {L.}~\bibnamefont
  {Vandenberghe}},\ }\href@noop {} {\emph {\bibinfo {title} {{Convex
  optimization}}}}\ (\bibinfo  {publisher} {Cambridge university press},\
  \bibinfo {year} {2004})\BibitemShut {NoStop}%
\bibitem [{\citenamefont {Horn}\ and\ \citenamefont
  {Johnson}(2013)}]{Horn2013}%
  \BibitemOpen
  \bibfield  {author} {\bibinfo {author} {\bibfnamefont {R.~A.}\ \bibnamefont
  {Horn}}\ and\ \bibinfo {author} {\bibfnamefont {C.~R.}\ \bibnamefont
  {Johnson}},\ }\href@noop {} {\emph {\bibinfo {title} {{Matrix analysis}}}},\
  \bibinfo {edition} {2nd}\ ed.\ (\bibinfo  {publisher} {Cambridge university
  press},\ \bibinfo {year} {2013})\BibitemShut {NoStop}%
\bibitem [{\citenamefont {Trefethen}\ and\ \citenamefont {{Bau
  III}}(1997)}]{Trefethen1997}%
  \BibitemOpen
  \bibfield  {author} {\bibinfo {author} {\bibfnamefont {L.~N.}\ \bibnamefont
  {Trefethen}}\ and\ \bibinfo {author} {\bibfnamefont {D.}~\bibnamefont {{Bau
  III}}},\ }\href@noop {} {\emph {\bibinfo {title} {{Numerical linear
  algebra}}}}\ (\bibinfo  {publisher} {SIAM},\ \bibinfo {year}
  {1997})\BibitemShut {NoStop}%
\bibitem [{\citenamefont {Boyd}(2001)}]{boyd2001chebyshev}%
  \BibitemOpen
  \bibfield  {author} {\bibinfo {author} {\bibfnamefont {J.~P.}\ \bibnamefont
  {Boyd}},\ }\href@noop {} {\emph {\bibinfo {title} {{Chebyshev and Fourier
  spectral methods}}}},\ \bibinfo {edition} {2nd}\ ed.\ (\bibinfo  {publisher}
  {Dover},\ \bibinfo {address} {New York},\ \bibinfo {year} {2001})\BibitemShut
  {NoStop}%
\bibitem [{\citenamefont {Levy}\ \emph {et~al.}(2016)\citenamefont {Levy},
  \citenamefont {Derevyanko},\ and\ \citenamefont {Silberberg}}]{Levy2016}%
  \BibitemOpen
  \bibfield  {author} {\bibinfo {author} {\bibfnamefont {U.}~\bibnamefont
  {Levy}}, \bibinfo {author} {\bibfnamefont {S.}~\bibnamefont {Derevyanko}},\
  and\ \bibinfo {author} {\bibfnamefont {Y.}~\bibnamefont {Silberberg}},\
  }\bibfield  {title} {\bibinfo {title} {{Light Modes of Free Space}},\ }\href
  {https://doi.org/10.1016/bs.po.2015.10.001} {\bibfield  {journal} {\bibinfo
  {journal} {Progress in Optics}\ }\textbf {\bibinfo {volume} {61}},\ \bibinfo
  {pages} {237} (\bibinfo {year} {2016})}\BibitemShut {NoStop}%
\bibitem [{\citenamefont {Mazilu}\ \emph {et~al.}(2011)\citenamefont {Mazilu},
  \citenamefont {Baumgartl}, \citenamefont {Kosmeier},\ and\ \citenamefont
  {Dholakia}}]{Mazilu2011}%
  \BibitemOpen
  \bibfield  {author} {\bibinfo {author} {\bibfnamefont {M.}~\bibnamefont
  {Mazilu}}, \bibinfo {author} {\bibfnamefont {J.}~\bibnamefont {Baumgartl}},
  \bibinfo {author} {\bibfnamefont {S.}~\bibnamefont {Kosmeier}},\ and\
  \bibinfo {author} {\bibfnamefont {K.}~\bibnamefont {Dholakia}},\ }\bibfield
  {title} {\bibinfo {title} {{Optical Eigenmodes; exploiting the quadratic
  nature of the light-matter interaction}},\ }\href
  {https://doi.org/10.1364/OE.19.000933} {\bibfield  {journal} {\bibinfo
  {journal} {Optics Express}\ }\textbf {\bibinfo {volume} {19}},\ \bibinfo
  {pages} {933} (\bibinfo {year} {2011})}\BibitemShut {NoStop}%
\bibitem [{\citenamefont {Kosmeier}\ \emph {et~al.}(2011)\citenamefont
  {Kosmeier}, \citenamefont {Mazilu}, \citenamefont {Baumgartl},\ and\
  \citenamefont {Dholakia}}]{Kosmeier2011}%
  \BibitemOpen
  \bibfield  {author} {\bibinfo {author} {\bibfnamefont {S.}~\bibnamefont
  {Kosmeier}}, \bibinfo {author} {\bibfnamefont {M.}~\bibnamefont {Mazilu}},
  \bibinfo {author} {\bibfnamefont {J.}~\bibnamefont {Baumgartl}},\ and\
  \bibinfo {author} {\bibfnamefont {K.}~\bibnamefont {Dholakia}},\ }\bibfield
  {title} {\bibinfo {title} {{Enhanced two-point resolution using optical
  eigenmode optimized pupil functions}},\ }\href
  {https://doi.org/10.1088/2040-8978/13/10/105707} {\bibfield  {journal}
  {\bibinfo  {journal} {Journal of Optics}\ }\textbf {\bibinfo {volume} {13}},\
  \bibinfo {pages} {105707} (\bibinfo {year} {2011})}\BibitemShut {NoStop}%
\bibitem [{\citenamefont {Levi}(1965)}]{Levi1965}%
  \BibitemOpen
  \bibfield  {author} {\bibinfo {author} {\bibfnamefont {L.}~\bibnamefont
  {Levi}},\ }\bibfield  {title} {\bibinfo {title} {{Fitting a bandlimited
  signal to given points}},\ }\href {https://doi.org/10.1109/TIT.1965.1053777}
  {\bibfield  {journal} {\bibinfo  {journal} {IEEE Transactions on Information
  Theory}\ }\textbf {\bibinfo {volume} {11}},\ \bibinfo {pages} {372} (\bibinfo
  {year} {1965})}\BibitemShut {NoStop}%
\bibitem [{\citenamefont {Slepian}(1978)}]{Slepian1978}%
  \BibitemOpen
  \bibfield  {author} {\bibinfo {author} {\bibfnamefont {D.}~\bibnamefont
  {Slepian}},\ }\bibfield  {title} {\bibinfo {title} {{Prolate Spheroidal Wave
  Functions, Fourier Analysis, and Uncertainty—V: The Discrete Case}},\
  }\href {https://doi.org/10.1002/j.1538-7305.1978.tb02104.x} {\bibfield
  {journal} {\bibinfo  {journal} {Bell System Technical Journal}\ }\textbf
  {\bibinfo {volume} {57}},\ \bibinfo {pages} {1371} (\bibinfo {year}
  {1978})}\BibitemShut {NoStop}%
\bibitem [{\citenamefont {Gundu}\ \emph {et~al.}(2005)\citenamefont {Gundu},
  \citenamefont {Hack},\ and\ \citenamefont {Rastogi}}]{Gundu2005}%
  \BibitemOpen
  \bibfield  {author} {\bibinfo {author} {\bibfnamefont {P.~N.}\ \bibnamefont
  {Gundu}}, \bibinfo {author} {\bibfnamefont {E.}~\bibnamefont {Hack}},\ and\
  \bibinfo {author} {\bibfnamefont {P.}~\bibnamefont {Rastogi}},\ }\bibfield
  {title} {\bibinfo {title} {{'Apodized superresolution' - Concept and
  simulations}},\ }\href {https://doi.org/10.1016/j.optcom.2005.01.025}
  {\bibfield  {journal} {\bibinfo  {journal} {Optics Communications}\ }\textbf
  {\bibinfo {volume} {249}},\ \bibinfo {pages} {101} (\bibinfo {year}
  {2005})}\BibitemShut {NoStop}%
\bibitem [{\citenamefont {de~Juana}\ \emph {et~al.}(2003)\citenamefont
  {de~Juana}, \citenamefont {Oti}, \citenamefont {Canales},\ and\ \citenamefont
  {Cagigal}}]{Dejuana2003}%
  \BibitemOpen
  \bibfield  {author} {\bibinfo {author} {\bibfnamefont {D.~M.}\ \bibnamefont
  {de~Juana}}, \bibinfo {author} {\bibfnamefont {J.~E.}\ \bibnamefont {Oti}},
  \bibinfo {author} {\bibfnamefont {V.~F.}\ \bibnamefont {Canales}},\ and\
  \bibinfo {author} {\bibfnamefont {M.~P.}\ \bibnamefont {Cagigal}},\
  }\bibfield  {title} {\bibinfo {title} {{Design of superresolving continuous
  phase filters}},\ }\href {https://doi.org/10.1364/OL.28.000607} {\bibfield
  {journal} {\bibinfo  {journal} {Optics Letters}\ }\textbf {\bibinfo {volume}
  {28}},\ \bibinfo {pages} {607} (\bibinfo {year} {2003})}\BibitemShut
  {NoStop}%
\bibitem [{\citenamefont {Qin}\ \emph {et~al.}(2017)\citenamefont {Qin},
  \citenamefont {Huang}, \citenamefont {Wu}, \citenamefont {Teng},
  \citenamefont {Qiu},\ and\ \citenamefont {Hong}}]{Qin2017}%
  \BibitemOpen
  \bibfield  {author} {\bibinfo {author} {\bibfnamefont {F.}~\bibnamefont
  {Qin}}, \bibinfo {author} {\bibfnamefont {K.}~\bibnamefont {Huang}}, \bibinfo
  {author} {\bibfnamefont {J.}~\bibnamefont {Wu}}, \bibinfo {author}
  {\bibfnamefont {J.}~\bibnamefont {Teng}}, \bibinfo {author} {\bibfnamefont
  {C.~W.}\ \bibnamefont {Qiu}},\ and\ \bibinfo {author} {\bibfnamefont
  {M.}~\bibnamefont {Hong}},\ }\bibfield  {title} {\bibinfo {title} {{A
  Supercritical Lens Optical Label-Free Microscopy: Sub-Diffraction Resolution
  and Ultra-Long Working Distance}},\ }\href
  {https://doi.org/10.1002/adma.201602721} {\bibfield  {journal} {\bibinfo
  {journal} {Advanced Materials}\ }\textbf {\bibinfo {volume} {29}},\ \bibinfo
  {pages} {1602721} (\bibinfo {year} {2017})}\BibitemShut {NoStop}%
\bibitem [{\citenamefont {Dong}\ \emph {et~al.}(2017)\citenamefont {Dong},
  \citenamefont {Wong}, \citenamefont {Kim},\ and\ \citenamefont
  {Eleftheriades}}]{Dong2017}%
  \BibitemOpen
  \bibfield  {author} {\bibinfo {author} {\bibfnamefont {X.~H.}\ \bibnamefont
  {Dong}}, \bibinfo {author} {\bibfnamefont {A.~M.~H.}\ \bibnamefont {Wong}},
  \bibinfo {author} {\bibfnamefont {M.}~\bibnamefont {Kim}},\ and\ \bibinfo
  {author} {\bibfnamefont {G.~V.}\ \bibnamefont {Eleftheriades}},\ }\bibfield
  {title} {\bibinfo {title} {{Superresolution far-field imaging of complex
  objects using reduced superoscillating ripples}},\ }\href
  {https://doi.org/10.1364/OPTICA.4.001126} {\bibfield  {journal} {\bibinfo
  {journal} {Optica}\ }\textbf {\bibinfo {volume} {4}},\ \bibinfo {pages}
  {1126} (\bibinfo {year} {2017})}\BibitemShut {NoStop}%
\bibitem [{\citenamefont {Bethe}(1944)}]{Bethe1944}%
  \BibitemOpen
  \bibfield  {author} {\bibinfo {author} {\bibfnamefont {H.~A.}\ \bibnamefont
  {Bethe}},\ }\bibfield  {title} {\bibinfo {title} {{Theory of diffraction by
  small holes}},\ }\href {https://doi.org/10.1103/PhysRev.66.163} {\bibfield
  {journal} {\bibinfo  {journal} {Physical Review}\ }\textbf {\bibinfo {volume}
  {66}},\ \bibinfo {pages} {163} (\bibinfo {year} {1944})}\BibitemShut
  {NoStop}%
\bibitem [{\citenamefont {Rumelhart}\ \emph {et~al.}(1986)\citenamefont
  {Rumelhart}, \citenamefont {Hinton},\ and\ \citenamefont
  {Williams}}]{Rumelhart1986}%
  \BibitemOpen
  \bibfield  {author} {\bibinfo {author} {\bibfnamefont {D.~E.}\ \bibnamefont
  {Rumelhart}}, \bibinfo {author} {\bibfnamefont {G.~E.}\ \bibnamefont
  {Hinton}},\ and\ \bibinfo {author} {\bibfnamefont {R.~J.}\ \bibnamefont
  {Williams}},\ }\bibfield  {title} {\bibinfo {title} {{Learning
  representations by back-propagating errors}},\ }\href
  {https://doi.org/10.1038/320129a0} {\bibfield  {journal} {\bibinfo  {journal}
  {Nature}\ }\textbf {\bibinfo {volume} {323}},\ \bibinfo {pages} {533}
  (\bibinfo {year} {1986})}\BibitemShut {NoStop}%
\bibitem [{\citenamefont {LeCun}\ \emph {et~al.}(1989)\citenamefont {LeCun},
  \citenamefont {Boser}, \citenamefont {Denker}, \citenamefont {Henderson},
  \citenamefont {Howard}, \citenamefont {Hubbard},\ and\ \citenamefont
  {Jackel}}]{LeCun1989}%
  \BibitemOpen
  \bibfield  {author} {\bibinfo {author} {\bibfnamefont {Y.}~\bibnamefont
  {LeCun}}, \bibinfo {author} {\bibfnamefont {B.}~\bibnamefont {Boser}},
  \bibinfo {author} {\bibfnamefont {J.~S.}\ \bibnamefont {Denker}}, \bibinfo
  {author} {\bibfnamefont {D.}~\bibnamefont {Henderson}}, \bibinfo {author}
  {\bibfnamefont {R.~E.}\ \bibnamefont {Howard}}, \bibinfo {author}
  {\bibfnamefont {W.}~\bibnamefont {Hubbard}},\ and\ \bibinfo {author}
  {\bibfnamefont {L.~D.}\ \bibnamefont {Jackel}},\ }\bibfield  {title}
  {\bibinfo {title} {{Backpropagation Applied to Handwritten Zip Code
  Recognition}},\ }\href {https://doi.org/10.1162/neco.1989.1.4.541} {\bibfield
   {journal} {\bibinfo  {journal} {Neural Computation}\ }\textbf {\bibinfo
  {volume} {1}},\ \bibinfo {pages} {541} (\bibinfo {year} {1989})}\BibitemShut
  {NoStop}%
\bibitem [{\citenamefont {Werbos}(1994)}]{Werbos1994}%
  \BibitemOpen
  \bibfield  {author} {\bibinfo {author} {\bibfnamefont {P.~J.}\ \bibnamefont
  {Werbos}},\ }\href@noop {} {\emph {\bibinfo {title} {{The Roots of
  Backpropagation}}}}\ (\bibinfo  {publisher} {John Wiley {\&} Sons, Inc.},\
  \bibinfo {year} {1994})\BibitemShut {NoStop}%
\bibitem [{\citenamefont {Nocedal}\ and\ \citenamefont
  {Wright}(2006)}]{Nocedal2006}%
  \BibitemOpen
  \bibfield  {author} {\bibinfo {author} {\bibfnamefont {J.}~\bibnamefont
  {Nocedal}}\ and\ \bibinfo {author} {\bibfnamefont {S.~J.}\ \bibnamefont
  {Wright}},\ }\href@noop {} {\emph {\bibinfo {title} {{Numerical
  Optimization}}}},\ \bibinfo {edition} {2nd}\ ed.\ (\bibinfo  {publisher}
  {Springer},\ \bibinfo {address} {New York, NY},\ \bibinfo {year}
  {2006})\BibitemShut {NoStop}%
\bibitem [{\citenamefont {Lalau-Keraly}\ \emph {et~al.}(2013)\citenamefont
  {Lalau-Keraly}, \citenamefont {Bhargava}, \citenamefont {Miller},\ and\
  \citenamefont {Yablonovitch}}]{Lalau-Keraly2013}%
  \BibitemOpen
  \bibfield  {author} {\bibinfo {author} {\bibfnamefont {C.~M.}\ \bibnamefont
  {Lalau-Keraly}}, \bibinfo {author} {\bibfnamefont {S.}~\bibnamefont
  {Bhargava}}, \bibinfo {author} {\bibfnamefont {O.~D.}\ \bibnamefont
  {Miller}},\ and\ \bibinfo {author} {\bibfnamefont {E.}~\bibnamefont
  {Yablonovitch}},\ }\bibfield  {title} {\bibinfo {title} {{Adjoint shape
  optimization applied to electromagnetic design}},\ }\href
  {https://doi.org/10.1364/oe.21.021693} {\bibfield  {journal} {\bibinfo
  {journal} {Optics Express}\ }\textbf {\bibinfo {volume} {21}},\ \bibinfo
  {pages} {21693} (\bibinfo {year} {2013})}\BibitemShut {NoStop}%
\bibitem [{\citenamefont {Ganapati}\ \emph {et~al.}(2014)\citenamefont
  {Ganapati}, \citenamefont {Miller},\ and\ \citenamefont
  {Yablonovitch}}]{Ganapati2014}%
  \BibitemOpen
  \bibfield  {author} {\bibinfo {author} {\bibfnamefont {V.}~\bibnamefont
  {Ganapati}}, \bibinfo {author} {\bibfnamefont {O.~D.}\ \bibnamefont
  {Miller}},\ and\ \bibinfo {author} {\bibfnamefont {E.}~\bibnamefont
  {Yablonovitch}},\ }\bibfield  {title} {\bibinfo {title} {{Light trapping
  textures designed by electromagnetic optimization for subwavelength thick
  solar cells}},\ }\href {https://doi.org/10.1109/JPHOTOV.2013.2280340}
  {\bibfield  {journal} {\bibinfo  {journal} {IEEE Journal of Photovoltaics}\
  }\textbf {\bibinfo {volume} {4}},\ \bibinfo {pages} {175} (\bibinfo {year}
  {2014})}\BibitemShut {NoStop}%
\bibitem [{\citenamefont {Marchand}(1978)}]{marchand1978gradient}%
  \BibitemOpen
  \bibfield  {author} {\bibinfo {author} {\bibfnamefont {E.}~\bibnamefont
  {Marchand}},\ }\href@noop {} {\emph {\bibinfo {title} {{Gradient index
  optics}}}}\ (\bibinfo  {publisher} {Academic},\ \bibinfo {address} {New
  York},\ \bibinfo {year} {1978})\BibitemShut {NoStop}%
\bibitem [{\citenamefont {Bertsekas}(2014)}]{bertsekas2014constrained}%
  \BibitemOpen
  \bibfield  {author} {\bibinfo {author} {\bibfnamefont {D.~P.}\ \bibnamefont
  {Bertsekas}},\ }\href@noop {} {\emph {\bibinfo {title} {{Constrained
  optimization and Lagrange multiplier methods}}}}\ (\bibinfo  {publisher}
  {Academic press},\ \bibinfo {year} {2014})\BibitemShut {NoStop}%
\bibitem [{\citenamefont {Taflove}\ \emph {et~al.}(2013)\citenamefont
  {Taflove}, \citenamefont {Oskooi},\ and\ \citenamefont
  {Johnson}}]{Taflove2013}%
  \BibitemOpen
  \bibfield  {author} {\bibinfo {author} {\bibfnamefont {A.}~\bibnamefont
  {Taflove}}, \bibinfo {author} {\bibfnamefont {A.}~\bibnamefont {Oskooi}},\
  and\ \bibinfo {author} {\bibfnamefont {S.~G.}\ \bibnamefont {Johnson}},\
  }\href@noop {} {\emph {\bibinfo {title} {{Advances in FDTD computational
  electrodynamics: photonics and nanotechnology}}}}\ (\bibinfo  {publisher}
  {Artech house},\ \bibinfo {year} {2013})\BibitemShut {NoStop}%
\bibitem [{\citenamefont {Oskooi}\ \emph {et~al.}(2010)\citenamefont {Oskooi},
  \citenamefont {Roundy}, \citenamefont {Ibanescu}, \citenamefont {Bermel},
  \citenamefont {Joannopoulos},\ and\ \citenamefont {Johnson}}]{Oskooi2010}%
  \BibitemOpen
  \bibfield  {author} {\bibinfo {author} {\bibfnamefont {A.~F.}\ \bibnamefont
  {Oskooi}}, \bibinfo {author} {\bibfnamefont {D.}~\bibnamefont {Roundy}},
  \bibinfo {author} {\bibfnamefont {M.}~\bibnamefont {Ibanescu}}, \bibinfo
  {author} {\bibfnamefont {P.}~\bibnamefont {Bermel}}, \bibinfo {author}
  {\bibfnamefont {J.~D.}\ \bibnamefont {Joannopoulos}},\ and\ \bibinfo {author}
  {\bibfnamefont {S.~G.}\ \bibnamefont {Johnson}},\ }\bibfield  {title}
  {\bibinfo {title} {{Meep: A flexible free-software package for
  electromagnetic simulations by the FDTD method}},\ }\href
  {https://doi.org/10.1016/j.cpc.2009.11.008} {\bibfield  {journal} {\bibinfo
  {journal} {Computer Physics Communications}\ }\textbf {\bibinfo {volume}
  {181}},\ \bibinfo {pages} {687} (\bibinfo {year} {2010})}\BibitemShut
  {NoStop}%
\bibitem [{\citenamefont {Luo}\ \emph {et~al.}(2010)\citenamefont {Luo},
  \citenamefont {Ma}, \citenamefont {So}, \citenamefont {Ye},\ and\
  \citenamefont {Zhang}}]{Luo2010}%
  \BibitemOpen
  \bibfield  {author} {\bibinfo {author} {\bibfnamefont {Z.~Q.}\ \bibnamefont
  {Luo}}, \bibinfo {author} {\bibfnamefont {W.~K.}\ \bibnamefont {Ma}},
  \bibinfo {author} {\bibfnamefont {A.}~\bibnamefont {So}}, \bibinfo {author}
  {\bibfnamefont {Y.}~\bibnamefont {Ye}},\ and\ \bibinfo {author}
  {\bibfnamefont {S.}~\bibnamefont {Zhang}},\ }\bibfield  {title} {\bibinfo
  {title} {{Semidefinite relaxation of quadratic optimization problems}},\
  }\href {https://doi.org/10.1109/MSP.2010.936019} {\bibfield  {journal}
  {\bibinfo  {journal} {IEEE Signal Processing Magazine}\ }\textbf {\bibinfo
  {volume} {27}},\ \bibinfo {pages} {20} (\bibinfo {year} {2010})}\BibitemShut
  {NoStop}%
\bibitem [{\citenamefont {Huang}\ and\ \citenamefont
  {Zhang}(2007)}]{Huang2007b}%
  \BibitemOpen
  \bibfield  {author} {\bibinfo {author} {\bibfnamefont {Y.}~\bibnamefont
  {Huang}}\ and\ \bibinfo {author} {\bibfnamefont {S.}~\bibnamefont {Zhang}},\
  }\bibfield  {title} {\bibinfo {title} {{Complex matrix decomposition and
  quadratic programming}},\ }\href {https://doi.org/10.1287/moor.1070.0268}
  {\bibfield  {journal} {\bibinfo  {journal} {Mathematics of Operations
  Research}\ }\textbf {\bibinfo {volume} {32}},\ \bibinfo {pages} {758}
  (\bibinfo {year} {2007})}\BibitemShut {NoStop}%
\bibitem [{\citenamefont {Mosk}\ \emph {et~al.}(2012)\citenamefont {Mosk},
  \citenamefont {Lagendijk}, \citenamefont {Lerosey},\ and\ \citenamefont
  {Fink}}]{Mosk2012}%
  \BibitemOpen
  \bibfield  {author} {\bibinfo {author} {\bibfnamefont {A.~P.}\ \bibnamefont
  {Mosk}}, \bibinfo {author} {\bibfnamefont {A.}~\bibnamefont {Lagendijk}},
  \bibinfo {author} {\bibfnamefont {G.}~\bibnamefont {Lerosey}},\ and\ \bibinfo
  {author} {\bibfnamefont {M.}~\bibnamefont {Fink}},\ }\bibfield  {title}
  {\bibinfo {title} {{Controlling waves in space and time for imaging and
  focusing in complex media}},\ }\href
  {https://doi.org/10.1038/nphoton.2012.88} {\bibfield  {journal} {\bibinfo
  {journal} {Nature Photonics}\ }\textbf {\bibinfo {volume} {6}},\ \bibinfo
  {pages} {283} (\bibinfo {year} {2012})}\BibitemShut {NoStop}%
\bibitem [{\citenamefont {Hansen}(1981)}]{Hansen1981}%
  \BibitemOpen
  \bibfield  {author} {\bibinfo {author} {\bibfnamefont {R.}~\bibnamefont
  {Hansen}},\ }\bibfield  {title} {\bibinfo {title} {{Fundamental limitations
  in antennas}},\ }\href {https://doi.org/10.1109/PROC.1981.11950} {\bibfield
  {journal} {\bibinfo  {journal} {Proceedings of the IEEE}\ }\textbf {\bibinfo
  {volume} {69}},\ \bibinfo {pages} {170} (\bibinfo {year} {1981})}\BibitemShut
  {NoStop}%
\bibitem [{\citenamefont {Tsang}\ \emph {et~al.}(2016)\citenamefont {Tsang},
  \citenamefont {Nair},\ and\ \citenamefont {Lu}}]{Tsang2016}%
  \BibitemOpen
  \bibfield  {author} {\bibinfo {author} {\bibfnamefont {M.}~\bibnamefont
  {Tsang}}, \bibinfo {author} {\bibfnamefont {R.}~\bibnamefont {Nair}},\ and\
  \bibinfo {author} {\bibfnamefont {X.~M.}\ \bibnamefont {Lu}},\ }\bibfield
  {title} {\bibinfo {title} {{Quantum theory of superresolution for two
  incoherent optical point sources}},\ }\href
  {https://doi.org/10.1103/PhysRevX.6.031033} {\bibfield  {journal} {\bibinfo
  {journal} {Physical Review X}\ }\textbf {\bibinfo {volume} {6}},\ \bibinfo
  {pages} {031033} (\bibinfo {year} {2016})}\BibitemShut {NoStop}%
\bibitem [{\citenamefont {Zhou}\ and\ \citenamefont {Jiang}(2019)}]{Zhou2019}%
  \BibitemOpen
  \bibfield  {author} {\bibinfo {author} {\bibfnamefont {S.}~\bibnamefont
  {Zhou}}\ and\ \bibinfo {author} {\bibfnamefont {L.}~\bibnamefont {Jiang}},\
  }\bibfield  {title} {\bibinfo {title} {{Modern description of Rayleigh's
  criterion}},\ }\href {https://doi.org/10.1103/PhysRevA.99.013808} {\bibfield
  {journal} {\bibinfo  {journal} {Physical Review A}\ }\textbf {\bibinfo
  {volume} {99}},\ \bibinfo {pages} {013808} (\bibinfo {year}
  {2019})}\BibitemShut {NoStop}%
\end{thebibliography}
\end{document}


\preprint{APS/123-QED}

\title{Supplementary Materials: Maximal Free-Space Concentration of Electromagnetic Waves}

\author{Hyungki Shim}
\affiliation{Department of Applied Physics and Energy Sciences Institute, Yale University, New Haven, Connecticut 06511, USA}
\affiliation{Department of Physics, Yale University, New Haven, Connecticut 06511, USA}
\author{Haejun Chung}
\affiliation{Department of Applied Physics and Energy Sciences Institute, Yale University, New Haven, Connecticut 06511, USA}
\author{Owen D. Miller}
\affiliation{Department of Applied Physics and Energy Sciences Institute, Yale University, New Haven, Connecticut 06511, USA}

\date{\today}

\maketitle

\tableofcontents

\section{Modal-Decomposition Bounds}
In this section we expand on the modal-decomposition bounds discussed in the main text, and we derive the analytical bounds for a planar decomposition. We start with the same optimization problem as in the main text,
\begin{equation}
    \begin{aligned}
        & \underset{\mu,\psi}{\text{maximize}} & & |\mu^\dagger \psi(\xv_0)|^2 \\
        & \text{subject to} & & \psi(\xv_\mcC) = 0, \\
        & & & \int |\psi(\xv)|^2 \leq 1, \label{eq:opt}
    \end{aligned}
\end{equation}
where $\mu$ is a unit six-vector oriented along the direction of optimal polarization.

\subsection{General Bounds}
Now we want to consider the case in which the field $\psi$ is expanded in some modal basis. We will index the modal basis with a continuous variable $\kv$, which is the Fourier wavevector in a 2D plane or 3D spherical surface, and a discrete variable $i$, which can correspond e.g. to a polarization index, in which case the field can be decomposed into modes $u_i(\kv;\xv)$ with coefficients $\psit_i(\kv)$:
\begin{align}
    \psi(\xv) = \sum_i \int \psit_i(\kv) u_i(\kv;\xv) \,{\rm d}\kv. \label{eq:modes}
\end{align}
Similarly, we assume the modes are normalized such that
\begin{align}
    \int u_i^\dagger (\kv;\xv) u_j(\kv';\xv) \,{\rm d}\xv = \delta(\kv - \kv') \delta_{ij},\label{eq:ortho}
\end{align}
in which case it is straightforward to integrate $u_j(\kv';\xv)$ against \eqref{modes} to determine the modal coefficients $\psit$:
\begin{align}
    \psit_i(\kv) = \int u_i^\dagger(\kv;\xv) \psi(\xv) \,{\rm d}\xv.\label{eq:mcoeffs}
\end{align}
To enforce the zero-field condition in \eqref{opt}, we will follow the same procedure as in the main text, enforcing the condition against some discrete set of basis functions (a different set that are defined only on the contour) $\phi_i(\xv)$. To derive an analytical bound we will require only that the field be zero as measured against the first basis function, i.e.,
\begin{align}
    \int_{\mcC} \phi_0^\dagger(\xv) \psi(\xv) \,{\rm d}\xv = 0.
\end{align}
If we replace $\psi$ with its modal decomposition, we can rewrite this zero-field condition as
\begin{align}
    \sum_i \int {\rm d}\kv \, \left[ \int_{\mcC} {\rm d}\xv \, \phi_0^\dagger(\xv) u_i(\kv;\xv) \right] \psit_i(\kv) = 0.
\end{align}
The inner integral is the conjugate transpose of the inverse decomposition of $\phi_0$, and is thus the conjugate transpose of the $i$-polarized modal coefficients of $\phi_0$, which we will call $\phit$, in analogy with \eqref{mcoeffs}. Then the zero-field condition is simply $\sum_i \int \phit^\dagger(\kv) \psit_i(\kv)\,{\rm d}\kv = 0$. Making the modal-decomposition replacement everywhere in \eqref{opt}, it can be written:
\begin{equation}
    \begin{aligned}
        & \underset{\mu,\psit_i}{\text{maximize}} & & \left|\sum_i \int \psit_i(\kv) u_i(\kv;\xv_0) \,{\rm d}\kv\right|^2 \\
        & \text{subject to} & & \sum_i \int {\rm d}\kv \, \phit^\dagger(\kv) \psit_i(\kv) = 0, \\
        & & & \int \left|\psit(\kv)\right|^2 \,{\rm d}\kv \leq 1, \label{eq:optk}
    \end{aligned}
\end{equation}
where in the zero-field constraint we take $\phi_0 = 0$ for points not on the zero-field contour $\mcC$. Just as in the main text, we can simplify the optimization problem by defining two scalar fields
\begin{equation}
\begin{aligned}
    \psi_{0,i}(\kv) &= \mu^T \cc{u}_i(\kv;\xv_0) \\
    \psi_1(\kv) &= \phit(\kv),
    \label{eq:psi01}
\end{aligned}
\end{equation}
where an asterisk denotes complex conjugation. Then, by any standard discretization scheme, we can write \eqref{optk} as the finite-dimensional problem
\begin{equation}
    \begin{aligned}
        & \underset{\psit}{\text{maximize}} & & \left| \psi_0^\dagger \psit  \right|^2 \\
        & \text{subject to} & & \psi_1^\dagger \psit = 0, \\
        & & & \psit^\dagger \psit \leq 1. \label{eq:optd}
    \end{aligned}
\end{equation}
As shown in the main text, this optimization problem is bounded above by the value,
\begin{align}
    I \leq \psi_0^\dagger \psi_0 - \frac{\left|\psi_0^\dagger \psi_1\right|^2}{\psi_1^\dagger \psi_1}.
    \label{eq:I2bound}
\end{align}

\subsection{Planar bounds}
The prototypical scenario in which one would want to apply the modal-decomposition bounds is to an optical beam in which light is focused at a single point, with a spot size defined by a circle (ring) in some transverse plane, and with the modes comprising all propagating plane waves. Then the modes are:
\begin{align}
    u_m(\kv;\xv) = \frac{\polm_m}{2\pi} e^{i\kv_m\cdot\xv}
    \label{eq:pw}
\end{align}
where we use $\kv$ to denote the wavevector of each plane wave, and $m$ is the modal index that additionally accounts for the two polarizations $\polm_m$ per wavevector. The modal fields of \eqref{pw} incorporate both the electric and magnetic fields as long as $\polm_m$ is a six-vector, and given the $1/2\pi$ prefactor they satisfy the orthogonality relation of \eqref{ortho}. (Note that the $1/2\pi$ prefactor is really the square of two $1/\sqrt{2\pi}$ prefactors in the unitary angular-frequency Fourier transforms in two dimensions.) Given the modes, we just need to compute $\psi_0(\kv)$ and $\psi_1(\kv)$ from \eqref{psi01}, from which the intensity bound is given by \eqref{I2bound}. The field at the origin (max-intensity point) is
\begin{align}
    \psi_{0,i}(\kv) = \frac{\polm_i^\dagger \mu}{2\pi}.
\end{align}
Given the electric/magnetic symmetry in the modal fields there is no advantage to maximizing the electric or magnetic fields or some linear combination, and thus we can choose the unit vector $\mu$ arbitrarily to align with the electric field along a nominal $\hat{\xv}$ direction. We will introduce the six-vector $\hat{x} = \begin{pmatrix} \hat{\xv} \\ \vect{0} \end{pmatrix}$, giving
\begin{align}
    \psi_{0,i}(\kv) = \frac{\polm_i^\dagger \hat{x}}{2\pi}.
\end{align}

Meanwhile, $\psi_1$ is given by the Fourier transform of the function $\psi_0$ that is a basis function for the zero-field condition. In a plane, one would want zero field on a circle of radius $R$, in which case a sensible basis against which to measure the field on the circle would be the complex exponentials, i.e.
\begin{align}
    \phi_n(\xv) = \polm_n e^{in\theta}
\end{align}
where $\theta$ is the angle at a location on the circle, $n$ is a non-negative integer, and $\polm_n$ is a polarization six-vector. (Note that technically $\phi$ should have two indices, one for the radial variations and one for polarization, but we will not include that since we will only consider a single basis function.) Here we will choose the same polarization for the first basis function, i.e., $\phi_0(\xv) = \hat{x}$. Then we can write $\psi_1$ as
\begin{align}
    \psi_1(\kv) &= \phit(\kv) \nonumber \\
                &= \int u_i^\dagger(\kv;\xv) \phi_0(\xv) \,{\rm d}\xv \nonumber \\
                &= \frac{\polm_i^\dagger \hat{x}}{2\pi} \int_{r=R} e^{-i\kv_i\cdot\xv} \,{\rm d}\xv. \label{eq:psi1v0}
\end{align}
The integral in \eqref{psi1v0} can be simplified by transforming both $\xv$ and $\kv$ to polar coordinates: $x = r\cos\theta$, $y = r\sin\theta$, $k_x = \rho\cos\theta'$, and $k_y = \rho\sin\theta$. Then the integral is
\begin{align}
    R \int_0^{2\pi} e^{iR\rho\left(\cos\theta\cos\theta' + \sin\theta\sin\theta'\right)} \,{\rm d}\theta &= R\int_0^{2\pi} e^{iR\rho\cos(\theta-\theta')} \,{\rm d}\theta \nonumber \\
                                                                                                         &= R\int_0^{2\pi} e^{iR\rho\cos\theta} \,{\rm d}\theta \nonumber \\
                                                                                                         &= 2\pi R J_0(R\rho),
\end{align}
where $J_0$ is the zeroth-order Bessel function, and the second equality follows because the endpoints of an integral can be shifted arbitrarily when integrating a periodic function over its period. The field $\psi_1$ is the two-dimensional Fourier transform of a ring, which is the zeroth-order Bessel function. The polarization term $\polm_i^\dagger \hat{x}$ is slightly more complicated and we will not solve for it just yet, leaving
\begin{align}
    \psi_1(\kv) = (\polm_i^\dagger \hat{x}) R J_0(R\rho).
    \label{eq:psi1eq}
\end{align}
To determine the bound, we need to derive three quantities: $\psi_0^\dagger \psi_0$, $\psi_0^\dagger \psi_1$, and $\psi_1^\dagger \psi_1$. First is $\psi_0^\dagger \psi_0$:
\begin{align}
    \psi_0^\dagger \psi_0 &= \sum_i \int \left| \psi_{0,i}(\kv) \right|^2 \,{\rm d}\kv \nonumber \\
                          &= \frac{1}{4\pi^2} \int \int \left[\sum_i (\polm_i^\dagger \hat{x})^2\right] \,{\rm d}k_x {\rm d}k_y \nonumber
\end{align}
To determine the polarization-dependent term, we can use the fact that the two electric-field polarizations for every wavevector are orthogonal to the wavevector itself. Thus, $(\hat{\kv_i}\cdot\hat{\xv})^2 + \sum_i (\polm_i^\dagger\hat{x})^2  = 1$, since the total $x$-component of three orthogonal unit vectors must 1, independent of their orientation. Rearranging, we have that $\sum_i (\polm_i^\dagger \hat{x})^2 = 1 - (\hat{\kv_i} \cdot \hat{\xv})^2$, and we can simplify the integral
\begin{align}
    \int \int \left[ 1 - \frac{k_x^2}{k^2} \right] {\rm d}k_x {\rm d}k_y \nonumber &= \int_0^{k} \int_0^{2\pi} \left[ 1 - \frac{\rho^2 \cos^2\theta}{k^2} \right] \rho \,{\rm d}\rho {\rm d}\theta. \nonumber \\
                                                                                             &= \left[ \pi k^2 - \frac{\pi}{4} k^2 \right] = \frac{3\pi k^2}{4}.
\end{align}
Thus we have
\begin{empheq}[box=\widefbox]{align}
    \psi_0^\dagger \psi_0 = \frac{3k^2}{16\pi}.
\end{empheq}

Next, we determine the value of $\psi_0^\dagger \psi_1$. This term has precisely the same $\sum_i (\polm_i^\dagger\hat{x})^2$ term in it, so we will use the same procedure as above to simplify the integral to:
\begin{align}
    \psi_0^\dagger \psi_1 &= \frac{R}{2\pi} \int_0^k \int_0^{2\pi} \rho J_0(R\rho) \left[ 1 - \frac{\rho^2 \sin^2\theta}{k^2} \right] \,{\rm d}\rho {\rm d}\theta.
\end{align}
The first term in the integral is given by $\int\int \rho J_0(R\rho) = 2\pi k J_1(kR) / R$. The second term is given by $\int\int \rho^3 J_0(R\rho) = (\pi k^2/R^2) \left[ 2J_2(kR) - kR J_3(kR) \right]$. Hence, the total term is given by
\begin{align}
    \psi_0^\dagger \psi_1 = k \left[ J_1(kR) - \frac{J_2(kR)}{kR} + \frac{1}{2} J_3(kR) \right].
\end{align}
We can use the recurrence relation $J_3(x) = 4J_2(x)/2 - J_1(x)$ to rewrite this term as
\begin{empheq}[box=\widefbox]{align}
    \psi_0^\dagger \psi_1 = k \left[ \frac{J_1(kR)}{2} + \frac{J_2(kR)}{kR} \right].
\end{empheq}

Finally, we determine the value of $\psi_1^\dagger \psi_1$, which again has the same polarization term, and is given by 
\begin{align}
    \psi_1^\dagger \psi_1 = R^2 \int_0^k \int_0^{2\pi} \rho \left[ J_0(R\rho) \right]^2 \left( 1 - \frac{\rho^2 \sin^2\theta}{k^2} \right) \,{\rm d}\rho {\rm d}\theta.
\end{align}
The first integral is given by
\begin{align}
    2\pi \int_0^k R\rho \left[ J_0(R\rho) \right]^2 \,{\rm d}(R\rho) = \pi k^2 R^2 \left\{ \left[ J_0(kR) \right]^2 + \left[ J_1(kR)\right]^2 \right\}
\end{align}
The second integral is given by
\begin{align}
    \frac{\pi}{R^2} \int_0^{kR} v^3 \left[ J_0(v) \right]^2 \,{\rm d}v = \pi k^4 R^2 \left\{ \frac{1}{6} \left[ J_0(kR) \right]^2 + \frac{1}{3} \frac{J_0(kR) J_1(kR)}{kR} + \frac{1}{6} \left(1 - \frac{2}{k^2 R^2} \right) \left[ J_1(kR) \right]^2 \right\}
\end{align}
Hence $\psi_1^\dagger \psi_1$ is given by
\begin{empheq}[box=\widefbox]{align}
    \psi_1^\dagger \psi_1 = \pi k^2 R^2 \left\{ \frac{5}{6} \left[ J_0(kR) \right]^2 + \frac{5}{6} \left[ J_1(kR) \right]^2 - \frac{J_0(kR) J_1(kR)}{3kR} + \frac{\left[J_1(kR)\right]^2}{3k^2R^2} \right\}.
\end{empheq}
Finally, we can write the bound:
\begin{empheq}[box=\widefbox]{align}
    I \leq \frac{3 k^2}{16\pi} - \frac{1}{\pi R^2} \frac{\left[ \frac{J_1(kR)}{2} + \frac{J_2(kR)}{kR} \right]^2}{\frac{5}{6} \left[ J_0(kR) \right]^2 + \frac{5}{6} \left[ J_1(kR) \right]^2 - \frac{J_0(kR) J_1(kR)}{3kR} + \frac{\left[J_1(kR)\right]^2}{3k^2R^2}}.
    \label{eq:IboundMB}
\end{empheq}
For small spot sizes $R$, i.e. $kR \ll 1$, a Taylor expansion of the bound gives
\begin{empheq}[box=\widefbox]{align}
    I \leq \frac{13}{13824\pi} k^6 R^4 \qquad \qquad kR \ll 1.
    \label{eq:IboundMBTE}
\end{empheq}
We can divide this bound by $I_0 = \psi_0^\dagger \psi_0$ to get a bound normalized by the unconstrained optimal value,
\begin{empheq}[box=\widefbox]{align}
    \frac{I}{I_0} \leq \frac{13}{2592} k^4 R^4 \qquad \qquad kR \ll 1.
    \label{eq:IboundMBTEnorm}
\end{empheq}
Finally, the one assumption we made to ultimately get to \eqrefrange{IboundMB}{IboundMBTEnorm} was that the optimal polarization is in the $\hat{\xv}$ direction. By symmetry, there is no difference between $\hat{\xv}$, $\hat{\vect{y}}$, or any linear combination thereof. The perpendicular polarization, $\hat{\vect{z}}$, would intuitively reduce performance (losing e.g. normal-incidence waves from the basis), but we can verify this intuition. The first term, $\psi_0^\dagger \psi_0$, is now given by
\begin{align}
    \psi_0^\dagger \psi_0 = \frac{1}{4\pi^2} \int \int \left[\sum_i (\polm_i^\dagger \hat{z})^2\right] \,{\rm d}k_x {\rm d}k_y 
    = \frac{1}{4\pi^2} \int \int \left[ 1 - \frac{k_z^2}{k^2} \right] {\rm d}k_x {\rm d}k_y 
    = \frac{1}{2\pi k^2} \int_0^{k} \rho^3 \,{\rm d}\rho = \frac{k^2}{8\pi}.
\end{align}
The second term, $\psi_0^\dagger \psi_1$, is given by
\begin{align}
    \psi_0^\dagger \psi_1 = \frac{R}{k^2} \int_0^k \rho^3 J_0(R\rho) \,{\rm d}\rho = \frac{1}{R} \left[ 2J_2(kR) - kR J_3(kR) \right].
\end{align}
Lastly, the third term is given by
\begin{align}
    \psi_1^\dagger \psi_1 = \frac{2\pi R^2}{k^2} \int_0^k \rho^3 \left[ J_0(R\rho) \right]^2 \,{\rm d}\rho {\rm d}\theta = \frac{\pi}{3} \left[ (kR)^2 (J_0(kR))^2 + 2(kR) J_0(kR) J_1(kR) + (k^2 R^2 - 2) (J_1(kR))^2 \right].
\end{align}
Putting these terms together, the $z$-polarized bound is
\begin{align}
    I \leq \frac{k^2}{8\pi} - \frac{3}{\pi R^2} \frac{\left[2J_2(kR) - (kR)J_3(kR)\right]^2}{(kR)^2 J_0(kR)^2 + 2(kR)J_0(kR)J_1(kR) + (k^2R^2 - 2)J_1(kR)^2}.
    \label{eq:IboundMBZ}
\end{align}
For small $kR$, the bound simplifies to
\begin{align}
    I \leq \frac{k^6 R^4}{2304\pi},
\end{align}
which is smaller than the equivalent bound for the $x$-polarization, \eqref{IboundMBTE}. Indeed, one can verify computationally that the general bound for the $z$-polarization, \eqref{IboundMBZ}, is always smaller than the equivalent bound for $x$-polarization, \eqref{IboundMB}, generally by about a factor of 2. Hence, the transverse polarization is optimal, and the bounds of \eqrefrange{IboundMB}{IboundMBTEnorm} are the global intensity bounds.

\subsection{1D Line: Equivalence to PSWF Approach}
Consider a two-dimensional scalar beam-propagation problem in which one wants to concentrate light as tightly as possible on a cross-sectional line. Then the general approach from the main text still applies, with \eqrefrange{IboundMB}{IboundMBTE} of this SM still providing bounds; to derive the 1D bounds, we simply need to identify $\psi_0$ and $\psi_1$. (There is no longer an $i$ index since it is a scalar problem.) The modal fields are identical to those of the three-dimensional case, but now without any polarization factor and with only one factor of $1/\sqrt{2\pi}$ due to the single dimension:
\begin{align}
    u(k;x) &= \frac{1}{\sqrt{2\pi}} e^{ikx},
\end{align}
leading to $\psi_0$ and $\psi_1$ values of
\begin{align}
    \psi_0(k) &= \frac{1}{\sqrt{2\pi}} \\
    \psi_1(k) &= \frac{1}{2} \int \cc{u}(k;x) \left[\delta(x-R) + \delta(x+R)\right] \,{\rm d}x \nonumber \\
              &= \frac{\cos(kR)}{\sqrt{2\pi}}.
\end{align}
Then the three bound terms are
\begin{align}
    \psi_0^\dagger \psi_0 &= \int_{-k}^k |\psi_0|^2 \,{\rm d}k = \frac{k}{\pi} \\ 
    \psi_0^\dagger \psi_1 &= \frac{1}{2\pi} \int_{-k}^k \cos(kR) \,{\rm d}k = \frac{\sin(kR)}{\pi R} \\
    \psi_1^\dagger \psi_1 &= \frac{1}{2\pi} \int_{-k}^k \cos^2(kR) \,{\rm d}k = \frac{k}{2\pi} \left[1 + \frac{\sin(2kR)}{2kR}\right].
\end{align}
Inserting these three terms into the bound of \eqref{I2bound}, we have
\begin{empheq}[box=\widefbox]{align}
    I \leq \frac{k}{\pi} \left[1 - 2 \frac{\sinc^2(kR)}{1 + \sinc(2kR)}\right]
    \label{eq:IboundMB1d}
\end{empheq}
If we denote the unconstrained maximum intensity as $I_0$, then
\begin{align}
    I_0 = \frac{k}{\pi},
\end{align}
and the ratio of $I_{\max}$ to $I_0$ is
\begin{empheq}[box=\widefbox]{align}
    \frac{I_{\rm max}}{I_0} = 1 - 2 \frac{\sinc^2(kR)}{1 + \sinc(2kR)}.
    \label{eq:IboundMB1dnorm}
\end{empheq}

We can show that these bounds, \eqreftwo{IboundMB1d}{IboundMB1dnorm}, are equivalent to those that would be derived by the classical prolate-spheroidal-wave-function approach that dates to the work of Slepian, Polak, and others~\cite{Slepian1961,Slepian1978}. Apparently the most general bounds for the scalar 1D problem are those derived in \citeasnoun{Ferreira2006}, although they primarily tackle the case in which one wants a signal that varies from +1 to -1 over some arbitrarily small distance. Here we will apply their analysis to the imaging problem, in which one wants a large intensity at the origin and then zeros at symmetric points around the focus. In following their approach (and those of earlier authors~\cite{Courant1953,Levi1965}), instead of maximizing the intensity at the origin, we will find the signal with minimum total energy such that the field amplitude at the origin equals one. (The two problems are equivalent, with solutions that are inverses of each other.)

The statement of the problem is as follows:
\begin{equation}
\begin{aligned}
     & \underset{\psit(k)}{\text{min}} & & \int |\psit(k)|^2 \,{\rm d}k \label{eq:qform} \\
     & \text{subject to}               & & \psi(-R) = 0 \\
     & & & \psi(0) = 1 \\
     & & & \psi(R) = 0.
\end{aligned}
\end{equation}
We can write this in a more general form, inserting the specific values later, as
\begin{equation}
\begin{aligned}
     & \underset{\psit(k)}{\text{min}} & & \int |\psit(k)|^2 \,{\rm d}k \label{eq:qform} \\
     & \text{subject to}               & & \psi(x_i) = \psi_i, \qquad i=0,1,2
\end{aligned}
\end{equation}
where $x_0 = -R$, $x_1 = 0$, $x_2 = R$, and $\psi_0 = 0$, $\psi_1 = 1$, and $\psi_2 = 0$. The field constraint can be written
\begin{align}
    \frac{1}{\sqrt{2\pi}} \int \psit(k) e^{ikx_i} = \psi_i.
    \label{eq:fc}
\end{align}
The Lagrangian for this problem is
\begin{align}
    \mathcal{L} = \int |\psit(k)|^2 \,{\rm d}k + \sum_i \lambda_i \left[ \Re \left(\frac{1}{\sqrt{2\pi}} \int \psit(k) e^{ikx_i}\,{\rm d}k \right) - \psi_i \right],
\end{align}
where the $\lambda_i$ are the Lagrange multipliers, and we can drop the constraints on the imaginary part of $\psit(k)$, which simply go to zero in the minimization. Then the extremal condition can be found either by independently varying the real and imaginary parts of $\psit(k)$, or, more conveniently, by using the CR calculus~\cite{KreutzDelgado2009} and formally treating $\psit(k)$ and $\cc{\psit}(k)$ as independent variables. In the latter case, we can only look at the variations in $\cc{\psit}(k)$, which give
\begin{align}
    \frac{\partial \mathcal{L}}{\partial \cc{\psit}(k)} = \psit(k) + \sum_i \lambda_i \frac{1}{2\sqrt{2\pi}} e^{-ikx_i} = 0.
\end{align}
Thus we have
\begin{align}
    \psit(k) = \sum_i \lambda_i e^{-ikx_i},
    \label{eq:psit_opt}
\end{align}
where we use the freedom in the Lagrange multipliers to absorb the negative sign and constant prefactor. We can find the (renormalized) Lagrange multipliers by inserting this expression into the field constraint, \eqref{fc}, giving
\begin{align}
    \frac{1}{\sqrt{2\pi}} \sum_j \lambda_j \int e^{ik(x_i - x_j)} \,{\rm d}k = \psi_i.
\end{align}
A straightforward calculation shows that $\int e^{ik(x_i - x_j)} = 2k \sinc(k(x_i-x_j))$, such that we can rewrite this expression as
\begin{align}
    \sum_j \underbrace{\sinc(k(x_i - x_j))}_{M_{ij}} \lambda_j = \underbrace{\sqrt{\frac{\pi}{2}} \frac{1}{k} \psi_i}_{a_i},
\end{align}
where we have defined the matrix $M$ and vector $a$ such that this can be succinctly written
\begin{align}
    M \lambda = a.
\end{align}
The Lagrange multipliers are the solution of this $3\times 3$ matrix equation, which then define the optimal $\psit(k)$ by \eqref{psit_opt}. Note that $M$ is precisely the matrix that defines the eigenequation for the discrete prolate spheroidal wave functions~\cite{Slepian1978,Ferreira2006}. The total energy in the minimum-energy solution is then given by
\begin{align}
    \int |\psit(k)|^2 \,{\rm d}k &= \sum_{ij} \lambda_i \lambda_j \int e^{ik(x_i - x_j)} \,{\rm d}k \nonumber \\
                                 &= 2k \sum_{ij} \lambda_i \lambda_j \sinc(k(x_i - x_j)) \nonumber \\
                                 &= 2k \lambda^T M \lambda \nonumber \\
                                 &= 2k a^T M^{-1} a.
\end{align}
Now we can use the fact that only the second element of $a$ is nonzero (since $\psi_0$ and $\psi_2$ are 0, to write
\begin{align}
    \int |\psit(k)|^2 \,{\rm d}k = \frac{\pi}{k} \left(M^{-1}\right)_{22},
    \label{eq:totenergy}
\end{align}
where $(M^{-1})_{22}$ refers to the second row and second column of the matrix $M^{-1}$. Since we know the $x_i$ values are $-R$, $0$, and $R$, the matrix $M$ is given by
\begin{align}
    M = \begin{pmatrix}
        1 & \sinc(kR) & \sinc(2kR) \\
        \sinc(kR) & 1 & \sinc(2kR) \\
        \sinc(2kR) & \sinc(kR) & 1
    \end{pmatrix}.
\end{align}
The value of $(M^{-1})_{22}$ is
\begin{align}
    \left(M^{-1}\right)_{22} = \frac{1}{|M|} \left| 
    \begin{pmatrix}
        M_{11} & M_{13} \\
        M_{31} & M_{33}
    \end{pmatrix} \right|.
\end{align}
The two determinants are given by
\begin{align}
    \left|M\right| = 1 + 2\sinc^2(kR) \left[\sinc(2kR) - 1\right] - \sinc^2(2kR),
    \label{eq:detM}
\end{align}
and
\begin{align}
    \left| \begin{pmatrix}
        M_{11} & M_{13} \\
        M_{31} & M_{33}
    \end{pmatrix} \right| = 1 - \sinc^2(2kR).
    \label{eq:det22}
\end{align}
The ratio of \eqref{det22} to \eqref{detM} can be shown to be
\begin{align}
    \frac{1}{1 - 2\frac{\sinc^2(kR)}{1+\sinc(2kR)}}.
\end{align}
Hence the minimum energy is given by exactly this expression multiplied by $\pi/k$, as seen from \eqref{totenergy}. This is the minimum energy for a unit amplitude; by contrast, the \emph{maximum amplitude} for a \emph{fixed energy} (equal to 1) is given by the inverse of the minimum-energy solution (as can be shown by rescaling). Thus the PSWF approach gives a maximum amplitude of
\begin{empheq}[box=\widefbox]{align}
    I \leq \frac{k}{\pi} \left[1 - 2\frac{\sinc^2(kR)}{1+\sinc(2kR)}\right],
\end{empheq}
which is exactly the same as that predicted by our approach.

Hence, we have shown that our approach indeed reduces exactly to the PSWF-based Fourier-analysis bound in the 1D scalar limit.

\section{Aperture-dependent bounds}
\subsection{Optimal solutions}
Starting from the rank-two optimization problem given in the main text,
\begin{equation}
\begin{aligned}
     & \underset{\pol,\nu}{\text{maximize}} & & \nu^\dagger \Pv \gamma_\pol \gamma_\pol^\dagger \Pv \nu \label{eq:qform} \\
     & \text{subject to}               & & \nu^\dagger \Pv \nu \leq 1,
\end{aligned}
\end{equation}
where $\pol$ is a six-component polarization vector and $\Pv$ is a projection matrix, which is  Hermitian and idempotent ($\Pv \Pv = \Pv$) such that $\xi = \left(\mathbf{I} - \tG_\mcC^\dagger \left(\tG_\mcC \tG_\mcC^\dagger\right)^{-1} \tG_\mcC\right) \nu = \mathbf{P} \nu$. \Eqref{qform} is equivalent to a Rayleigh-quotient optimization, which is maximized by the largest eigenvalue of the generalized eigenproblem
\begin{align}
\Pv \gamma_\pol \gamma_\pol^\dagger \Pv \nu = \lambda \Pv \nu,
\end{align}
whose eigenvector is given by $\nu = \Pv \gamma_\pol / \|\Pv \gamma_\pol\|$ (satisfying the constraint in \eqref{qform} with largest magnitude), with the nonzero eigenvalue $\gamma_\pol^\dagger \Pv \gamma_\pol = \gamma_\pol^{\dag}  \left(\mathbf{I} - \tG_\mcC^\dagger \left(\tG_\mcC \tG_\mcC^\dagger\right)^{-1} \tG_\mcC\right) \gamma_\pol$, using the property that $\Pv \nu = \nu$. It is guaranteed to be the only nonzero eigenvalue, since $\Pv \gamma_\pol \gamma_\pol^\dagger \Pv$ is a rank one matrix, and any other eigenvectors would have to be orthogonal to $\Pv \gamma_\pol$ such that the matrix times the eigenvector is necessarily zero. Using the definition $\gamma_\pol = \tG_0^\dagger \pol$, we can write the upper bound to intensity $I$ in terms of the dyadic Green's functions $\Gamma$~\cite{Chew1995}:
\begin{align}
    I \leq \pol^\dagger \left[ \tG_0  \tG_0^\dagger - \tG_0 \tG_\mcC^\dagger \left(\tG_\mcC \tG_\mcC^\dagger \right)^{-1} \tG_\mcC \tG_0^\dagger \right]\pol.
    \label{eq:Ibound}
\end{align}

If we loosen the zero-field constraint and instead constrain only the zeroth-order, $\pol$-polarized mode ($\phi_0^\dagger \tG_\mcC \xi = 0$ where $\phi_0$ is a 6-vector), we can simplify the upper bound on intensity. By replacing $\tG_\mcC \rightarrow \phi_0^\dagger \tG_\mcC$, \eqref{Ibound} reduces to
\begin{align}
    I \leq \pol^\dagger \left[ \tG_0  \tG_0^\dagger - \tG_0 \tG_\mcC^\dagger \phi_0 \left(\phi_0^\dagger \tG_\mcC \tG_\mcC^\dagger \phi_0 \right)^{-1} \phi_0^\dagger \tG_\mcC \tG_0^\dagger \right]\pol.
    \label{eq:Ibound2}
\end{align}
Defining $\psi_0 = \gamma_\pol = \tG_0^\dagger \pol$ and $\psi_1 = \tG_\mcC^\dagger \phi_0$, \eqref{Ibound2} reduces to an expression in terms of only the two new fields $\psi_0$ and $\psi_1$,
\begin{align}
    I \leq \psi_0^\dagger \psi_0 - \frac{|\psi_0^\dagger \psi_1|^2}{\psi_1^\dagger \psi_1}. \label{eq:Imax}
\end{align}
To find the Strehl ratio, we need to divide \eqref{Imax} by the diffraction-limited (i.e. no zero-field constraint) intensity, which is simply $\psi_0^{\dag} \psi_0$ (corresponding to the projection matrix $\proj$ being an identity matrix). Hence the upper bound to the Strehl ratio, which we denote $S_{\textrm{max}}$, is
\begin{align}
S_{\textrm{max}} =  1 - \frac{|\psi_0^\dagger \psi_1|^2}{\psi_0^\dagger \psi_0 \, \psi_1^\dagger \psi_1}. \label{eq:optSgen} 
\end{align}
The (normalized) optimal effective current is $\xi_{\textrm{opt}} = \Pv \nu = \Pv \psi_0 / \|\Pv \psi_0\|$ (where $\Pv= \mathbf{I} - \frac{ \psi \psi_1^\dagger}{\psi_1^\dagger \psi_1}$),
\begin{align}
\xi_{\rm opt}=\xi_0 \Bigg[\psi_0 - \frac{ \psi_1^\dagger \psi_0  }{\psi_1^\dagger \psi_1}\psi_1 \Bigg]. \label{eq:optcurrentgen} 
\end{align}	
where $\xi_0= 1 / \sqrt{\psi_0^\dagger \psi_0 - |\psi_0^\dagger \psi_1|^2 / \psi_1^\dagger \psi_1}$ such that $\xi_0^\dagger \xi_0 = 1$.  (the same analysis holds for the general bound in \eqref{Ibound}, but for simplicity we have shown the Strehl ratio bound and normalized optimal currents for the simpler bound of \eqref{Imax}).

\subsection{Far-zone asymptotic solutions} \label{sec:farzone}
Encoding the radiation from electric and magnetic currents in free space, the dyadic Green's function $\Gamma$~\cite{Chew1995} simplifies considerably in the far zone---the aperture and focal spot sizes are negligible compared to the distance between aperture and focal spot, such that the distance between any point on the spot and aperture $\approx z$. 

Recalling the definition of $\psi_0$ and $\psi_1$ from the main text and adopting a Fourier basis (with coordinates on the aperture denoted by $\xv'$),
\begin{align}
\psi_0 =  \tens{\Gamma}^\dagger(\xv=0,\xv') \pol, \quad \psi_1 =  \int _{\textrm{ring}}\tens{\Gamma}^\dagger(\xv,\xv') \pol \label{eq:recipfield}
\end{align}
In the far zone, the aperture-focus distance $z$ greatly exceeds the physical dimensions of the aperture and focal spot, such that $\psi_0$ can be written as
\begin{align}
\psi_0 = \left( \frac{k e^{ikz}}{4\pi icz} \right)^*  \underbrace{\begin{pmatrix} 1&0&0&0&-1&0\\ 0&1&0&1&0&0\\ 0&0&0&0&0&0\\ 0&1&0&1&0&0\\-1&0&0&0&1&0 \\ 0&0&0&0&0&0 \end{pmatrix} }_{K} \pol = \frac{A e^{-ikz}}{z} \tilde{\pol} \label{eq:psi0}
\end{align}
where $A \equiv \frac{i k}{4\pi c}$ is fixed for a given wavelength and $\tilde{\pol} \equiv K \pol$ encoding the directions of $\psi_0$ and $\psi_1$ (same direction as $\psi_0$ in far zone). 

To determine $\psi_1$, we need to keep the correction term from aperture and focal spot size in the exponent, since even small changes in the exponent can affect the field significantly. Denoting the coordinate on the aperture and spot-size ring as $(x',y')$ and $(x,y)$ respectively (with respective radius of $r$ and $\rho_0$), the exact distance between the two points is $\sqrt{(x-x')^2 + (y-y')^2 + z^2} = z \sqrt{1  -2 \frac{xx' + yy'}{z^2} + \frac{\rho_0^2 + r^2}{z^2} } $. Inserting this expression in the exponent appearing in $\psi_1$,
\begin{align}
\psi_1 =  \frac{A}{z} \tilde{\pol} \int_{\textrm{ring}} e^{-ikz \sqrt{ 1  -2 \frac{xx' + yy'}{z^2} + \frac{\rho_0^2 + r^2}{z^2} } }.
\end{align}
In the far zone, we can neglect the last quadratic term in the exponent above. Since $z \gg \rho_0, r$, we can Taylor expand the exponential, $ \sqrt{ 1  -2 \frac{xx' + yy'}{z^2}} \approx 1 - \frac{xx' + yy'}{z^2} $. Then, we switch to polar coordinates with the angle on the aperture and focal spot defined as $\theta$ and $\phi$ respectively. It follows that $xx' + yy' = r \rho_0 \cos(\theta - \phi)$ so that 
\begin{align}
\psi_1 = \frac{A e^{-ikz} }{z} \tilde{\pol} \int_{0}^{2\pi} \rho_0 e^{-i \frac{k r\rho_0}{z} \cos(\theta - \phi)} \,{\rm d}\phi. \label{eq:psi1angle}
\end{align}
Using the reciprocity-based intuition explained in the main text, we know that the (fictitious) current sources placed on the focal spot are circularly symmetric. This means we can set $\theta=0$, as $\psi_1$ will only depend on the radial coordinate of the aperture. From the integral representation of the Bessel function of the first kind~\cite{abramowitz_stegun_1974}, $J_0 (u) = \frac{1}{2\pi} \int_0^{2\pi} e^{-iu \cos v} \,{\rm d}v$, the final expression for $\psi_2$ only depends on the radial coordinate:
\begin{align}
\psi_1 =  \frac{A e^{-ikz} }{z}  2\pi  \rho_0 J_0 \left (\frac{kr\rho_0}{z} \right) \tilde{\pol} = 2\pi  A \rho_0 J_0 \left (\frac{kr\rho_0}{z} \right) \frac{e^{-ikz}}{z} \tilde{\pol} \label{eq:psi1}.
\end{align}
Having calculated $\psi_0$ and $\psi_1$, it is now straightforward to derive the bound on maximal focusing intensity, Strehl ratio and the optimal effective current. Using the relations $\int_0^u u' J_0(u') \,{\rm d}u' = u J_1(u)$, $\int_0^u u' [J_0(u')]^2 \,{\rm d}u' = u^2 \left \{ [J_0(u)]^2 + [J_1(u)]^2 \right \}/ 2$ and rescaling the limits of integration from 0 to 1 (with the aperture radius $R$ taken out of the limits), the maximum intensity is given by inserting \eqreftwo{psi0}{psi1} into \eqref{Imax} (we choose units such that $c=1$): 
\begin{align}
    I \leq  \frac{\pi R^2 |A|^2}{z^2}- \frac{2\pi R^2 |A|^2}{z^2} \frac{\left | \int_0^1 J_0(kR\rho_0 r/z ) \, r \,{\rm d}r \right | ^2}{ \int_0^1 [J_0(kR\rho_0 r/z)]^2 \, r \,{\rm d}r} = \frac{k^2 R^2}{16 \pi z^2} - \frac{1}{4 \pi \rho_0^2} \frac{\left[J_1(kR\rho_0/z)\right]^2}{\left[J_0(kR\rho_0/z)\right]^2 + \left[J_1(kR\rho_0/z)\right]^2}.
    \label{eq:ImaxFZ}
\end{align}
Defining a normalized spot-size radius $\spot = \frac{k R \rho_0}{z}$, the Strehl ratio in \eqref{optSgen} reduces to:
\begin{align}
S_{\textrm{max}} = 1 - \frac{\left | \int_0^1 J_0(\spot r) \, r \,{\rm d}r \right | ^2}{ \int_0^1 \, r \,{\rm d}r  \int_0^1 [J_0(\spot r)]^2 \, r \,{\rm d}r} =    1 - \frac{4}{\spot^2} \frac{[J_1(\spot)]^2}{[J_0(\spot)]^2 + [J_1(\spot)]^2}. \label{eq:optS}
\end{align}

An asymptotic bound of $S_{\rm max}$ for small spot sizes ($\spot \ll 1$) turns out to be quartic in $\spot$ to lowest order, as we show below. Invoking the power series representation of $J_0(\spot)$ and $J_1(\spot)$ and keeping up to quartic order in $\spot$,
\begin{align}
S_{\rm max} = 1 - \frac{4}{\spot^2} \frac{(\spot/2 )^2 ( 1- \spot^2 / 8 + \spot^4 / 192)^2}{(1- \spot^2 / 4 + \spot^4 / 64)^2 + (\spot/2 )^2 ( 1- \spot^2 / 8 + \spot^4 / 192)^2} = \frac{ \spot^4}{192}, \quad \spot \ll 1. \label{eq:optSsmall}
\end{align}

Similarly, inserting the far-zone fields of \eqreftwo{psi0}{psi1} into \eqref{optcurrentgen} ($r$ denotes the radial coordinate on the aperture), 
\begin{align}
\xi_{\textrm{opt}}(r) =\xi_0 \frac{A e^{-ikz} }{z} \Bigg[ 1 -  \frac{2}{\spot} \frac{J_1(\spot)}{[J_0(\spot)]^2 + [J_1(\spot)]^2}J_0(\spot r) \Bigg] \tilde{\pol}.   \label{eq:optcurrent}
\end{align}
Based on \eqref{optcurrent}, we can also deduce the optimal field profile that achieves the bound in \eqref{optS}. Defining the normalized radial coordinate on the focal plane as $\fspot = \frac{k R \rho}{z}$ (where $\rho$ is the focal-plane radial coordinate),
\begin{align} 
| \psi(\fspot) | &= \psi_0 \Bigg[ \frac{J_1(\fspot)}{\fspot} - \frac{2}{\spot} \frac{J_1(\spot)}{[J_0(\spot)]^2 + [J_1(\spot)]^2} \int_0^1 J_0(\spot r) J_0(\fspot r) \, r \,{\rm d}r \Bigg] \nonumber \\
&= \psi_0 \Bigg[ \frac{J_1(\fspot)}{\fspot} - \frac{2}{\spot} \frac{J_1(\spot)}{[J_0(\spot)]^2 + [J_1(\spot)]^2} \frac{\spot J_0(\fspot)J_1(\spot) - \fspot J_0(\spot)J_1(\fspot)}{\spot^2 - \fspot^2} \Bigg] \quad (\textrm{0 if $\fspot = \spot$}) \label{eq:optfield}
\end{align}
where $\psi_0$ is a normalization factor. Near the diffraction limit, the ideal field profile resembles an Airy disk pattern, with a sharply focused center peak and negligible sidelobe intensity. As the spot size decreases, the sidelobe intensity begins to increase at the expense of focal intensity. 

\section{Far-zone bounds for focal sphere}
Instead of maximizing intensity for a given focal spot, we can consider a focal sphere centered around the peak intensity. While the same reasoning as in \secref{farzone} holds, we modify $\psi_1$ in \eqref{recipfield} to involve integration around a sphere of radius $\rho_0$ (we also adopt the zeroth-order spherical harmonic basis, which is a constant).

With $z_0$ denoting the $z$-coordinate on the sphere relative to the peak intensity, the distance between points on the aperture and focal sphere is $\sqrt{(x-x')^2 + (y-y')^2 + (z+z_0)^2} = z \sqrt{1  -2 \frac{xx' + yy'-zz_0}{z^2} + \frac{\rho_0^2 + r^2}{z^2} }$. Dropping the quadratic terms in the far zone, we can Taylor expand the exponential, $ \sqrt{ 1  -2 \frac{xx' + yy'-zz_0}{z^2}} \approx 1 - \frac{xx' + yy'-zz_0}{z^2} $. Switching to spherical polar coordinates with the sphere parametrized by $(\rho_0, \theta, \phi)$ (the angle on the aperture denoted by $\theta_{\textrm{ap}}$), we have that $xx' + yy' -zz_0 = r \rho_0 \cos(\theta_{\textrm{ap}} - \phi) - z\rho_0 \cos\theta$. In analogy to \eqref{psi1angle}, we arrive at the following expression for $\psi_1$:
\begin{align}
\psi_1 = \frac{A e^{-ikz} }{z} \tilde{\pol} \int_{\phi=0}^{2\pi} \int_{\theta=0}^{\pi} \rho_0 e^{-i \frac{k \rho_0}{z} \left[ r \cos(\theta_{\textrm{ap}} - \phi) - z \cos\theta \right]} \sin\theta \,{\rm d}\theta \,{\rm d}\phi.
\end{align}
Again, reciprocity tells us that $\theta_{\textrm{ap}}$ can be set to 0. This enables us to do the $\phi$-integral:
\begin{align}
\psi_1 = \frac{2\pi A e^{-ikz} }{z} \tilde{\pol} \int_{0}^{\pi} \rho_0 e^{i k \rho_0  \cos\theta} J_0 \left (\frac{kr\rho_0}{z} \sin\theta \right) \sin\theta \,{\rm d}\theta. \label{eq:psi13D}
\end{align}
\Eqref{psi13D} is a general expression valid in the far zone. The derivations for maximum intensity, Strehl ratio, optimal currents and field profile mimic those laid out in \secref{farzone}, but the form of \eqref{psi13D} renders them somewhat messier and less intuitive. For the sake of clarity (and also of interest for highly localized, subwavelength focal sphere), we instead focus on the scaling behavior of Strehl ratio for extremely small spot sizes. In such a regime ($\spot \ll 1$), we can further expand $J_0$ in the integrand, which results in spherical bessel functions $j_0$ and $j_1$:
\begin{align}
\psi_1 = \frac{2\pi A e^{-ikz} }{z} \tilde{\pol} \int_{0}^{\pi} \rho_0 e^{i k \rho_0  \cos\theta} \left[ 1 - \frac{1}{4}  \left (\frac{kr\rho_0}{z} \sin\theta \right) ^2 \right] \sin\theta \,{\rm d}\theta= \frac{2 \pi A e^{-ikz} }{z} \tilde{\pol} \left [ 2j_0(k\rho_0) - \frac{k\rho_0 r^2}{z^2}j_1(k\rho_0) \right]. 
\end{align}
Inserting the above expression into \eqref{optSgen} gives us Strehl ratio (after rescaling the limits of integration):
\begin{align}
S_{\textrm{max}} = 1 - \frac{\left |  \int_0^1 \left [ 2j_0(k\rho_0) - \frac{k\rho_0 R^2}{z^2}j_1(k\rho_0)r^2 \right] \, r \,{\rm d}r \right | ^2}{ \int_0^1 \, r \,{\rm d}r  \int_0^1  \left [ 2j_0(k\rho_0) - \frac{k\rho_0 R^2}{z^2}j_1(k\rho_0)r^2 \right]^2  \, r \,{\rm d}r} =    1 - \frac{2\left[ j_0(k\rho_0) - \frac{\spot R}{4z} j_1(k\rho_0) \right] ^2}{2[j_0(k\rho_0)]^2 - \frac{\spot R}{z}j_0(k\rho_0)j_1(k\rho_0) + \frac{\spot^2 R^2}{6z^2}[j_1(k\rho_0)]^2  }. 
\end{align}
To simplify the analysis, we make another assumption that the focal sphere has subwavelength radius such that $k\rho_0<1$, allowing us to Taylor expand the spherical Bessel functions. Denoting the normalized sphere radius as $\spot = \frac{k R \rho_0}{z}$, and defining $x = k\rho_0$ (focal radius relative to wavelength), we expand up to quartic order in $\spot$ or $x$ (the asymptotic bound turns out to be quartic):
\begin{align}
S_{\textrm{max}} = 1 - \frac{2\left[ (1-x^2 / 6 + x^4 / 120) - \frac{R}{4z} \spot (x/3 - x^3 / 30) \right] ^2}{2[ (1-x^2 / 6 + x^4 / 120)^2 - \frac{R}{z}\spot (1-x^2 / 6 + x^4 / 120)(x/3 - x^3 / 30) + \frac{ R^2}{6z^2}\spot ^2(x/3 - x^3 / 30) ^2  }. 
\end{align}
Although not obvious from the above expression, straightforward but tedious algebra and repeated Taylor expansion show that the leading order term is quartic just in $\spot$ and that other quadratic terms and a quartic term in $x$ all cancel out:
\begin{align}
S_{\rm max} = \frac{ \spot^4}{432}, \quad \spot \ll 1, \label{eq:optSsmall3D}
\end{align}
As \eqref{optSsmall3D} shows, the maximum Strehl ratio for a focal sphere also scales quartically with the normalized focal radius $\spot$.

\section{2D Far-zone bounds}
Mirroring the derivation presented in \secref{farzone}, we can bound the maximum intensity for the 2D problem where we have translational invariance along one direction (chosen to be the $y$-axis in what follows). Here we do not constrain only the zeroth-order mode as in the previous two sections, but solve for the zero-field constraint exactly, as the zero-field contour involves just two evaluation points.

Working in the $xz$-coordinate, we appropriately modify the Green's function taking the symmetry into account (electric Green's function explicitly shown in Eq.~(39) of \citeasnoun{DGF-notes}, albeit with $z$-axis chosen as symmetry direction). Using the asymptotic form for the Hankel function $H_0^{(1)}(u) \sim \sqrt {\frac{2}{\pi u}}e^{i(u-\frac{\pi}{4})}$ for large $u$ and neglecting the transverse $x$-coordinates relative to the aperture-focus distance $z$,
\begin{align}
\psi_0 =  \frac{k e^{-i(kz-\frac{\pi}{4})}}{4c} \sqrt {\frac{2}{k \pi z}} \underbrace{\begin{pmatrix} 1&0&0&1\\ 0&1&-1&0\\ 0&-1&1&0\\ 1&0&0&1\end{pmatrix} }_{K} \pol = \frac{A e^{-ikz}}{\sqrt{z}} \tilde{\pol} \label{eq:2dpsi0}
\end{align}
where $A \equiv \frac{\sqrt{2k/\pi}}{4c}e^{\frac{i\pi}{4}}$ is fixed for a given wavelength and $\tilde{\pol} \equiv K \pol$ encoding the directions of $\psi_0$ and $\psi_1$ (same direction as $\psi_0$ in far zone). 

Adopting the same notation as in \secref{farzone} (except that now we do not consider the $y$-coordinate), we can write down the expression for $\psi$, which now involves just two points $x = [\rho_0, -\rho_0]$ instead of all points around a spot-size ring:
\begin{align}
\psi_1 = \frac{A}{\sqrt{z}} \tilde{\pol} \left[ e^{-ikz \sqrt{ 1  -\frac{2\rho_0 x'}{z^2} + \frac{\rho_0^2 + {x'}^2}{z^2} } } + e^{-ikz \sqrt{ 1  + \frac{2\rho_0 x'}{z^2} + \frac{\rho_0^2 + {x'}^2}{z^2} } } \right].
\end{align}
Dropping the quadratic terms in the exponent in the far zone and Taylor expanding the remaining term, we obtain
\begin{align}
\psi_1 = \frac{A e^{-ikz}}{\sqrt{z}} \tilde{\pol} \left[ e^{i \frac{kx'\rho_0}{z} } + e^{-i \frac{kx'\rho_0}{z} } \right]=  \frac{2A e^{-ikz}}{\sqrt{z}} \cos \left( \frac{kx'\rho_0}{z}\right) \tilde{\pol}.  
\end{align}
Using the fields $\psi_0$ and $\psi_1$ obtained above, we can calculate the maximum intensity ($R$ is the half-length of the aperture):
\begin{align}
I^{2D} \leq \frac{2R \, |A|^2}{z} -  \frac{2R \, |A|^2}{z}  \frac{\left | \int_0^1 \cos(kR\rho_0x'/z) \,{\rm d}x' \right | ^2}{ \int_0^1 \cos^2(kR\rho_0x'/z) \,{\rm d}x'}  = \frac{kR}{4\pi z} - \frac{1}{\pi \rho_0}  \frac{\sin^2(kR\rho_0/z)}{\sin(2kR\rho_0/z) + 2kR\rho_0/z}  \label{eq:ImaxFZ2d}
\end{align}
Note that the unit here differs from that in \eqref{ImaxFZ} by a length dimension, as the aperture considered here is essentially a one-dimensional array of effective currents (extended infinitely along the $y$-axis) in contrast to a closed surface for the 3D case. 

Denoting the normalized spot size width as $\spot = \frac{k R \rho_0}{z}$, the maximum Strehl ratio is as follows:
\begin{align}
S_{\rm max}^{2D}  = 1 - \frac{\left | \int_0^1 \cos(\spot x) \,{\rm d}x \right | ^2}{ \int_0^1  \,{\rm d}x  \int_0^1 \cos^2(\spot x) \,{\rm d}x} = 1 - \frac{4}{\spot} \frac{\sin^2(\spot)}{\sin(2\spot) + 2\spot}. \label{eq:optS2d}
\end{align}

For small spot sizes ($\spot \ll 1$), an asymptotic bound of $S_{\rm max}^{2D}$ again emerges as quartic in $\spot$:
\begin{align}
S_{\rm max}^{2D} =1 - \frac{4}{\spot} \frac{(1 - \spot^3 / 3! + \spot^5 / 5!)^2}{[1 - (2\spot)^3 / 3! + (2\spot)^5 / 5!] + 2\spot}  = \frac{ \spot^4}{45}, \quad \spot \ll 1, \label{eq:optSsmall2d}
\end{align}
We can also derive the optimal effective current ($x$ denotes the aperture coordinate):
\begin{align}
\xi_{\textrm{opt}}^{2D}(x)=\frac{A e^{-ikz}}{\sqrt{z}}  \Bigg[ 1 - 4 \frac{\sin(\spot)}{\sin(2\spot) + 2\spot} \cos(\spot x) \Bigg] \tilde{\pol}.  \label{eq:optcurrent2d}
\end{align}
Defining the normalized coordinate on the focal plane as $\fspot = \frac{k R \rho}{z}$ (where $\rho$ is the focal-plane coordinate),
 the optimal field profile
\begin{align}
| \psi^{2D}(\fspot) | = \psi_0 \Bigg[ \frac{\sin(\fspot)}{\fspot}  - 4 \frac{\sin(\spot)}{\sin(2\spot) + 2\spot} \frac{\spot \sin(\spot)\cos(\fspot) - \fspot \cos(\spot)\sin(\fspot) }{\spot^2 - \fspot^2} \Bigg] \quad (\textrm{0 if $\fspot = \spot$}) \label{eq:optfield2d}
\end{align}
where $\psi_0$ is a normalization factor, resembles an Airy-like pattern (sinc function) near the diffraction limit, with increasing sidelobe relative to focal intensity for smaller spot sizes.

\section{Strehl ratio reconstruction of experimental results in Fig. 4(b)}
\begin{figure} [t!]
    \includegraphics[width=0.5\linewidth]{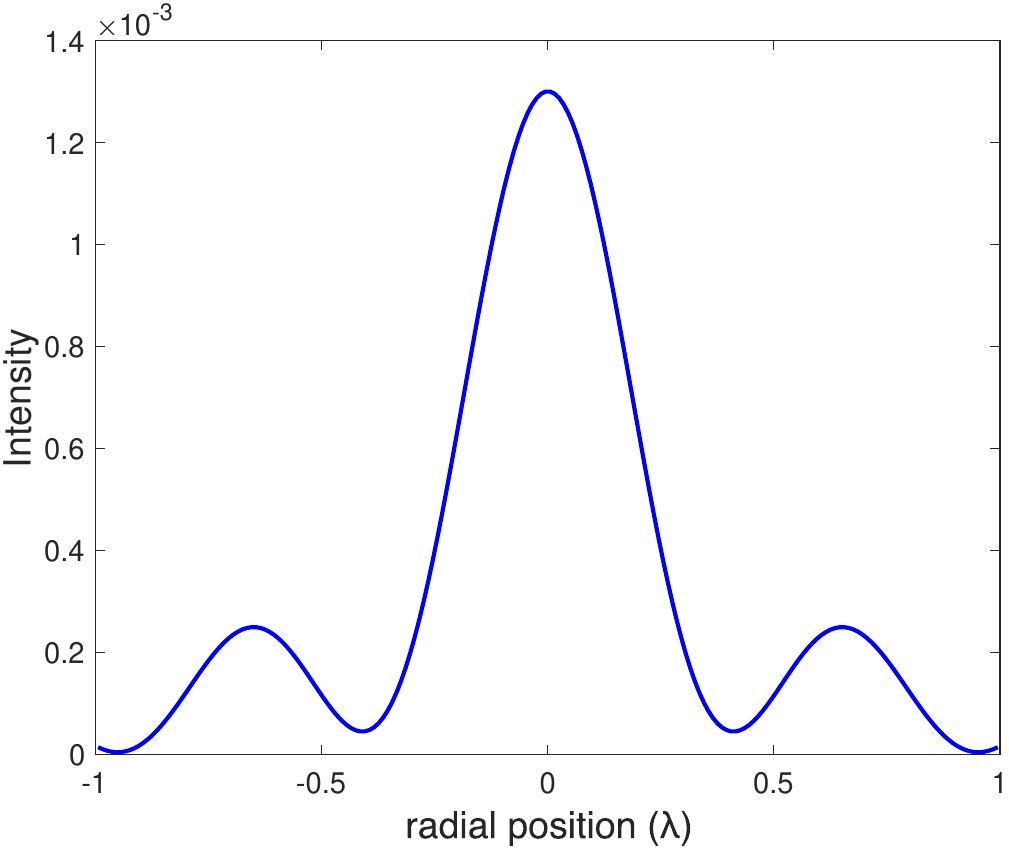} 
    \caption{Reconstructed intensity profile for the lens design in \citeasnoun{Rogers2013}, using effective currents to simulate the beam incident on the lens. The physical dimensions of the lens and focal distance are also fixed as in [ref], and our figure compares favorably with the experimental intensity profile in Fig. 5d of \citeasnoun{Rogers2013}. The intensity is normalized such that the peak intensity gives the Strehl ratio.}
    \label{fig:onsol} 
\end{figure}

To obtain the simulated results in Fig. 4b based on \citeasnoun{Dejuana2003,Gundu2005,Kosmeier2011,Rogers2012,Rogers2013,Qin2017,Dong2017}, we have reconstructed the various lens designs by discretizing an aperture with the same physical dimensions. The incident monochromatic waves passing through these lenses were then simulated by appropriately choosing the effective currents on the aperture (with $800\times800=640000$ equally spaced grid points, each grid allowing for transverse electric and magnetic current---longitudinal currents do not appear in the equivalence principle~\cite{Kong1975}). To calculate the Strehl ratio, we normalize the intensities from each lens by the unconstrained (no zero-field condition) maximum intensity from a circular aperture of the same radius. As an example, \figref{onsol} shows the reconstructed intensity profile for the lens in~\cite{Rogers2013} (scaled such that the peak intensity coincides with the Strehl ratio). There is close agreement in the spot sizes and peak widths (\figref{onsol} has full width at half maximum (FWHM) of 0.4$\lambda$) compared to the experimental result (Fig. 5d of \citeasnoun{Rogers2013}, FWHM of 0.42$\lambda$), the difference arising from discretization and experimental errors (and more pronounced further away from the peak, whose details are not important).

As a technical aside, most of these experiments are not carried out in the far zone and so one might be concerned about the effects of longitudinal polarization. However, the longitudinal fields do not contribute at the peak for these setups (but they can alter the overall intensity distribution across the focal plane). Hence, we need only consider the transverse (in-plane polarization) intensity for purposes of Strehl ratio calculation. To facilitate comparisons with the far-zone case, for which the longitudinal components vanish, we plot the transverse intensity in \figref{onsol}. Also, the normalized spot-size radius is modified to take the numerical aperture (NA) into account: $\spot = k \textrm{NA} \rho_0$, which reduces to the definition given in the main text ($\spot = k R \rho_0 / z$) in the far zone. However, numerical experiments suggest that our bound also holds in the near zone with an error of at most a few percent even for NA $\sim 0.5$. 

\section{Comparison of our bound to quartic transmission through a subwavelength hole}
We have shown in the main text that our bound on focal-point intensity reduces to the following expression for small spot sizes $\spot \ll 1$: 
\begin{align}
    S_{\rm max} = \spot^4 /192. \label{eq:smallS}
\end{align}

At first glance, \eqref{smallS} is related to the quartic dependence between the transmission through a subwavelength hole in an opaque screen and the hole size~\cite{PhysRev.66.163}. However, the physical origins are quite different. For the transmission through a subwavelength hole, the quartic dependence arises from Rayleigh scattering of small particles with size $a \ll \lambda$. Upon incident radiation, these particles behave as induced electric and magnetic dipoles, each scaling with amplitudes $\sim a^3$ (the subwavelength hole in Bethe’s paper can also be represented by such dipoles with same scaling behavior, as shown in Ref.~\cite[Sec.~6b]{PhysRev.66.163}).  Thus, the diffracted electric and magnetic fields in the far field of the hole will also scale as $\sim a^3$, resulting in the $a^6$ scaling behavior of the total scattered power/scattering cross section (characteristic of Rayleigh cross section). The $a^4$ transmission dependence then comes from normalizing the scattering cross section by the hole area (see Ref.~\cite{Genet2007} for example). 

On the other hand, the maximum focal intensity in \eqref{smallS} arises from the interference between constant-intensity fields radiating from a dipole at the origin and a current loop at the spot size radius. In the limit of small spot size, the first non-zero interference term in the fields is quadratic, and hence the maximum intensity is quartic in spot size. Moreover, the scattering phenomena of subwavelength apertures is dimension-dependent, whereas the quartic dependence of the focal-intensity scaling is dimension-independent (at least for one-dimensional, two-dimensional, and three-dimensional focusing). Our bounds on focal-intensity are captured by the dimensionless parameter $\spot = kR\rho_0 / z$---the actual spot-size radius $\rho_0$ is essentially normalized by the aperture radius $R$ and the aperture-focus distance $z$. 

\section{FWHM spot size and local wavevector normalization}
The diffraction-limited beam for the far zone scenario considered in the main text (circular aperture enclosing any shape of aperture) is an Airy disk. In order to render the FWHM spot size and local wavevector dimensionless (as used in Fig. 6 of the main text), we can normalize their values relative to those for the Airy peak. In this section, we provide explicit expressions for these normalization parameters.  

The Airy-pattern intensity on the focal plane, at radial position $\rho$, is given by~\cite{born2013principles} (normalized such that the intensity at the origin is 1): 
\begin{align}
I(\rho) = \left( \frac{2 J_1(kR\rho / z)}{kR\rho / z} \right)^2 \label{eq:airy}.
\end{align}
For convenience, we define the dimensionless quantity, $x = kR\rho / z $. We first focus on the FWHM spot size of the Airy disk (referring to it by the subscript $\textsubscript{\textrm{FWHM},A}$. By definition, 
we have that $I(x_{\textrm{FWHM},A}) = 1/2$, from which we can substitute \eqref{airy} to obtain the following equation:
\begin{align}
J_1(x_{\textrm{FWHM},A}) = x_{\textrm{FWHM},A}/{2\sqrt{2}}.
\end{align}
From the above equation, we obtain $x_{\textrm{FWHM},A} = 1.61633...$, which gives the diffraction-limited FWHM spot size of $\rho_{\textrm{FWHM},A}= \frac{1.61633...z}{kR}$.  

Similarly, we denote the local wavevector of the Airy disk by $k_{{\rm loc},A}$, whose value at the origin is given by:
\begin{align}
k_{{\rm loc},A}(\rho=0) = \sqrt {-\nabla^2 | I(x) |^2 \big\rvert_{0}}  \label{eq:klocdef}.
\end{align}
Since the Airy intensity is radially symmetric, the Laplacian operator reduces to $\frac{\mathrm{d}^2}{\mathrm{d}r^2} = \left(\frac{kR}{z}\right)^2 \frac{\mathrm{d}^2}{\mathrm{d}x^2}$ (the gradient term vanishes at the origin for the Airy disk). From \eqref{airy}, we obtain the following expression for the second derivative:  
\begin{align}
 \frac{\mathrm{d}^2 I}{\mathrm{d}x^2} =24 \frac{J_1(x)}{x^4} - 16 \frac{\left (J_0(x)-J_2(x)\right)J_1(x)}{x^3} + 2 \frac{\left (J_0(x)-J_2(x)\right)^2 + J_1(x)\left (J_3(x)-3 J_1(x)\right)}{x^2}
\end{align}
from which we obtain the Laplacian as $\lim_{x \to 0} \frac{\mathrm{d}^2 I}{\mathrm{d}x^2} = -\frac{1}{8} \left(\frac{kR}{z}\right)^2$. Upon using this result to \eqref{klocdef}, the local wavevector at the focal point for the Airy disk is $k_{{\rm loc},A}(x=0) = \frac{kR}{\sqrt{2}z}$.

\section{Small-but-nonzero field constraints}
In this section we aim to understand how the bounds change when the field at a specified spot size does not necessarily have to go to zero, but rather is small by some metric.

As discussed in the main text, the rank-two versions of all of the bounds that we derive can be written in the form
\begin{equation*}
\begin{aligned}
     & \underset{\psi}{\text{maximize}} & & \psi^\dagger \psi_0 \psi_0^\dagger \psi  \\
     & \text{subject to}               & & \psi^\dagger \psi_1 \psi_1^\dagger \psi = 0 \\
     & & & \psi^\dagger \psi = 1,
\end{aligned}
\end{equation*}
where the first-line term is the intensity at the origin at the second-line left-hand-side term is the total intensity on the spot-size contour. We showed in the main text that the optimum to this problem is attained for $\psi = v$, where 
\begin{align}
    v = \psi_0 - \frac{\psi_1^\dagger \psi_0}{|\psi_1|^2} \psi_1.
\end{align}
The vector $v$ is orthogonal to $\psi_1$ (it is the projection of $\psi$ onto $\psi_0$ in the subspace orthogonal to $\psi_1$), and the maximum intensity for the zero-field constraint, which we denote here as $I_0$, is given by
\begin{align}
    I_0 = v^\dagger v = \psi_0^\dagger \psi_0 - \frac{|\psi_0^\dagger \psi_1|^2}{|\psi_1|^2}.
\end{align}
In this case of a zero-field constraint, we did not have to be specific about the meaning of $\psi^\dagger \psi_1 \psi_1^\dagger \psi$; it could be the \emph{total} intensity integrated along the zero-field contour, it could have been an average, it could have been a discrete sum, etc.

Now that we want to allow for nonzero intensities, however, we must be specific about the precise value, and we will take $\psi^\dagger \psi_1 \psi_1^\dagger \psi$ to represent the \emph{average} intensity around the zero-field contour, which will allow for direct comparison to the intensity at the origin. (Note that this is a rescaled version of the $\psi_1$ that we compute in \eqref{psi1eq}.) For a physically intuition figure of merit, we consider a scenario in which we allow the average intensity along the zero-field contour to be some fraction $f$ of the originally optimal focusing intensity $I_0$, which would be represented by the constraint
\begin{align}
    \psi^\dagger \psi_1 \psi_1^\dagger \psi \leq f I_0.
\end{align}
We want to answer the following question: given this loosening of the constraint, how much can the focal intensity increase? We will see that the increase cannot be substantial, and thus that our zero-field results give realistic constraints even for nonzero constraints.

The new optimization problem is
\begin{equation}
\begin{aligned}
     & \underset{\psi}{\text{maximize}} & & \psi^\dagger \psi_0 \psi_0^\dagger \psi  \\
     & \text{subject to}               & & \psi^\dagger \psi_1 \psi_1^\dagger \psi \leq f I_0 \\
     & & & \psi^\dagger \psi = 1,
     \label{eq:optnew}
\end{aligned}
\end{equation}
We know that $\psi_0$ and $\psi_1$ are not orthogonal, and in fact are very nearly parallel. And that the optimal field with the original zero-field constraint, $v$, was orthogonal to $\psi_1$. Thus we can write
\begin{align}
    \psi_0 = a \psi_1 + v,
\end{align}
where the first term indicates the portion of $\psi_0$ that aligns with $\psi_1$ while the second term is the portion of $\psi_0$ orthogonal to $\psi_1$. If we assume that solving the unconstrained problem yields nearly equal intensities whether at the origin or averaged along the zero-field contour (very reasonable due to the homogeneity of free space), then $\psi_0^\dagger \psi_0 = \psi_1^\dagger \psi_1$, which implies that $|a| \lesssim 1$. We can bound the metric of \eqref{optnew} by
\begin{align}
    \psi^\dagger \psi_0 \psi_0^\dagger \psi &= \psi^\dagger (a\psi_1 + v)(a\psi_1+v)^\dagger \psi \nonumber \\
                                            &= |a|^2 \psi^\dagger \psi_1 \psi_1^\dagger \psi + 2 \Re \left[a \psi^\dagger \psi_1 v^\dagger \psi\right] + \psi^\dagger v v^\dagger \psi \nonumber \\
                                            &\leq |a|^2 f I_0  + 2 |a| \sqrt{\psi^\dagger \psi_1 \psi_1^\dagger \psi}\sqrt{\psi^\dagger v v^\dagger \psi}  + I_0 \nonumber \\ \nonumber 
                                            &\leq \left(|a|^2 f + 2|a|\sqrt{f} +1\right) I_0.
\end{align}
We see that the focusing intensity increases only linearly with the spot-size intensity. Thus, dramatic increases in focusing intensity are not possible simply by allowing small, nonzero spot-size fields. (For example, getting even a factor three intensity enhancement at the origin requires a spot-size field that is 50\% of the original \emph{focusing} intensity.)

\section{Target field constraints}
Instead of enforcing the field to vanish on some contour $\mathcal{C}$, we can also explore the bounds on focal intensity subject to a desired target field $\psi_0(\xv)$ on $\mathcal{C}$ produced by effective currents $\xi_0$. Framed as an optimization problem,
\begin{equation}
    \begin{aligned}
        & \underset{\xi}{\text{maximize}} & & I(\xv=0) = \xi^\dagger \tG_0^\dagger \tG_0 \xi \\
        & \text{subject to}       & &  \psi(\xv) \big\rvert_{\mathcal{C}} = \tG_\mcC \xi = \psi_0(\xv) \big\rvert_{\mathcal{C}} =   \tG_\mcC \xi_0  \quad \textrm{and} \quad \xi^\dagger \xi \leq 1. \label {eq:targetopt}
    \end{aligned}
\end{equation}
We can rewrite the equality constraint above in terms of the difference $x = \xi - \xi_0$ as $\tG_\mcC x = 0$. As the (differences of) currents $x$ satisfies the zero-field condition, we subsume the equality constraint by projecting $x$ onto the subspace of all currents that generate zero field ($\mathbf{P}$ is an orthogonal projection matrix that projects onto the null space of $\tG_\mcC$):
\begin{align}
x = \left(\mathbf{I} - \tG_\mcC^\dagger \left(\tG_\mcC \tG_\mcC^\dagger\right)^{-1} \tG_\mcC\right) z = \mathbf{P} z. \label {eq:projx}
\end{align}
Now, we can substitute \eqref{projx} into \eqref{targetopt} to obtain the following optimization problem, which is quadratic in the (complex) variable $z$:
\begin{equation}
    \begin{aligned}
        & \underset{\xi}{\text{maximize}} & & \left(\mathbf{P} z + \xi_0 \right)^\dagger \tG_0^\dagger \tG_0 \left(\mathbf{P} z + \xi_0 \right) \\
        & \text{subject to}       & &  \left(\mathbf{P} z + \xi_0 \right)^\dagger \left(\mathbf{P} z + \xi_0 \right) \leq 1   \label {eq:targetopt_red}
    \end{aligned}
\end{equation}
The objective function and the constraint above can also be written down explicitly in full quadratic form:
\begin{align}
\left(\mathbf{P} z + \xi_0 \right)^\dagger \tG_0^\dagger \tG_0 \left(\mathbf{P} z + \xi_0 \right) = z^\dagger  \mathbf{P} \tG_0^\dagger \tG_0 \mathbf{P} z + 2\Re \left( \xi_0^\dagger \tG_0^\dagger \tG_0  \mathbf{P} z \right) +  \xi_0^\dagger \tG_0^\dagger \tG_0 \xi_0,
\end{align}
\begin{align}
\left(\mathbf{P} z + \xi_0 \right)^\dagger \left(\mathbf{P} z + \xi_0 \right) = z^\dagger \mathbf{P} z + 2\Re \left( \xi_0^\dagger \mathbf{P} z \right) +  \xi_0^\dagger  \xi_0 \leq 1.
\end{align}
\Eqref{targetopt_red} is a complex-variable QCQP with one constraint.  As described in the main text, one can use semidefinite relaxation (SDR)~\cite{Luo2010} in optimization theory to obtain global bounds for the relaxed version of the original problem. In fact, for complex-variable quadratic problems with three or fewer constraints (as in our case here), SDR bounds are ``tight,'' i.e. attainable (strictly zero duality gap)~\cite{Huang2007b}. Hence, one can solve for the exact bound as well as the resulting field profile of \eqref{targetopt_red} in polynomial time.

\clearpage
\section{Table of permittivity values for metasurface designs in Fig. 5(b,c,d)}

\begin{center}
\begin{table}[htp]
\caption{Permittivity values for $G=0.21$ grayscale metalens design. Each grid point shown in this table occupies an area of $(\lambda/20)^2$. A mirror symmetry is applied between 100th-grid and 101st-grid.}
\begin{tabular}{ |c|c|c|c|c|c|c|c|c|c|c|c|c|c|c|c|c|c|c|c| } 
\hline
width$\setminus$height&1&2&3&4&5&6&7&8&9&10&11&12&13&14&15&16&17&18&19  \\
\hline
1 &  1.53&  1.61&  2.17&  2.17&  2.09&  1.77&  1.69&  2.25&  2.01&  2.01&  2.01&  1.93&  1.93&  1.93&  1.04&  2.81&  1.53&  1.37&  1.00\\ 
2 &  1.00&  1.00&  1.00&  2.25&  2.17&  2.09&  1.77&  2.25&  1.05&  2.01&  2.01&  2.01&  2.01&  1.04&  1.04&  2.01&  1.29&  1.45&  1.13\\ 
3 &  1.00&  1.00&  1.00&  1.00&  2.25&  2.17&  2.01&  3.29&  1.00&  2.01&  2.01&  2.01&  2.01&  2.01&  3.77&  1.00&  1.21&  1.45&  1.85\\ 
4 &  1.00&  1.00&  1.00&  1.00&  2.17&  2.17&  2.09&  4.49&  1.00&  1.29&  2.01&  2.01&  2.01&  2.01&  3.61&  2.57&  2.01&  2.01&  2.01\\ 
5 &  2.09&  3.61&  3.53&  2.25&  2.25&  2.25&  2.17&  2.97&  1.00&  1.00&  2.01&  2.09&  2.09&  2.17&  2.09&  2.09&  2.73&  2.09&  2.09\\ 
6 &  2.41&  2.41&  5.05& 10.33&  3.13&  4.25&  2.25&  2.25&  2.33&  2.49&  2.81&  2.89&  5.13&  3.45&  3.13&  2.41&  2.57&  2.65&  2.97\\ 
7 &  2.41&  2.49&  2.57&  6.97&  3.45&  3.53&  2.33&  2.41&  2.49&  2.49&  2.57&  2.65&  2.65&  4.89&  3.29&  2.49&  2.49&  2.57&  2.57\\
\hline
\end{tabular} 
\end{table}
\end{center}

\clearpage
\begin{center}
\begin{tabular}{ |c|c|c|c|c|c|c|c|c|c|c|c|c|c|c|c|c|c|c|c| } 
\hline
width$\setminus$height&1&2&3&4&5&6&7&8&9&10&11&12&13&14&15&16&17&18&19  \\
\hline
51 &  2.01&  2.01&  2.01&  1.77&  1.69&  1.37&  1.04&  1.13&  1.29&  2.01&  2.01&  2.01&  1.77&  2.09&  2.65&  2.57&  1.93&  2.01&  2.01\\ 
52 &  2.01&  2.01&  2.01&  1.93&  1.93&  1.00&  1.00&  1.00&  1.04&  1.77&  2.01&  2.01&  2.25&  2.97&  2.73&  1.04&  1.21&  1.53&  2.01\\ 
53 &  2.01&  2.01&  2.01&  2.01&  2.01&  1.00&  1.00&  1.00&  1.00&  1.21&  2.01&  2.01&  2.01&  2.09&  6.65&  1.77&  1.04&  1.04&  2.01\\ 
54 &  2.09&  2.01&  2.01&  2.01&  2.17&  1.77&  1.77&  1.13&  1.00&  1.00&  1.53&  2.01&  2.01&  2.01&  1.00&  3.61&  2.25&  1.00&  1.00\\ 
55 &  2.09&  2.09&  2.01&  2.01&  2.33&  1.77&  1.93&  2.01&  1.00&  1.00&  2.17&  2.01&  2.01&  2.01&  1.00&  1.00&  1.00&  1.00&  1.45\\ 
56 &  2.01&  2.09&  2.01&  2.41&  2.57&  2.17&  2.09&  2.25&  2.09&  1.00&  2.01&  2.01&  2.01&  1.04&  1.13&  1.00&  1.00&  1.00&  1.69\\ 
57 &  1.53&  1.00&  2.01&  2.33&  2.41&  3.77&  2.65&  3.61&  2.25&  2.09&  2.49&  3.69&  2.01&  1.00&  1.00&  2.01&  1.04&  1.04&  1.04\\
\hline
\end{tabular} 
\end{center}

\clearpage
\begin{center}
\begin{tabular}{ |c|c|c|c|c|c|c|c|c|c|c|c|c|c|c|c|c|c|c|c|c| } 
\hline
width$\setminus$height&20&21&22&23&24&25&26&27&28&29&30&31&32&33&34&35&36&37&38&39  \\
\hline
1 &  2.01&  2.01&  2.01&  1.04&  1.13&  1.45&  1.53&  1.37&  1.13&  1.00&  1.00&  1.00&  1.05&  1.69&  2.01&  2.09&  2.25&  2.33&  2.33&  2.41\\ 
2 &  2.01&  2.01&  2.01&  1.93&  1.37&  1.53&  1.37&  1.61&  1.69&  1.00&  1.00&  1.00&  1.53&  1.77&  2.97&  3.85&  2.41&  2.25&  2.25&  2.17\\ 
3 &  2.01&  2.01&  2.01&  2.89&  1.29&  1.53&  2.25&  2.01&  1.13&  1.00&  1.00&  1.00&  1.69&  2.17&  2.41&  2.41&  2.41&  2.17&  2.01&  2.01\\ 
4 &  2.01&  2.01&  2.01&  2.01&  2.17&  3.13&  2.17&  1.77&  1.00&  1.85&  1.00&  1.00&  1.77&  2.25&  2.41&  2.49&  2.41&  2.01&  2.01&  1.93\\ 
5 &  2.09&  2.09&  3.13&  2.17&  2.01&  2.81&  2.65&  2.17&  2.01&  1.53&  1.00&  1.00&  2.01&  2.25&  2.33&  2.33&  2.17&  2.01&  1.93&  2.17\\ 
6 &  2.73&  2.65&  2.49&  2.33&  2.33&  2.41&  3.05&  3.05&  2.81&  1.00&  1.00&  1.29&  2.25&  2.33&  2.33&  2.25&  1.36&  1.77&  2.17&  2.25\\ 
7 &  2.49&  2.49&  2.49&  2.49&  2.49&  2.41&  2.41&  2.41&  2.41&  2.09&  2.25&  2.41&  2.33&  2.33&  2.25&  2.89&  1.93&  1.04&  2.33&  2.33\\
\hline
\end{tabular} 
\end{center}

\clearpage
\begin{center}
\begin{tabular}{ |c|c|c|c|c|c|c|c|c|c|c|c|c|c|c|c|c|c|c|c|c| } 
\hline
width$\setminus$height&20&21&22&23&24&25&26&27&28&29&30&31&32&33&34&35&36&37&38&39  \\
\hline
51 &  2.01&  2.01&  4.81&  2.57&  2.57&  1.93&  2.01&  2.01&  2.01&  2.01&  2.65&  2.49&  2.25&  2.33&  2.01&  2.01&  2.01&  4.09&  1.61&  1.00\\ 
52 &  2.01&  2.01&  2.33&  2.65&  2.65&  2.41&  2.17&  2.01&  2.01&  2.01&  1.53&  1.77&  2.09&  2.01&  2.01&  2.81&  3.29&  9.21&  3.21&  2.17\\ 
53 &  2.01&  2.01&  2.17&  4.17&  4.65&  2.17&  2.09&  2.01&  2.01&  2.01&  1.04&  2.09&  2.09&  2.01&  2.01&  2.01&  2.01&  2.01& 12.09&  3.29\\ 
54 &  2.01&  2.01&  1.13&  1.04&  2.33&  2.01&  2.01&  2.01&  2.01&  1.69&  2.49&  2.33&  1.61&  2.01&  2.01&  2.01&  2.01&  2.01&  2.09&  5.37\\ 
55 &  1.04&  1.93&  1.04&  1.21&  2.01&  2.01&  2.01&  2.01&  2.01&  2.33&  3.53&  2.41&  2.73&  2.01&  2.01&  2.01&  2.01&  2.01&  2.09&  2.01\\ 
56 &  1.00&  1.53&  1.05&  1.05&  2.01&  2.01&  2.01&  2.01&  2.01&  2.01&  2.65&  3.77&  2.57&  1.77&  2.01&  2.01&  2.01&  2.01&  2.01&  2.01\\ 
57 &  1.00&  2.01&  1.85&  1.04&  2.01&  2.01&  2.01&  2.01&  2.01&  2.01&  2.01&  4.73&  2.49&  1.37&  2.01&  2.01&  2.01&  2.01&  2.01&  2.01\\
\hline
\end{tabular} 
\end{center}

\begin{center}
\begin{table}
\caption{Permittivity values for for $G=0.52$ metalens design. Each grid point shown in this table occupies an area of $(\lambda/20)^2$. A mirror symmetry is applied between 230th-grid and 231st-grid. The value ``1'' means permittivity value of 1.44 while the value ``0'' means free space.}
\begin{tabular}{ |c|c|c|c|c|c|c|c|c|c|c|c|c|c|c|c|c|c|c|c|c|c|c|c|c|c|c|c|c|c|c|c|c|c|c|c|c|c|c|c| } 
\hline
width$\setminus$height&1&2&3&.&.&.&.&.&.&.&.&.&.&.&.&.&.&.&.&.&.&.&.&.&.&.&.&.&.&.&.&.&.&.&.&.&.&.&39   \\
\hline
1 &1&1&1&1&1&1&1&1&1&1&1&1&1&1&1&1&1&1&1&1&1&1&1&1&1&1&1&1&1&1&1&1&1&1&1&1&1&1&1\\ 
2 &1&1&1&1&1&1&1&1&1&1&1&1&1&1&1&1&1&1&1&1&1&1&1&1&1&1&1&1&1&1&1&1&1&1&1&1&1&1&1\\ 
3 &1&1&1&1&1&1&1&1&1&1&1&1&1&1&1&1&1&1&1&1&1&1&1&1&1&1&1&1&1&1&1&1&1&1&1&1&1&1&1\\ 
4 &1&1&1&1&1&1&1&1&1&1&1&1&1&1&1&1&1&1&1&1&1&1&1&1&1&1&1&1&1&1&1&1&1&1&1&1&1&1&1\\ 
5 &1&1&1&1&1&1&1&1&1&1&1&1&1&1&1&1&1&1&1&1&1&1&1&1&1&1&1&1&1&1&1&1&1&1&1&1&1&1&1\\ 
6 &1&1&1&1&1&1&1&1&1&1&1&1&1&1&1&1&1&1&1&1&1&1&1&1&1&1&1&1&1&1&1&1&1&1&1&1&1&1&1\\ 
7 &1&1&1&1&1&1&1&1&1&1&1&1&1&1&1&1&1&1&1&0&0&1&1&1&1&1&1&1&1&1&1&1&1&1&1&1&1&1&1\\
\hline
\end{tabular} 
\end{table}
\end{center}

\begin{center}
\begin{tabular}{ |c|c|c|c|c|c|c|c|c|c|c|c|c|c|c|c|c|c|c|c|c|c|c|c|c|c|c|c|c|c|c|c|c|c|c|c|c|c|c|c| } 
\hline
width$\setminus$height&1&2&3&.&.&.&.&.&.&.&.&.&.&.&.&.&.&.&.&.&.&.&.&.&.&.&.&.&.&.&.&.&.&.&.&.&.&.&39   \\
\hline
57 &0&0&0&0&0&0&0&0&0&0&0&0&0&0&0&0&0&0&0&0&0&0&0&0&0&0&0&0&0&0&0&0&0&0&0&0&0&0&0\\ 
58 &0&0&0&0&0&0&0&0&0&0&0&0&0&0&0&0&0&0&0&0&0&0&0&0&0&0&0&0&0&0&0&0&0&0&0&0&0&0&0\\ 
59 &0&0&0&0&0&0&0&0&0&0&0&0&0&0&0&0&0&0&0&0&0&0&0&0&0&0&0&0&0&0&0&0&0&0&0&0&0&0&0\\ 
60 &0&0&0&0&0&0&0&0&0&0&0&0&0&0&0&0&0&0&0&0&0&0&0&0&0&0&0&0&0&0&0&0&0&0&0&0&0&0&0\\ 
61 &0&0&0&0&0&0&0&0&0&0&0&0&0&0&0&0&0&0&0&0&0&0&0&0&0&0&0&0&0&0&0&0&0&0&0&0&0&0&0\\ 
62 &0&0&0&0&0&0&0&0&0&0&0&0&0&0&0&0&0&0&0&0&0&0&0&0&0&0&0&0&0&0&0&0&0&0&0&0&0&0&0\\ 
63 &0&0&0&0&0&0&0&0&0&0&0&0&0&0&0&0&0&0&0&0&0&0&0&0&0&0&0&0&0&0&0&0&0&0&0&0&0&0&0\\
\hline
\end{tabular} 
\end{center}

\begin{center}
\begin{tabular}{ |c|c|c|c|c|c|c|c|c|c|c|c|c|c|c|c|c|c|c|c|c|c|c|c|c|c|c|c|c|c|c|c|c|c|c|c|c|c|c|c| } 
\hline
width$\setminus$height&1&2&3&.&.&.&.&.&.&.&.&.&.&.&.&.&.&.&.&.&.&.&.&.&.&.&.&.&.&.&.&.&.&.&.&.&.&.&39   \\
\hline
115 &1&1&1&1&1&1&1&1&1&1&1&1&1&1&1&1&1&1&1&1&1&1&1&1&1&1&1&1&1&1&1&1&1&1&1&1&1&1&1\\ 
116 &1&1&1&1&1&1&1&1&1&1&1&1&1&1&1&1&1&1&1&1&1&1&1&1&1&1&1&1&1&1&1&1&1&1&1&1&1&1&1\\ 
117 &1&1&1&1&1&1&1&1&1&1&1&1&1&1&1&1&1&1&1&1&1&1&1&1&1&1&1&1&1&1&1&1&1&1&1&1&1&1&1\\ 
118 &1&1&1&1&1&1&1&1&1&1&1&1&1&1&1&1&1&1&1&1&1&1&1&1&1&1&1&1&1&1&1&1&1&1&1&1&1&1&1\\ 
119 &1&1&1&1&1&1&1&1&1&1&1&1&1&1&1&1&1&1&1&1&1&1&1&1&1&1&1&1&1&1&1&1&1&1&1&1&1&1&1\\ 
120 &1&1&1&1&1&1&1&1&1&1&1&1&1&1&1&1&1&1&1&1&1&1&1&1&1&1&1&1&1&1&1&1&1&1&1&1&1&1&1\\ 
121 &1&1&1&1&1&1&1&1&1&1&1&1&1&1&1&1&1&1&1&1&1&1&1&1&1&1&1&1&1&1&1&1&1&1&1&1&1&1&1\\
\hline
\end{tabular} 
\end{center}

\begin{center}
\begin{tabular}{ |c|c|c|c|c|c|c|c|c|c|c|c|c|c|c|c|c|c|c|c|c|c|c|c|c|c|c|c|c|c|c|c|c|c|c|c|c|c|c|c| } 
\hline
width$\setminus$height&1&2&3&.&.&.&.&.&.&.&.&.&.&.&.&.&.&.&.&.&.&.&.&.&.&.&.&.&.&.&.&.&.&.&.&.&.&.&39   \\
\hline
175 &0&0&0&0&0&0&0&0&0&0&0&0&0&0&0&0&0&0&0&0&0&0&0&0&0&0&0&0&0&0&0&0&0&0&0&0&0&0&0\\ 
176 &0&0&0&0&0&0&0&0&0&0&0&0&0&0&0&0&0&0&0&0&0&0&0&0&0&0&0&0&0&0&0&0&0&0&0&0&0&0&0\\ 
177 &0&0&0&0&0&0&0&0&0&0&0&0&0&0&0&0&0&0&0&0&0&0&0&0&0&0&0&0&0&0&0&0&0&0&0&0&0&0&0\\ 
178 &0&0&0&0&0&0&0&0&0&0&0&0&0&0&0&0&0&0&0&0&0&0&0&0&0&0&0&0&0&0&0&0&0&0&0&0&0&0&0\\ 
179 &0&0&0&0&0&0&0&0&0&0&0&0&0&0&0&0&0&0&0&0&0&0&0&0&0&0&0&0&0&0&0&0&0&0&0&0&0&0&0\\ 
180 &0&0&0&0&0&0&0&0&0&0&0&0&0&0&0&0&0&0&0&0&0&0&0&0&0&0&0&0&0&0&0&0&0&0&0&0&0&0&0\\ 
181 &0&0&0&0&0&0&0&0&0&0&0&0&0&0&0&0&0&0&0&0&0&0&0&0&0&0&0&0&0&0&0&0&0&0&0&0&0&0&0\\
\hline
\end{tabular} 
\end{center}

\begin{center}
\begin{table}
\caption{Permittivity values for $G=0.985$ metalens design. Each grid point shown in this table occupies an area of $(\lambda/20)^2$. A mirror symmetry is applied between 230th-grid and 231st-grid. The value ``1'' means permittivity value of 1.69 while the value ``0'' means free space.}
\begin{tabular}{ |c|c|c|c|c|c|c|c|c|c|c|c|c|c|c|c|c|c|c|c|c|c|c|c|c|c|c|c|c|c|c|c|c|c|c|c|c|c|c|c| } 
\hline
width$\setminus$height&1&2&3&.&.&.&.&.&.&.&.&.&.&.&.&.&.&.&.&.&.&.&.&.&.&.&.&.&.&.&.&.&.&.&.&.&.&.&39 \\
\hline
1 &0&0&0&0&1&1&0&0&0&1&1&1&1&1&0&0&0&0&0&0&0&0&0&0&0&0&0&0&0&0&0&0&0&0&0&0&0&0&0\\ 
2 &0&1&1&1&1&1&1&1&1&1&1&1&1&1&1&1&1&1&1&0&0&1&1&0&1&1&0&0&1&1&1&1&1&1&1&1&1&1&0\\ 
3 &0&1&1&1&1&1&1&1&1&1&1&1&1&1&1&1&1&0&0&1&1&1&0&0&0&1&1&0&1&1&1&1&1&1&1&1&1&1&0\\ 
4 &1&0&1&1&1&1&0&1&1&1&1&1&1&1&1&1&1&0&0&0&0&0&0&0&1&1&1&1&0&1&1&1&1&1&1&1&1&1&0\\ 
5 &1&1&1&1&0&0&1&0&0&0&1&1&1&1&0&0&0&0&0&0&0&0&0&1&1&1&1&1&1&0&0&1&1&1&1&1&1&1&0\\ 
6 &1&1&0&0&0&0&0&0&0&0&1&1&0&0&0&0&0&0&0&0&0&0&0&1&1&1&0&0&0&0&0&1&1&1&1&1&1&1&1\\ 
7 &1&1&0&0&0&0&0&0&0&0&0&0&0&0&0&0&0&0&0&0&0&0&1&1&1&0&0&0&0&0&0&1&1&1&1&1&1&1&1\\
\hline
\end{tabular} 
\end{table}
\end{center}

\begin{center}
\begin{tabular}{ |c|c|c|c|c|c|c|c|c|c|c|c|c|c|c|c|c|c|c|c|c|c|c|c|c|c|c|c|c|c|c|c|c|c|c|c|c|c|c|c| } 
\hline
width$\setminus$height&1&2&3&.&.&.&.&.&.&.&.&.&.&.&.&.&.&.&.&.&.&.&.&.&.&.&.&.&.&.&.&.&.&.&.&.&.&.&39   \\
\hline
57 &1&1&1&1&1&1&1&1&1&1&1&1&1&1&1&1&1&1&1&1&1&1&1&1&1&1&1&1&1&1&1&1&1&1&1&1&1&1&1\\ 
58 &1&1&1&1&1&1&1&1&1&1&1&1&1&1&1&1&1&1&1&1&1&1&1&1&1&1&1&1&1&1&1&1&1&1&1&1&1&1&1\\ 
59 &1&1&1&1&1&1&1&1&1&1&1&1&1&1&1&1&1&1&1&1&1&1&1&1&1&1&1&1&1&1&1&1&1&1&1&1&1&1&1\\ 
60 &1&1&1&1&1&1&1&1&1&1&1&1&1&1&1&1&1&1&1&1&1&1&1&1&1&1&1&1&1&1&1&1&1&1&1&1&1&1&1\\ 
61 &1&1&1&1&1&1&1&1&1&1&1&1&1&1&1&1&1&1&1&1&1&1&1&1&1&1&1&1&1&1&1&1&1&1&1&1&1&1&1\\ 
62 &1&1&1&1&1&1&1&1&1&1&1&1&1&1&1&1&1&1&1&1&1&1&1&1&1&1&1&1&1&1&1&1&1&1&1&1&1&1&1\\ 
63 &1&1&1&1&1&1&1&1&1&1&1&1&1&1&1&1&1&1&1&1&1&1&1&1&1&1&1&1&1&1&1&1&1&1&1&1&1&1&1\\
\hline
\end{tabular} 
\end{center}

\begin{center}
\begin{tabular}{ |c|c|c|c|c|c|c|c|c|c|c|c|c|c|c|c|c|c|c|c|c|c|c|c|c|c|c|c|c|c|c|c|c|c|c|c|c|c|c|c| } 
\hline
width$\setminus$height&1&2&3&.&.&.&.&.&.&.&.&.&.&.&.&.&.&.&.&.&.&.&.&.&.&.&.&.&.&.&.&.&.&.&.&.&.&.&39   \\
\hline
115 &1&1&1&1&1&1&1&0&0&1&0&0&0&0&0&0&0&0&0&0&0&0&1&1&0&0&0&0&0&1&0&0&0&0&0&0&1&1&0\\ 
116 &1&1&1&1&1&1&1&1&0&0&0&0&0&0&0&0&0&0&0&0&0&0&1&0&0&0&0&0&0&1&0&0&1&0&1&0&1&1&0\\ 
117 &1&1&1&1&1&1&1&1&1&0&0&0&0&0&0&0&0&0&0&0&0&0&0&0&0&0&0&0&0&1&1&0&1&1&1&1&1&1&0\\ 
118 &0&1&1&0&0&0&1&1&1&0&0&0&0&0&0&0&0&0&0&0&0&0&0&0&0&0&0&0&0&1&1&1&1&1&1&1&0&1&0\\ 
119 &0&1&1&0&0&0&0&0&0&0&0&0&0&0&0&1&1&0&0&0&0&0&0&0&0&0&0&0&0&0&1&1&1&1&1&1&0&0&0\\ 
120 &0&0&0&0&0&0&0&0&0&0&0&0&0&0&0&1&1&0&0&0&0&0&0&0&0&0&0&0&0&1&1&1&1&1&1&1&0&0&0\\ 
121 &0&0&0&0&0&0&0&0&0&0&0&0&0&0&0&0&0&0&0&0&0&0&0&0&0&0&1&1&0&0&0&0&1&0&0&0&0&0&0\\
\hline
\end{tabular} 
\end{center}

\begin{center}
\begin{tabular}{ |c|c|c|c|c|c|c|c|c|c|c|c|c|c|c|c|c|c|c|c|c|c|c|c|c|c|c|c|c|c|c|c|c|c|c|c|c|c|c|c| } 
\hline
width$\setminus$height&1&2&3&.&.&.&.&.&.&.&.&.&.&.&.&.&.&.&.&.&.&.&.&.&.&.&.&.&.&.&.&.&.&.&.&.&.&.&39   \\
\hline
174 &1&1&1&1&1&1&1&0&0&0&0&1&1&1&1&1&0&0&0&0&0&0&0&0&0&0&0&0&0&0&0&0&0&1&1&1&1&1&1\\ 
175 &0&0&0&1&1&1&1&1&0&1&1&1&1&1&1&1&0&0&0&0&0&0&0&1&1&1&1&1&1&0&0&0&1&1&1&1&1&1&1\\ 
176 &1&0&1&1&1&1&1&1&0&0&0&0&0&1&1&1&1&0&0&0&0&0&1&1&1&1&1&1&1&1&1&1&1&1&1&1&1&1&1\\ 
177 &1&1&1&1&1&1&1&1&1&0&0&1&1&1&1&1&1&1&0&0&0&1&1&1&1&1&1&0&0&0&1&1&1&1&1&1&0&0&0\\ 
178 &1&1&1&1&1&1&1&1&1&1&1&1&1&1&1&1&1&1&1&1&1&1&1&1&1&1&1&1&1&1&1&1&1&1&1&1&1&1&1\\ 
179 &1&1&1&1&1&1&1&1&1&1&1&1&1&1&1&1&1&1&1&1&1&1&1&1&1&1&1&1&1&1&1&1&1&1&1&1&1&1&1\\ 
180 &1&1&1&1&1&1&1&1&0&0&0&0&0&0&1&1&0&0&0&0&0&0&1&1&1&0&0&1&1&1&1&1&1&1&1&1&1&1&1\\
\hline
\end{tabular} 
\end{center}

\providecommand{\noopsort}[1]{}\providecommand{\singleletter}[1]{#1}%